%% file: MACSJ0717_kSZ.tex
\documentclass[twocolumn,traditabstract]{aa}  

\usepackage{fixltx2e}
\usepackage[english]{babel}
\usepackage{graphicx,amsmath}
\usepackage{epstopdf}
\usepackage{epsf,color}
\usepackage[mathscr]{eucal}
\usepackage{amsmath}
\usepackage{amssymb,amsfonts}
\usepackage{natbib}
\usepackage{graphicx}
\usepackage{txfonts}
\usepackage{dsfont}
\definecolor{Mygreen}{rgb}{0.75, 0.0, 0.0}
\definecolor{Mypink}{rgb}{1.0, 0.0, 0.5}
\definecolor{Myred}{rgb}{0.7, 0.0, 0.0}
\usepackage[breaklinks, citecolor=blue, linkcolor=Myred, urlcolor=Myred, colorlinks=true, debug, baseurl=' ']{hyperref}
\usepackage{float} 
\usepackage{color}

\bibpunct{(}{)}{;}{a}{}{,}
\begin{document}
\title{Mapping the kinetic Sunyaev-Zel'dovich effect toward \mbox{MACS~J0717.5+3745} with NIKA}
\input{listeauthors}

\date{Received \today \ / Accepted --}
\abstract {Measurement of the gas velocity distribution in galaxy clusters provides insight into the physics of mergers, through which large scale structures form in the Universe. Velocity estimates within the intracluster medium (ICM) can be obtained via the Sunyaev-Zel'dovich (SZ) effect, but its observation is challenging both in term of sensitivity requirement and control of systematic effects, including the removal of contaminants. In this paper we report resolved observations, at 150 and 260~GHz, of the SZ effect toward the triple merger \mbox{MACS~J0717.5+3745} ($z=0.55$), using data obtained with the NIKA camera at the IRAM 30m telescope. Assuming that the SZ signal is the sum of a thermal (tSZ) and a kinetic (kSZ) component and by combining the two NIKA bands, we extract for the first time a resolved map of the kSZ signal in a cluster. The kSZ signal is dominated by a dipolar structure that peaks at $-5.1$ and $+3.4 \sigma$, corresponding to two subclusters moving respectively away and toward us and coincident with the cold dense X-ray core and a hot region undergoing a major merging event. We model the gas electron density and line-of-sight velocity of \mbox{MACS~J0717.5+3745} as four subclusters. Combining NIKA data with X-ray observations from XMM-{\it Newton} and {\it Chandra}, we fit this model to constrain the gas line-of-sight velocity of each component, and we also derive, for the first time, a velocity map from kSZ data (i.e. that is model-dependent). Our results are consistent with previous constraints on the merger velocities, and thanks to the high angular resolution of our data, we are able to resolve the structure of the gas velocity. Finally, we investigate possible contamination and systematic effects with a special care given to radio and submillimeter galaxies. Among the sources that we detect with NIKA, we find one which is likely to be a high redshift lensed submillimeter galaxy.}
\titlerunning{kSZ mapping toward MACS~J0717.5+3745}
\authorrunning{R. Adam, I. Bartalucci, G.W. Pratt et al.}
\keywords{Techniques: high angular resolution -- Galaxies: clusters: individual: \mbox{MACS~J0717.5+3745}; intracluster medium}
\maketitle

\section{Introduction}\label{sec:Introduction}
The assembly of massive clusters of galaxies represents the last step of the hierarchical formation of structures in the Universe. Galaxy clusters arise from smooth accretion of surrounding material, but also from the merging of subclusters and groups, at the intersection of filaments in the cosmic web \citep[see][for a review]{Kravtsov2012}. Merging clusters are not only useful as large scale astrophysical laboratories \citep[e.g.,][]{Clowe2006}, but are important from a cosmological point of view \citep[e.g.,][]{Voit2005}. The physical processes involved in mergers significantly affect the cluster observables via the distribution of dark matter, the physics of the intracluster plasma, and the dynamics of galaxies \citep[e.g.,][]{Ferrari2005,Maurogordato2011}, i.e., all the components of the clusters. These observables are in turn essential to allow us to infer the masses of clusters when used as a cosmological probe \citep[see][for a review]{Allen2011}. In this context, measuring the collision velocity of merging systems provides insights of the cluster physics, such as shocks in the intracluster medium (ICM), the kinematics of the mergers, the hydrodynamical interactions between cluster cores and less dense shock heated regions, and the relation between the baryonic matter and the dark matter. In addition, such velocity measurements provide a complementary cosmological probe to constrain structure formation \citep{Bhattacharya2007} and the study of the properties of the velocity field is a subject of active research \citep[e.g.,][using numerical simulations]{Hahn2015}. 

Measuring the peculiar velocities of galaxy clusters is a difficult task, in particular when looking at objects beyond the local Universe. Indeed, classical methods that are based on redshift surveys allow us to reconstruct the internal galaxy kinematics of clusters, but require an independent distance measurement to subtract the Hubble flow at objects' positions in order to measure their true peculiar velocities \citep[e.g.,][]{Tully1977}. They are therefore limited to small redshifts as the errors scale with distance. 

The Sunyaev-Zel'dovich effect \citep[SZ,][]{Sunyaev1972}, on the other hand, provides a way to probe the ICM gas bound in clusters without suffering from cosmological dimming \citep{birkinshaw1999,carlstrom2002,kitayama2014}. Its observation is therefore independent of redshift as long as the angular resolution of the observations is sufficient to resolve the clusters. We can distinguish between the thermal SZ effect \citep[tSZ,][]{Sunyaev1972}, for which the cosmic microwave background (CMB) photons are spectrally distorted by the electronic thermal pressure of the ICM, and the kinetic SZ effect \citep[kSZ,][]{Sunyaev1980}, arising from the CMB Doppler shift induced by the bulk motion of the cluster electrons. While the tSZ effect is related to the integrated pressure along the line-of-sight and can be used as a mass proxy in cosmological studies \citep[see, e.g.,][]{Planck2015XXIV,Bleem2015,Hasselfield2013}, the kSZ effect is sensitive to the integrated line-of-sight electron density and the gas velocity with respect to the CMB reference frame. The kSZ signal is subdominant to the tSZ unless the gas velocity reaches a few tenths of a percent of the speed of light \citep[1000 km/s, see e.g.,][]{birkinshaw1999}. In addition, the spectral dependence of the kSZ is the same as that of the CMB, so that they can only be separated from each other using the difference in spatial distribution of the respective signals. Direct observation of the kSZ signal within clusters therefore requires both high sensitivity and high angular resolution observations \citep{Haehnelt1996}, which makes it particularly challenging to detect in individual clusters (see attempts on individual clusters by e.g., \citealt{Holzapfel1997,Benson2003,Sayers2015} and the first clear detection on a statistical sample by \citealt{Hand2012}). 

The cluster of galaxies \mbox{MACS~J0717.5+3745} at $z=0.55$ is a striking example of a merging system and is an excellent target with which to measure the kSZ effect. A detailed review of the previous analyses of \mbox{MACS~J0717.5+3745} can be found in \cite{Mroczkowski2012} and \cite{Sayers2013}, and we summarize some essential features here. The cluster is located at the end of a filamentary structure \citep[e.g.,][]{Ebeling2004,Jauzac2012} and appears to be an extremely disturbed system. It is composed of four optically-identified main subclusters \citep{Ma2009}, which coincide with peaks in the observed strong lensing surface mass distribution \citep[see e.g.,][]{Zitrin2009,Diego2015,Limousin2015,Kawamata2016}. The X-ray emitting gas is very disturbed \citep[e.g.,][]{Ebeling2007,Ma2009} and the spectroscopic temperature reaches up to $\gtrsim 20$ keV \citep{Ma2009}, indicating a scenario where two objects are undergoing a major merging event. \cite{Edge2003} discovered the presence of powerful radio emission near the high temperature region, indicating the presence of non-thermal component \citep[see also, e.g.,][]{vanWeeren2009}. \mbox{MACS~J0717.5+3745} was also observed using 30 GHz interferometric SZ data by the Sunyaev-Zel'dovich Array (SZA), indicating that the cluster is a hot and massive system \citep{LaRoque2003}. \cite{Mroczkowski2012} reported high angular resolution SZ observations at 90 GHz using MUSTANG (The MUltiplexed SQUID/TES Array at Ninety GHz) at the Green Bank Telescope (GBT). The data revealed small scale structures in the pressure distribution, as expected from a merger scenario.

The velocity distribution of \mbox{MACS~J0717.5+3745} was first studied using optical spectroscopy \citep{Ma2009}, showing that the galaxy groups have exceptionally large relative line-of-sight velocities, providing a more detailed picture of the merger kinematics. Using Bolocam SZ data obtained at the Caltech Submillimeter Observatory (CSO) at 140 and 268~GHz, \cite{Mroczkowski2012} reported the first indication of the presence of kSZ signal in \mbox{MACS~J0717.5+3745}. \cite{Sayers2013} then reported the first significant detection of the kSZ signal in a single cluster, \mbox{MACS~J0717.5+3745}, after collecting additional data with Bolocam and refining the former analysis. As the Bolocam beams are larger than the target subclusters, the limited angular resolution of these data is a factor impacting the kSZ detection significance \citep[see][for more details]{Sayers2013}.

In this paper, we report deep, high angular resolution observations ($<20$ arcsec) of \mbox{MACS~J0717.5+3745} at 150 and 260~GHz, obtained at the Institut de Radio Astronomie Millim\'etrique (IRAM) 30m telescope using the New IRAM KID Arrays \citep[NIKA,][]{Monfardini2011,Calvo2013,Adam2014,Catalano2014}. The resolved nature of the observations allows us for a robust treatment of the contaminating astrophysical signal and the data are used to produce a map of the kSZ signal toward the cluster. Combining X-ray spectroscopic temperature measurements with NIKA, we also fit a parametric model to constrain the gas line-of-sight velocity distribution.

The paper is organized as follows. In Section \ref{sec:sunyaev_Zel_dovich_observations_and_data_reduction}, we present the SZ observations of \mbox{MACS~J0717.5+3745}. The removal of the astrophysical contamination is presented and discussed in Section \ref{sec:astrophysical_contamination_of_the_Sunyaev_Zel_dovich_signal}. In Section \ref{sec:X_ray_data_reduction}, we detail the X-ray data reduction. In Section \ref{sec:A_map_of_the_kinetic_Sunyaev_Zel_dovich_signal}, we present the mapping of the kSZ effect in \mbox{MACS~J0717.5+3745} and we use our data together with X-ray temperatures to constrain the gas line-of-sight velocity in Section \ref{sec:constraint_on_the_gas_line_of_sight_velocity_distribution_of_MACSJ0717}. Throughout this paper we assume a flat $\Lambda$CDM cosmology according to the latest {\it Planck} results \citep{Planck2015XIII} with $H_0 = 67.8$ km s$^{-1}$ Mpc$^{-1}$, $\Omega_M = 0.308$, and $\Omega_{\Lambda} = 0.692$.

\section{Sunyaev-Zel'dovich observations and data reduction}\label{sec:sunyaev_Zel_dovich_observations_and_data_reduction}
\subsection{The Sunyaev-Zel'dovich effect}
The SZ effect is the spectral distortion of the CMB due to the inverse Compton scattering of CMB photons off energetic electrons in galaxy clusters. Only the thermal and the kinetic SZ effects have been observed so far (induced by a thermal electron population and the bulk motion of electrons with respect to the CMB, respectively), but in principle the SZ effect can also arise from the CMB interaction with more exotic electron populations, such as those originating from dark matter annihilation \citep{Colafrancesco2004} or ultra-relativistic electrons accelerated in radio lobes \citep{Colafrancesco2008}. These other contributions are expected to be small and, in this paper, we assume that only the tSZ and the kSZ effects contribute to the SZ signal.

The tSZ effect follows the characteristic frequency dependence given by \citep{birkinshaw1999}
\begin{equation}
	f(x,T_e) = \frac{x^4 e^x}{\left(e^x - 1\right)^2} \left(x \ {\rm coth} \left(\frac{x}{2}\right) - 4\right) \left(1+\delta_{\rm tSZ}(x,T_e) \right),
\label{eq:f_x_tsz}
\end{equation}
where $x=\frac{h \nu}{k_{\rm B} T_{\rm CMB}}$ is the dimensionless frequency, with $\nu$ the observing frequency, $h$ Planck's constant, $k_{\rm B}$ Boltzmann's constant, and the CMB temperature $T_{\rm CMB} = 2.725$ K \citep{Fixsen2009}. The term $\delta_{\rm tSZ}(x,T_e)$ corresponds to relativistic corrections, which depend on the observing frequency and the electron temperature $T_e$, and which we compute using the work of \cite{Itoh2003}. These corrections are valid at a level of $\lesssim 5$\% up to 50 keV within the NIKA bands.

The spectral dependance of the kSZ effect follows that of the CMB in the non relativistic regime. It is given by \citep{birkinshaw1999}
\begin{equation}
	g(x, v_z, T_e) = \frac{x^4 e^x}{\left(e^x - 1\right)^2} \left(1+\delta_{\rm kSZ}(x,v_z,T_e) \right),
\label{eq:g_x_ksz}
\end{equation}
where $v_z$ is the gas line-of-sight velocity (positive for a cluster receding from the observer) and $\delta_{\rm kSZ}$ are the relativistic corrections that depend on the observing frequency, $T_e$, and $v_z$. We use the analytical formula provided by \cite{Nozawa2006} to compute relativistic corrections, which we expect to be more accurate than $\sim 10$\% up to 30 keV in the NIKA bands. The velocity dependence on these corrections is negligible because $v_z$ is always much smaller than the speed of light.

\begin{table}[!ht]
\caption{\footnotesize{Spectral conversion coefficients. The units are Jy/beam per unit of $y_{\rm tSZ}$ and Jy/beam per unit of $y_{\rm kSZ}$ for $I_0 \ f(\nu)$ and $I_0 \ g(\nu)$, respectively. The frequencies are $\nu_1 = 260$ GHz and $\nu_2 = 150$ GHz. The uncertainty on the NIKA bandpasses is about 2\% \citep[see][for more details]{Adam2014} and are included in the overall calibration budget (see Section \ref{sec:NIKA_observations_and_data_reduction}). The coefficients given here account for the atmospheric absorption based on the \cite{Pardo2001} model for 2 mm precipitable water vapor above the telescope, which slightly changes the effective NIKA bandpasses.}}
\begin{center}
\begin{tabular}{ccccc}
\hline
\hline
$T_e$ (keV) & $I_0 \ f(\nu_1)$ & $I_0 \ f(\nu_2)$ & $I_0 \ g(\nu_1)$ & $I_0 \ g(\nu_2)$ \\
\hline
1   &  3.76 & -11.63 &  7.47 & 12.40 \\
5   &  3.31 & -11.34 &  7.30 & 12.03 \\
10  &  2.83 & -11.00 &  7.11 & 11.62 \\
15  &  2.43 & -10.71 &  6.94 & 11.24 \\
20  &  2.06 & -10.38 &  6.81 & 10.90 \\
25  &  1.76 & -10.17 &  6.75 & 10.58 \\
\hline
\end{tabular}
\end{center}
\label{tab:sz_coefficients}
\end{table}

The observed change of specific intensity expressed as a function of $\nu$, $\Delta I_{\nu}$, with respect to the CMB, $I_0$, is then given by 
\begin{equation}
	\frac{\Delta I_{\nu}}{I_0} = f(\nu, T_e) \ y_{\rm tSZ} + g(\nu, T_e, v_z) \ y_{\rm kSZ},
\label{eq:dIsz}
\end{equation}
where the amplitude of the tSZ and the kSZ signal are given by
\begin{equation}
	y_{\rm tSZ} = \frac{\sigma_{\rm T}}{m_e c^2} \int P_e dl \equiv \frac{k_{\rm B} T_e}{m_e c^2} \tau
\label{eq:ytsz}
\end{equation}
and
\begin{equation}
	y_{\rm kSZ} = \sigma_T \int \frac{-v_z}{c} n_e dl \equiv \frac{-v_z}{c} \tau,
\label{eq:yksz}
\end{equation}
respectively. The parameter $\sigma_{\rm T}$ is the Thomson cross section, $m_e$ is the electron rest mass, $c$ is the speed of light, and $n_e$ is the electronic gas density. The kSZ and tSZ amplitudes can also be expressed using the optical depth, $\tau$, as given in equations \ref{eq:ytsz} (assuming constant temperature along the line-of-sight) and \ref{eq:yksz} (assuming constant velocity along the line-of-sight). As the ICM is a dilute medium, we assume the ideal gas law to hold and express the electronic pressure as $P_e = n_e k_{\rm B} T_e$. The functions $f(\nu, T_e)$ and $g(\nu, T_e, v_z)$ are integrated over the NIKA bands, and they are listed in Table \ref{tab:sz_coefficients} for a few electronic gas temperature values, accounting for the NIKA beam solid angle to express them in terms of Jy/beam. While the tSZ signal is always negative at 150~GHz and positive at 260~GHz, the kSZ signal can be either positive or negative, depending on the sign of the velocity along the line-of-sight, but its sign is the same at both NIKA frequencies. The kSZ maximum corresponds to the null of the tSZ signal and is located around 217~GHz if relativistic effects are negligible.

\subsection{NIKA observations and data reduction}\label{sec:NIKA_observations_and_data_reduction}
\begin{figure*}[h]
\centering
\includegraphics[width=0.49\textwidth]{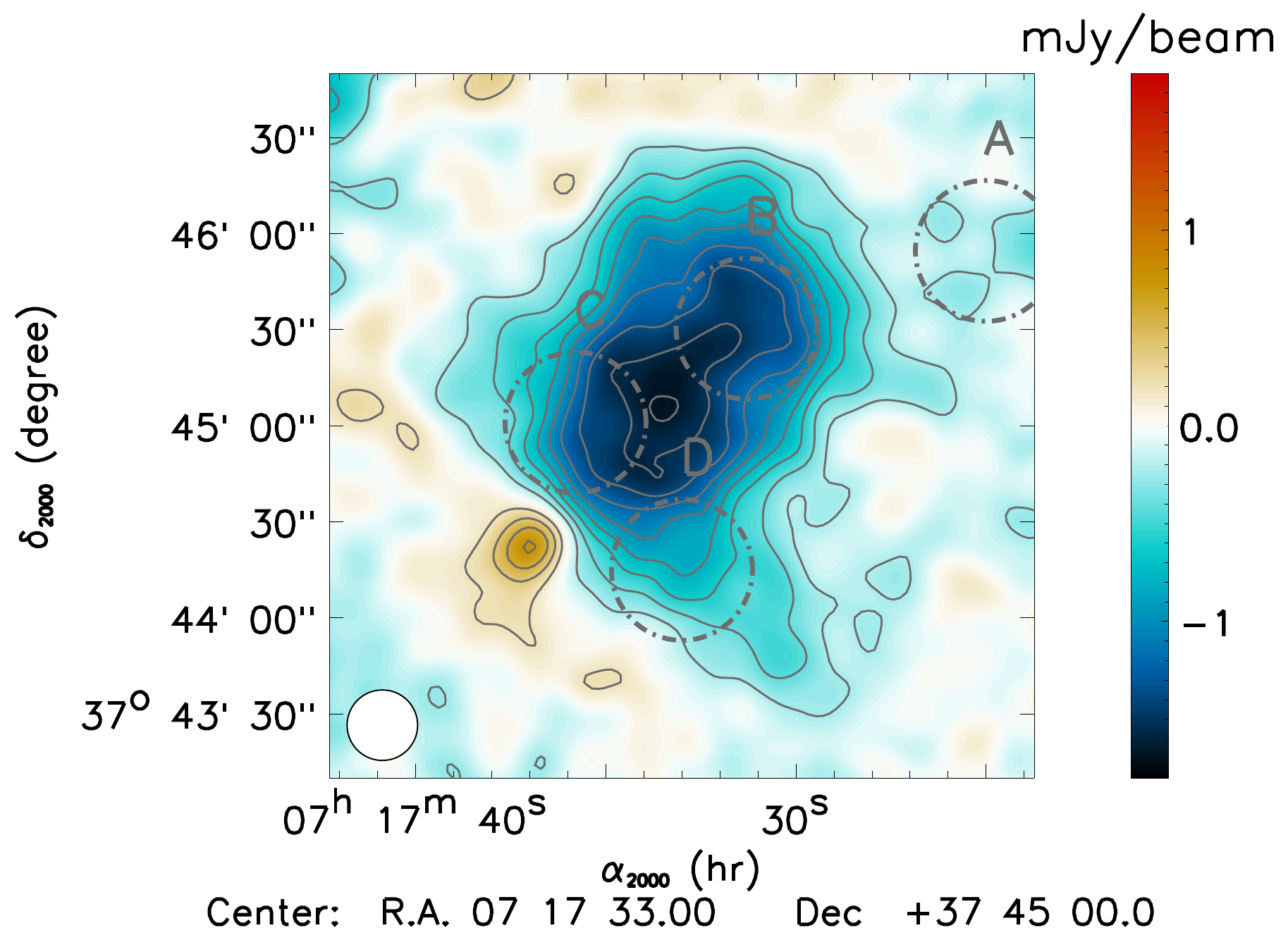}
\includegraphics[width=0.49\textwidth]{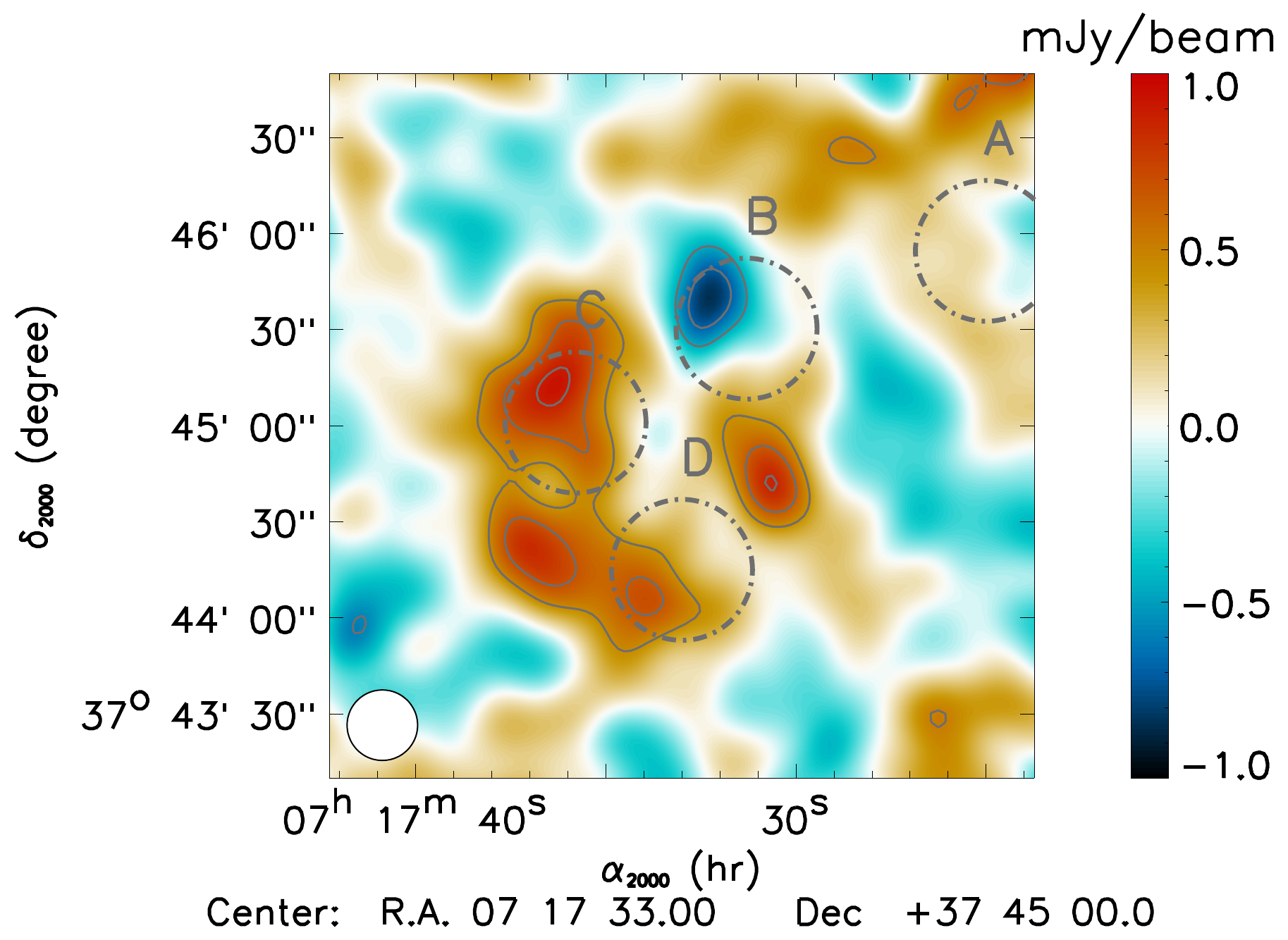}
\includegraphics[width=0.49\textwidth]{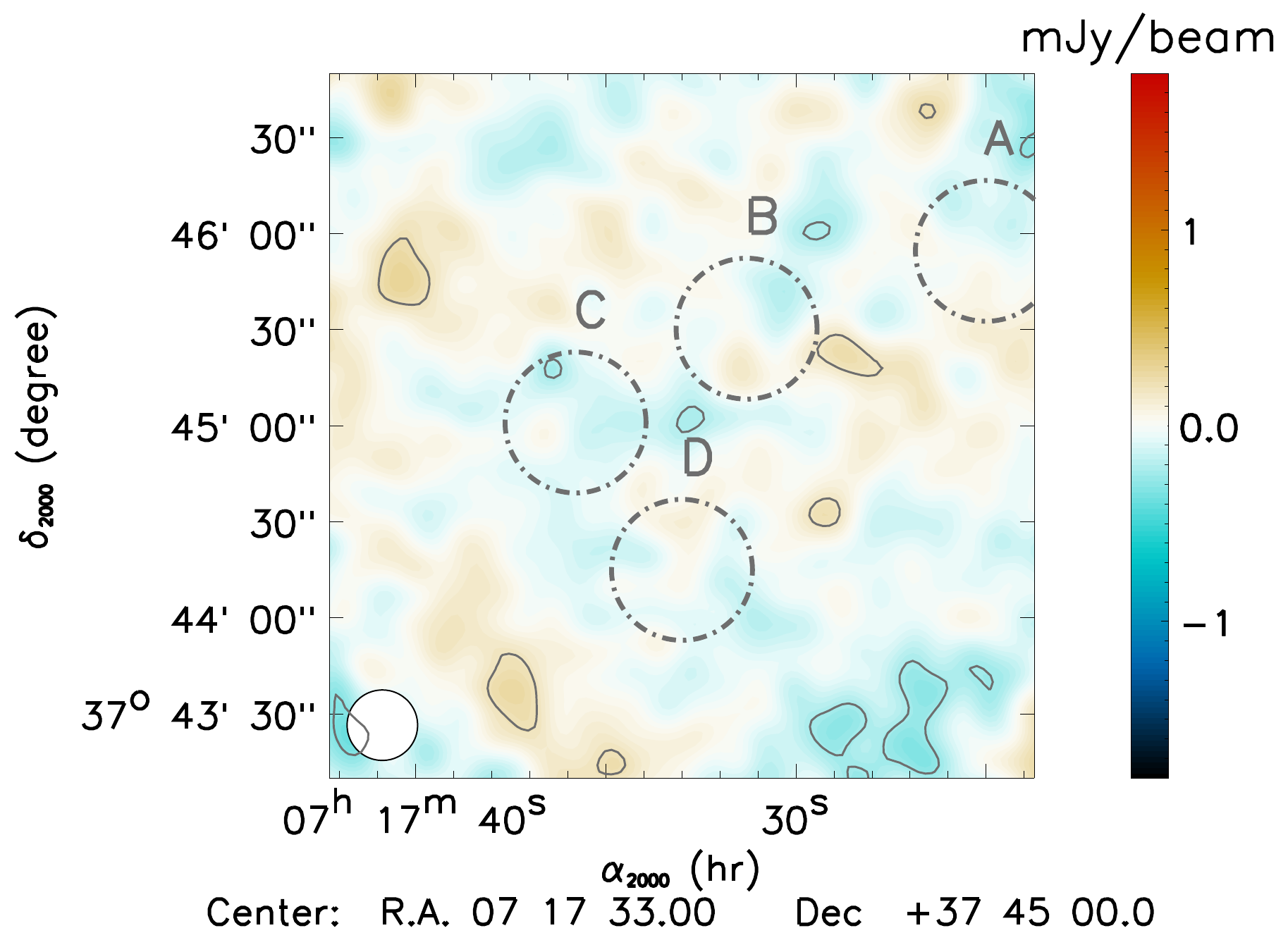}
\includegraphics[width=0.49\textwidth]{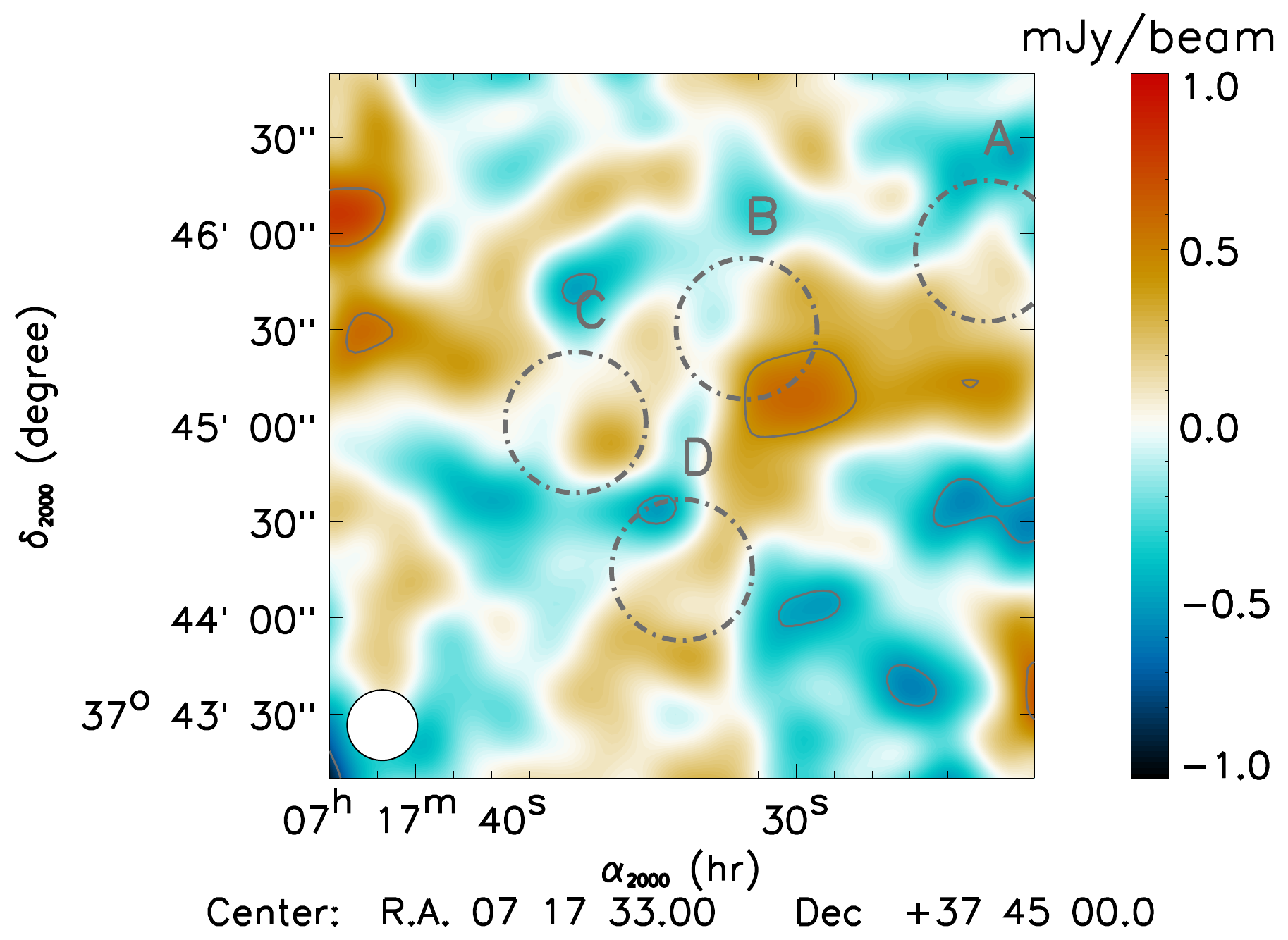}
\caption{\footnotesize{{\bf Top:} NIKA maps of \mbox{MACS~J0717.5+3745} at 150~GHz (left) and 260~GHz (right). The gray contours show the significance in units of standard deviation. They are multiples of $2 \sigma$ at 150~GHz and $1 \sigma$ at 260~GHz, starting at $\pm 2 \sigma$. Both maps have been smoothed to have the same effective resolution of 22 arcsec FWHM, as represented by the white circle on the bottom left corner of the maps. The regions defined in Table \ref{tab:coord_region} are shown as gray dot-dashed circles. {\bf Bottom:} Same as top-panel figures, in the case of half difference maps computed from two equivalent sub-samples. The steps between contours is $1 \sigma$ for both frequencies.}}
\label{fig:NIKA_raw_maps}
\end{figure*}
\mbox{MACS~J0717.5+3745} was observed for 7.26 hours in February 2014 and 5.85 hours in January and February 2015. Most of the observations took place under good conditions, with a mean zenith opacity of 0.116 at 260~GHz and 0.085 at 150~GHz, and a stable atmosphere. The pointing center was (R.A., Dec.)$_{\rm J2000}$ = (07:17:32.3, +37:44:47) for the 2014 data set and (R.A., Dec.)$_{\rm J2000}$ = (07:17:32.3, +37:45:10) for 2015. The scanning strategy was the same as that employed for other NIKA clusters as detailed in \cite{Adam2015,Adam2016}. Similarly, the detailed calibration procedure can be found in \cite{Adam2014,Adam2015}. This results in a root mean square pointing error of $\lesssim 3$ arcsec, the absolute calibration uncertainties are 7\% at 150 GHz and 12\% at 260~GHz, and the Gaussian beams FWHM were measured to be 18.2 and 12.0 arcsec at 150 and 260~GHz, respectively.

The data were reduced similarly and independently for the two bands, as described in \cite{Adam2015}. The removal of the correlated noise across the arrays, due mainly to the atmospheric contribution and correlated electronic noise, leads to the filtering of the astrophysical signal at scales that are larger than the NIKA field of view ($\sim 2$ arcmin). The resulting transfer function was computed using simulations. It is close to unity at small scales and vanishes smoothly at scales larger than the field of view. Apart from beam smoothing, the transfer function is the same at 150 and 260~GHz, which allows for direct combination of the two maps. The absolute zero level brightness of the maps is not measured by NIKA. In practice, the effective zero level is computed during the mapmaking procedure by setting the mean of the time ordered data to zero, taken in regions that are outside the cluster, flagged in an iterative manner. The flagged cluster region is based on the 150 GHz map only because of the higher signal-to-noise ratio, and corresponds to the $5 \sigma$ contours of Figure \ref{fig:NIKA_raw_maps}, top left panel. The beam efficiency correction as a function of elevation was optimized for extended emission \citep{Greve1998}, which can lead to up to $\sim 10$\% loss in the flux of point sources for very high or very low elevations. However, most of the data where taken between elevations of 35 and 62 degrees, with a mean of 49 degrees, so we expect this bias to be less than a few percent and thus assume it is negligible.

The instrumental noise and the atmospheric noise residuals are estimated as detailed in \cite{Adam2016}. In brief, we compute noise maps from the difference between two equivalent subsamples. Examples of such jackknife maps are provided in Figure \ref{fig:NIKA_raw_maps}. We can observe that the noise is correlated over spatial scales that are larger than the beam size, in particular at 260 GHz. This can give rise to spurious signal at the scale of the subclusters, and the method we use allows us to account for such correlations as detailed below. The noise maps are normalized by the integration time spent per pixel in order to generate homogeneous noise in the observed sky patch. The homogeneous noise is then used to compute its power spectra \cite[using the POKER software,][]{Ponthieu2011}, which is well described by a white component plus a pixel--pixel correlated contribution. We also check that the noise is Gaussian, and we do not observe deviations from gaussianity. This noise model allows us to simulate noise maps that account for the noise inhomogeneity and spatial correlations, and to compute the full noise covariance matrix from noise Monte Carlo realizations. The signal-to-noise ratio of the surface brightness maps are computed, at a given angular resolution (22 arcsec FWHM in the following), by dividing the signal by the root mean square of the noise realizations smoothed to the resolution of the signal map, and thus accounting for spatial correlations of the noise. The noise power spectrum is thus compared to the one of the signal, at the scales of the subclusters (after the filtering of the power on small scales), to infer the signal-to-noise ratio. We also estimate the cross-correlation between the 150 and 260~GHz noise using the cross spectra of the difference maps. We observe a small correlation due to the atmospheric noise residuals, which we expect because NIKA observes the sky simultaneously at 150 and 260~GHz. The corresponding correlated noise between the two bands corresponds to 0.22\% of the variance at 260~GHz, and 1.25\% of that at 150~GHz. However, this is negligible with respect to the cross correlation induced by the cosmic infrared background (CIB) contamination (see Section \ref{sec:Diffuse_galactic_emission_and_cosmological_background}) and we do not account for it.

\subsection{NIKA raw maps}\label{sec:NIKA_raw_maps}
The NIKA maps are shown in Figure \ref{fig:NIKA_raw_maps}. They have been smoothed with a Gaussian filter to have the same effective resolution of 22 arcsec FWHM. Due to the integration time spent per pixel, the noise is relatively constant within a radius of 1.5 arcmin around the map center, and increases radially toward the edges. At the 22 arcsec effective resolution, the noise root mean square at the center of the field is 97 $\mu$Jy/beam at 150~GHz and 226 $\mu$Jy/beam at 260~GHz, which accounts for instrumental noise, residual atmospheric noise, and the astrophysical noise as discussed in Section \ref{sec:Diffuse_galactic_emission_and_cosmological_background}. The root mean square noise on the flux of a point source located near the center is 0.14 mJy at 150~GHz and 0.58 mJy at 260~GHz. The positions of the four subcluster regions that we consider, A, B, C and D, are also shown in Figure \ref{fig:NIKA_raw_maps}. Various coordinates have been provided in the literature, obtained for example by fitting dark matter profiles on strong lensing data \citep[e.g.,][]{Limousin2012,Limousin2015}. Since we are interested in the gas distribution rather than the dark matter, we use the best-fit cored-component model coordinates of \cite{Limousin2015} as a starting point. These are adjusted slightly to match the peak distribution of the X-ray \textit{Chandra} wavelet smoothed map, which is used as a prior in determining the gas distribution (see Section \ref{sec:Spectroscopy_analysis} and the top right panel of Figure \ref{fig:Xray_all_maps}). These coordinates (see Table \ref{tab:coord_region}) are still consistent with the strong lensing analysis within the dispersion obtained from the different models by \cite{Limousin2015}. Unless otherwise stated, they are used only for display purpose in this paper and the detailed choice of these centers has no impact on the results that we present.

\begin{table}[]
\caption{\footnotesize{Definition of the location of the four subclusters in \mbox{MACS~J0717.5+3745}. These regions are defined using \cite{Limousin2015} strong lensing fit and the \textit{Chandra} X-ray surface brightness (see also Figure \ref{fig:tSZ_kSZ_multiL}). In the following, we use a radius of 22 arcsec for illustration.}}
\begin{center}
\begin{tabular}{ccc}
\hline
\hline
Subcluster & R.A. & Dec. \\
\hline
A & +07:17:25.0 & +37:45:54.6 \\
B & +07:17:31.3 & +37:45:30.3 \\
C & +07:17:35.8 & +37:45:01.0 \\
D & +07:17:33.0 & +37:44:15.0 \\
\hline
\end{tabular}
\end{center}
\label{tab:coord_region}
\end{table}

The 150~GHz map shows the presence of a diffuse negative signal associated with the SZ effect. The signal is dominated by the tSZ signal at this frequency, but the kSZ signal is also significant (see Section \ref{sec:A_map_of_the_kinetic_Sunyaev_Zel_dovich_signal}). The signal is non-zero, but very small in the direction of subcluster A, in agreement with its expected low gas content \citep[see][]{Ma2009}. The brightness in the direction of all other subclusters is significant with a peak between regions B and C, reaching $-18.4 \sigma$. However, we do not observe a peaked signal in the direction of each group, but rather a diffuse signal that does not allow the clear identification of the subclusters using this map. We also report the detection of a foreground radio galaxy on the southeast edge of the cluster (see Section \ref{sec:Radio_sources}).

At 260~GHz, we observe a diffuse positive signal around region C reaching up to $4.3 \sigma$ at the peak, but the surface brightness is negative in region B, with a peak at $-3.7 \sigma$. The signal-to-noise ratio is insufficient to detect the signal in region A at 260~GHz. We also detect three compact sources at this frequency (see Section \ref{sec:astrophysical_contamination_of_the_Sunyaev_Zel_dovich_signal}). Due to the relatively high astrophysical contamination observed in the 260~GHz map, its detailed interpretation requires the removal and the treatment of the point sources as will be discussed in Section \ref{sec:astrophysical_contamination_of_the_Sunyaev_Zel_dovich_signal}. Nevertheless, we observe an excess of signal around region C with respect to the 150~GHz tSZ expectation (see Table \ref{tab:sz_coefficients} for conversions between tSZ and surface brightness at 150 and 260 GHz), and a negative signal in region B that counters the positive tSZ signal expected at this frequency. While the former could be attributed to another astrophysical signal, it is not the case for the latter, which we will interpret as a kSZ signature in Section \ref{sec:A_map_of_the_kinetic_Sunyaev_Zel_dovich_signal}.

\section{Astrophysical contamination of the Sunyaev-Zel'dovich signal}\label{sec:astrophysical_contamination_of_the_Sunyaev_Zel_dovich_signal}
In addition to the SZ signal from the cluster itself, we expect galactic, extragalactic, and cluster associated astrophysical signal to contaminate our data. This Section describes how such contaminants are taken into account in the present paper.

\begin{figure*}[h]
\centering
\includegraphics[width=0.33\textwidth]{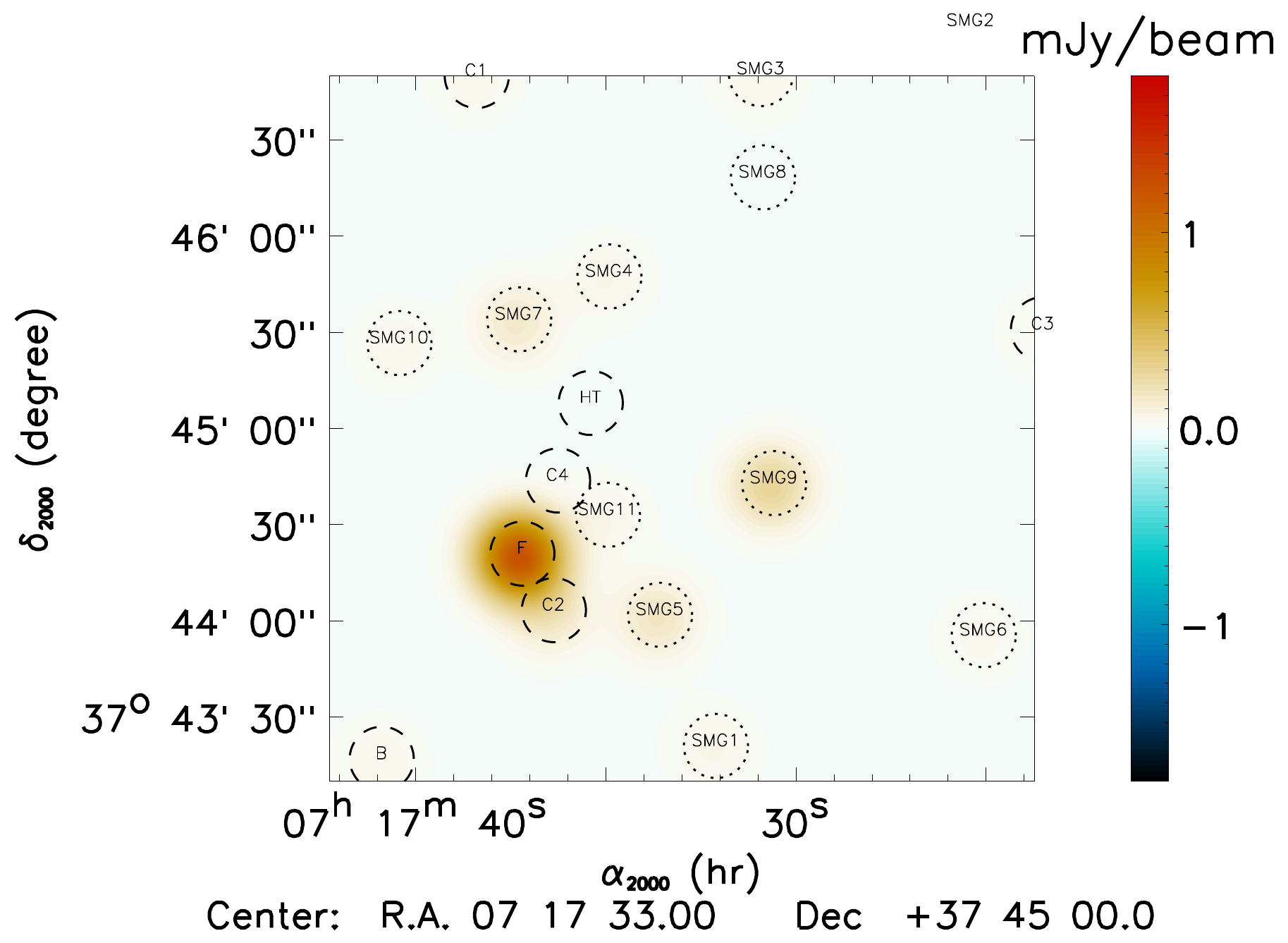}
\includegraphics[width=0.33\textwidth]{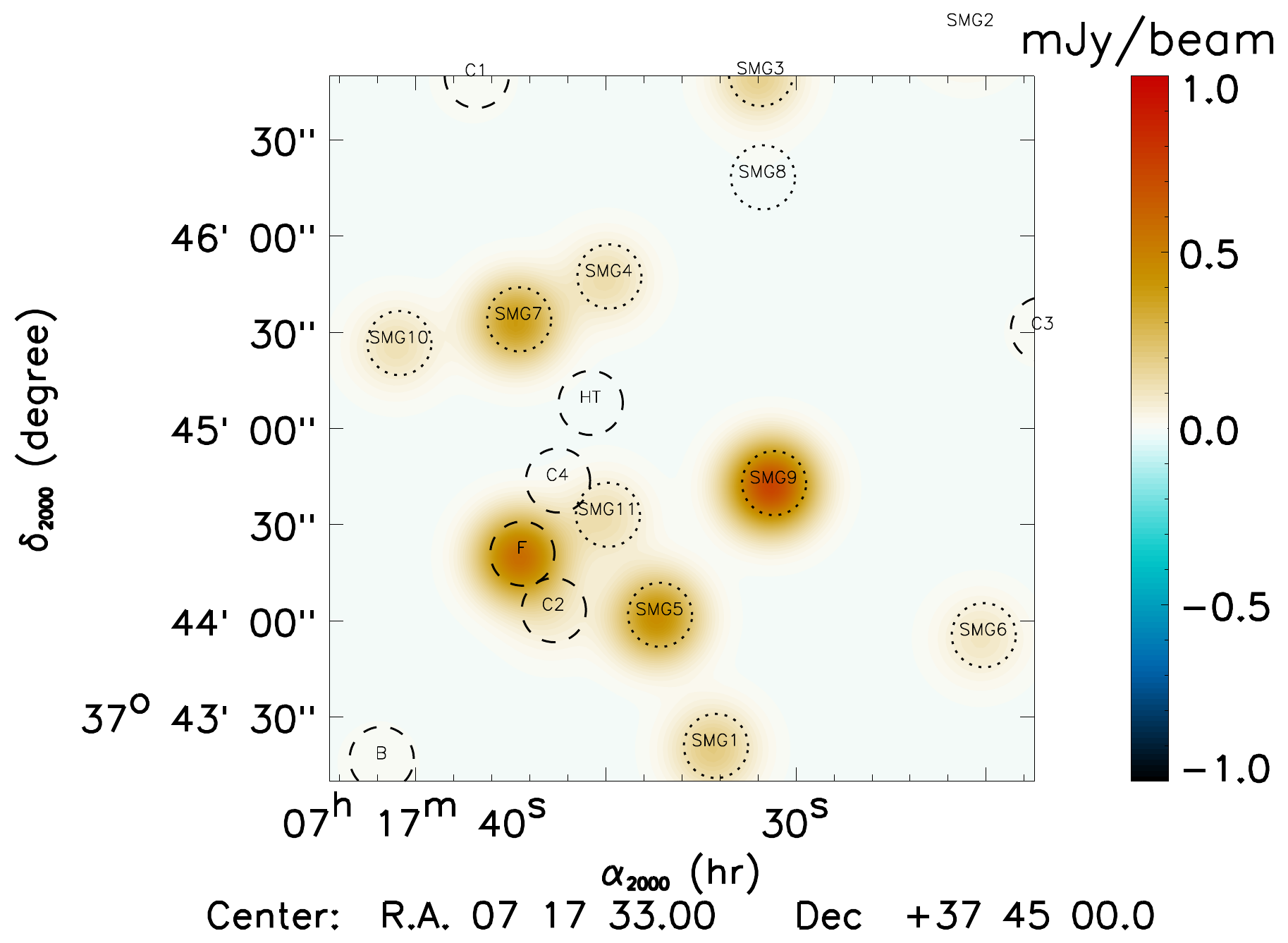}
\includegraphics[width=0.33\textwidth]{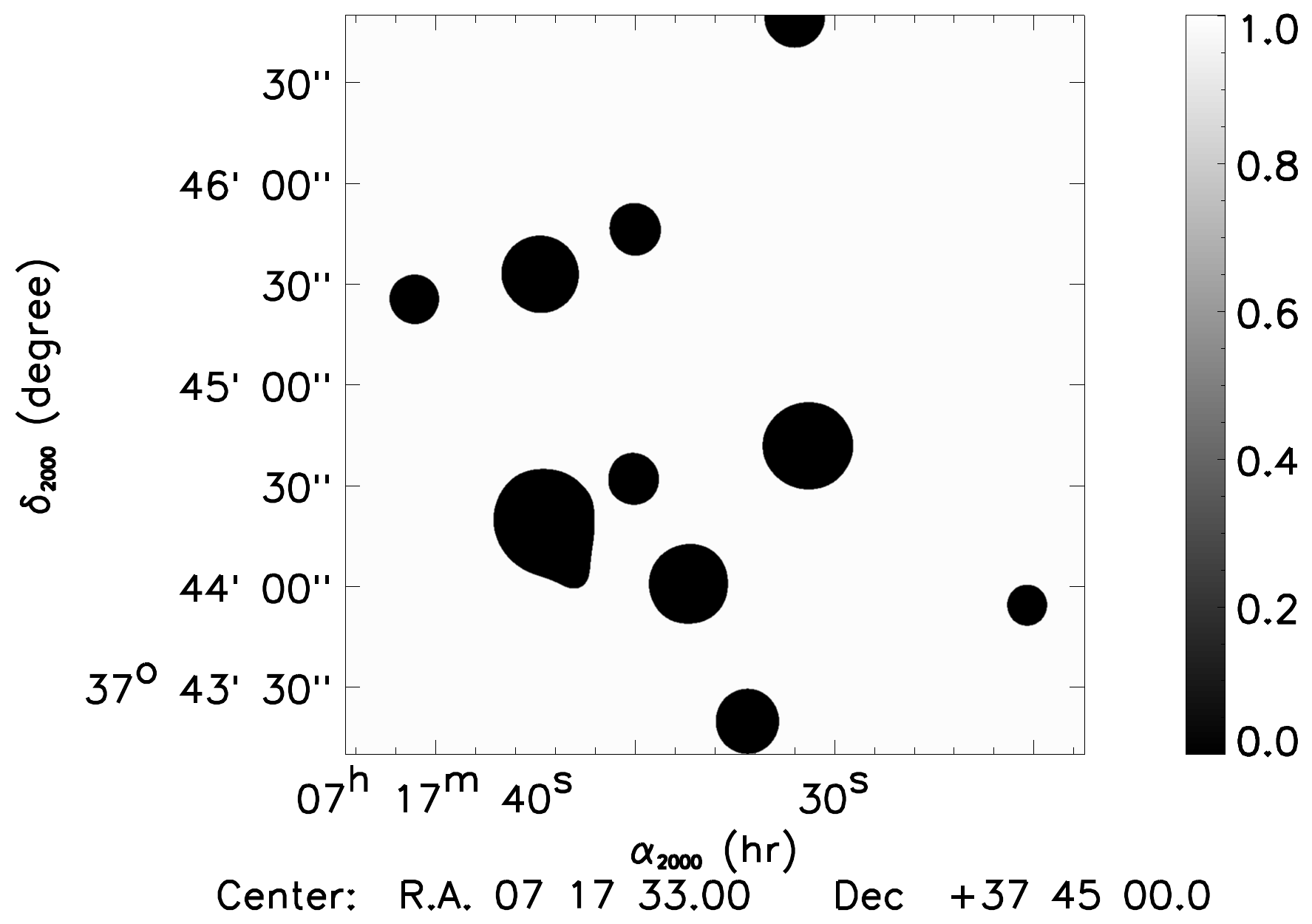}
\caption{\footnotesize{Point source model at 150~GHz (left) and 260~GHz (middle), and point source mask constructed from the 150 and 260~GHz point source models (right). The color scale of the models is the same as in Figure \ref{fig:NIKA_raw_maps}. We also provide the labels of each source on the figure, in dashed circles for radio sources and dotted circles for submillimeter sources. The point sources subtracted maps can be seen in Figure \ref{fig:best_fit_maps}.}}
\label{fig:point_source_modeling}
\end{figure*}

\subsection{Diffuse galactic emission and cosmological background}\label{sec:Diffuse_galactic_emission_and_cosmological_background}
We apply the approach used in \cite{Adam2016} in order to check that the CMB and the galactic emission (we consider synchrotron, free-free and thermal dust) are negligible in the case of \mbox{MACS~J0717.5+3745} NIKA data. The CIB was also considered in \cite{Adam2016}. Despite being the largest diffuse astrophysical noise contribution, it was shown to be negligible. The data we present here, however, are about seven times deeper, such that the CIB is no longer negligible and requires more careful investigation as we discuss below. To account for it, we produce CIB realizations, which we add to our Monte Carlo noise realizations as discussed in Section \ref{sec:NIKA_observations_and_data_reduction}. The modeling and removal of the submillimeter sources that are individually detected by \textit{Herschel} is treated separately as detailed in Section \ref{sec:Submillimeter_point_sources}.

We consider the clustering of dusty star forming galaxies, the shot noise from dusty star forming galaxies and the shot noise from radio sources. The clustering term from radio sources is negligible \citep{Hall2010}. The clustering of dusty star forming galaxies is computed using the CIB power spectrum measured by \cite{Planck2014XXX} at 143 and 217 GHz. It is modeled by a one halo term and a two halo term and extrapolated to the NIKA frequencies. The shot noise originates from the discrete nature of the CIB signal and is due to unresolved sources below the detection threshold of NIKA. For dusty star forming galaxies, it is computed using the model from \cite{Bethermin2012}, and we use the model of \cite{Tucci2011} for radio sources. We account for the beam smoothing and the large scale filtering implied by the NIKA processing. At the angular scales considered from the NIKA data, the shot noise dominates over the clustering by a factor of about five, and the shot noise from radio sources is negligible compared to that from dusty star forming galaxies at both NIKA frequencies. Using the results of \cite{Bethermin2013}, we assume that the CIB (clustering and shot noise) is completely correlated between the two NIKA bands. At our effective angular resolution of 22 arcsec, we find that the noise root mean square of the CIB is 47 $\mu$Jy/beam at 150~GHz and 130 $\mu$Jy/beam at 260~GHz. This boosts the noise in our maps by 22.1\% and 14.8\% at 260 and 150 GHz, respectively, and induces a noise correlation between the two bands. The uncertainty on the CIB model that we use is expected to be about 20\%. This translates to 7\% and 5\% uncertainty on the overall noise estimate at 260 and 150~GHz, respectively. Finally, we note that we expect CIB dimming due to the gravitational lensing by the cluster mass \citep{Zemcov2013}, but this effect was checked to be negligible in the case of our data.

\subsection{Radio sources}\label{sec:Radio_sources}
First, we consider the diffuse radio emission associated with the cluster. We extrapolate the surface brightness maps measured by NVSS \citep[NRAO VLA Sky Survey,][]{Condon1998} and FIRST \citep[Faint Images of the Radio Sky at Twenty Centimeters,][]{Becker1995} at an angular resolution of 45 arcsec and 5.4 arcsec, respectively, assuming a spectral index of -1.25 \citep{vanWeeren2009,Bonafede2009}. We find that the radio emission is negligible in both cases within the NIKA bands. Moreover, this constitutes an upper limit on the radio contamination because we expect the radio spectra to be steeper at higher frequencies due to energy loss of the synchrotron-emitting electrons \citep{Scheuer1968}. Therefore, we do not correct for the extrapolated signal and neglect its contribution.

In addition to the diffuse radio emission, a total of seven compact radio sources are identified in the NIKA field from the literature, at lower frequencies \citep{Condon1998,LaRoque2003,Coble2007,Bonafede2009,vanWeeren2009,Mroczkowski2012,PandeyPommier2013}. Their coordinates are listed in Table \ref{tab:radio_sources} and they are labeled on Figure \ref{fig:point_source_modeling}. We also consider the more recent work of \cite{vanWeeren2016} to add extra constraints on the flux densities of the sources. One of them, a foreground radio galaxy at $z=0.1546$ \citep{vanWeeren2009}, is significantly detected by NIKA at 150 and 260~GHz (source F). We assume the radio SED to follow a power law in order to model the contamination in the NIKA bands. Since we know that the field around \mbox{MACS~J0717.5+3745} is relatively crowded, and to avoid source confusion, we only use photometric data from which the observations are sensitive to scales smaller than 30 arcsec: FIRST, Very Large Array (VLA), Jansky Very Large Array (JVLA), Giant Metrewave Radio Telescope (GMRT), Owens Valley Radio Observatory / Berkeley Illinois Maryland Association (OVRO/BIMA) and MUSTANG/GBT. We also extract the flux density of the sources from the NIKA maps as detailed in \cite{Adam2016}, except for sources HT and C4, which are located near the peak of the SZ signal. The available flux densities used to constrain the SED model, including the NIKA measurements, are listed in Table \ref{tab:radio_sources}. In the case of source F, the spectrum is poorly described by our model over the full frequency coverage. Instead, the SED presents two distinct spectral indices at low ($<2$ GHz) and high ($>10$ GHz) frequencies. This could be due to significant free-free emission at high frequencies, in addition to the synchrotron at low frequencies, but might also indicate biases in the measurement of the flux density (obtained with different instruments) because of the source extension and location near the radio relic at low frequencies. Therefore, we assume that the power law model is valid at high frequency only and consider data at $\nu > 10$ GHz. Despite indications of steepening, the other sources are correctly described by a power law, even if our NIKA constraints are in tension with respect to the best-fit model for C1 and C2 (see Table \ref{tab:radio_sources}). In the case of sources C4, the limited data suggests to interpret the model prediction with caution. In addition, we emphasize that the NIKA photometry might be contaminated by submillimeter sources, as discussed in Section \ref{sec:Submillimeter_point_sources}, and possibly also by SZ signal. The results flux density prediction in the NIKA bands are listed in Table \ref{tab:radio_sources} and are used as discussed in the following. The source model, i.e. the 150 and 260 GHz best-fit flux densities of Table \ref{tab:radio_sources}, is given in Figure \ref{fig:point_source_modeling}, including submillimeter sources, as discussed in Section \ref{sec:Submillimeter_point_sources}.

\begin{table*}[h]
\caption{\footnotesize{Coordinates and flux densities of the seven compact radio sources identified in the NIKA field, including the NIKA measurements. The model predictions at 150 and 260 GHz are also provided.}}
\begin{center}
\begin{tabular}{c|ccccccc}
\hline
\hline
Label & HT & B & F & C1 & C2 & C3 & C4 \\
\hline
\hline
R.A. & +07:17:35.4$^{(a)}$ & +07:17:40.9$^{(a)}$ & +07:17:37.2$^{(a)}$ & +07:17:38.4$^{(e)}$ & +07:17:36.4$^{(b)}$ & +07:17:23.5$^{(e)}$ & +07:17:36.3$^{(b)}$ \\
Dec. & +37:45:08$^{(a)}$ & +37:43:17$^{(a)}$ & +37:44:21$^{(a)}$ & +37:46:50$^{(e)}$ & +37:44:03$^{(b)}$ & +37:45:31$^{(e)}$ & +37:44:44$^{(b)}$ \\
\hline
\hline
Freq. (GHz) & \multicolumn{6}{c}{Source flux density (mJy)} \\
 \hline
0.235 & 573$\pm$7$^{(b)}$ & 100.93$\pm$0.80$^{(b)}$ & 63.17$\pm$3.00$^{(b)}$ &  --  & 1.95$\pm$0.50$^{(b)}$ &  --  & 29.64$\pm$0.70$^{(b)}$ \\
0.610 & 109.8$\pm$3.2$^{(b)}$ & 44.79$\pm$2.60$^{(b)}$ & 30.29$\pm$0.50$^{(b)}$ &  --  & 0.98$\pm$0.05$^{(b)}$ &  --  & 0.15$\pm$0.05$^{(b)}$ \\
1.365 & 22.2$^{(a)}$ & 19.9$^{(a)}$ &  --  &  --  &  --  &  --  &  --  \\
1.435 & 22.4$^{(a)}$ & 18.8$^{(a)}$ & 6.46$\pm$0.15$^{(e)}$ & 1.37$\pm$0.15$^{(e)}$ &  --  & 2.03$\pm$0.15$^{(e)}$ &  --  \\
1.485 & 20.0$^{(a)}$ & 18.3$^{(a)}$ &  --  &  --  &  --  &  --  &  --  \\
1.500 & 17.797$\pm$0.029$^{(f)}$ &  --  & 10.505$\pm$0.071$^{(f)}$ & 1.686$\pm$0.012$^{(f)}$ & 0.442$\pm$0.009$^{(f)}$ & 2.013$\pm$0.016$^{(f)}$ &  --  \\
1.665 & 18.1$^{(a)}$ & 18.3$^{(a)}$ &  --  &  --  &  --  &  --  &  --  \\
3.000 & 7.058$\pm$0.014$^{(f)}$ &  --  & 8.356$\pm$0.036$^{(f)}$ & 1.049$\pm$0.004$^{(f)}$ & 0.578$\pm$0.003$^{(f)}$ & 1.274$\pm$0.008$^{(f)}$ &  --  \\
4.535 & 4.7$^{(a)}$ & 7.0$^{(a)}$ &  --  &  --  &  --  &  --  &  --  \\
4.885 & 3.9$^{(a)}$ & 6.4$^{(a)}$ &  --  &  --  &  --  &  --  &  --  \\
5.500 & 3.149$\pm$0.009$^{(f)}$ &  --  & 7.132$\pm$0.024$^{(f)}$ & 0.848$\pm$0.008$^{(f)}$ & 0.446$\pm$0.005$^{(f)}$ & 0.811$\pm$0.009$^{(f)}$ &  --  \\
8.460 & 0.7$^{(a)}$ & 1.9$^{(a)}$ &  --  &  --  &  --  &  --  &  --  \\
28.5 &  --  & 2.28$\pm$0.25$^{(d)}$ & 3.29$\pm$0.19$^{(d)}$ &  --  &  --  &  --  &  --  \\
30 &  --  & 1.60$\pm$0.17$^{(d)}$ & 2.70$\pm$0.17$^{(d)}$ &  --  &  --  &  --  &  --  \\
90 &  --  &  --  & 2.8$\pm$0.2$^{(c)}$ &  --  &  --  &  --  &  --  \\
150 &    N.A. &     0.32$\pm$    0.29$^{(g)}$ &     2.30$\pm$    0.18$^{(g)}$ &     0.80$\pm$    0.26$^{(g)}$ &     0.90$\pm$    0.18$^{(g)}$ &     0.31$\pm$    0.21$^{(g)}$ &  N.A. \\
260 &    N.A. &    -0.24$\pm$    1.15$^{(g)}$ &     1.82$\pm$    0.72$^{(g)}$ &     0.81$\pm$    1.04$^{(g)}$ &     0.44$\pm$    0.73$^{(g)}$ &     0.39$\pm$    0.86$^{(g)}$ &  N.A. \\
\hline
150 (model) &   0.0184$\pm$  0.0002 &   0.130$\pm$  0.002 &   2.552$\pm$  0.043 &   0.118$\pm$  0.004 &   0.317$\pm$  0.016 &   0.086$\pm$  0.003 &   negligible \\
260 (model) &   0.0089$\pm$  0.0001 &   0.080$\pm$  0.001 &   2.295$\pm$  0.043 &   0.093$\pm$  0.004 &   0.311$\pm$  0.018 &   0.063$\pm$  0.003 &   negligible \\
\hline
\end{tabular}
\end{center}
{\small {\bf Notes.} $^{(a)}$ VLA; \cite{Bonafede2009}. $^{(b)}$ GMRT; \cite{PandeyPommier2013}. $^{(c)}$ MUSTANG; \cite{Mroczkowski2012}. $^{(d)}$ OVRO/BIMA; \cite{LaRoque2003,Coble2007}. $^{(e)}$ FIRST; \cite{Condon1998}. $^{(f)}$ JVLA; \cite{vanWeeren2016}. $^{(g)}$ NIKA; this work.}
\label{tab:radio_sources}
\end{table*}

\subsection{Submillimeter point sources}\label{sec:Submillimeter_point_sources}
As mentioned in Section \ref{sec:NIKA_raw_maps}, submillimeter sources significantly contaminate our data, even at 150~GHz. We use the approach detailed in \cite{Adam2016} in order to subtract their contribution. To do so, in addition to the NIKA observation itself, we use the SPIRE \citep[Spectral and Photometric Imaging REceiver,][]{Griffin2010} catalog produced by \cite{Sayers2013} to clean their Bolocam images. Eleven sources are identified in the 4 arcmin $\times$ 4 arcmin region around the cluster and we report their flux densities in Table \ref{tab:submm_sources}. The gray body model which we fit to the data, excluding the NIKA sources located near the SZ signal, allow us to better constrain the excepted flux densities in the NIKA bands \citep[see][for more details]{Adam2016}. The corresponding template to be subtracted from our data is provided in Figure \ref{fig:point_source_modeling} together with the radio contribution. Most of the model predictions are fully compatible with the NIKA observations. Only SMG11 at 150~GHz differ by more than $3 \sigma$, and we note that we suspect the NIKA measurement of SMG11 to be strongly contaminated by SZ signal because of its location. The SPIRE data are also well described by the model and we do not observe significant excess in the residual between the model and the data for any of the sources.

In addition to the radio source, F, the excess seen in the NIKA 260~GHz map coincide with SMG09 (northwest of subcluster D) and SMG05 (south of subcluster D). The corresponding SED peaks at relatively low frequencies, indicating that these sources could be high redshift lensed galaxies. This is particularly true for SMG09, for which the SED peaks at about 600 GHz (500 $\mu$m). Therefore, we search for counterparts to the NIKA detected sources in strong lensing HST data set \citep[Hubble Space Telecope, Frontier Field campaign, see][]{Diego2015}. By requiring that the projected distance is less than 5 arcsec between our identified sources and that of HST, we find two possible candidates for SMG09 at $z=4.5$ (R.A. +07:17:31.269, Dec. +37:44:41.10) at 2.0 arcsec from our coordinates, and $z=3.0$ (R.A. +07:17:31.082, Dec. +37:44:42.36) at 2.5 arcsec from our coordinates \citep[ID25.1 and ID62.1, respectively,][]{Diego2015}. The source SMG09 is, therefore, very likely to be a high redshift submillimeter galaxy lensed by \mbox{MACS~J0717.5+3745}, or even a blend of two galaxies. In the case of SMG05, no counterpart is found, even if we increase our threshold radius to 15 arcsec.

\begin{table*}[h]
\caption{\footnotesize{Coordinates, measured flux densities and predicted flux densities in the NIKA bands for the eleven submillimeter sources identified in our field. The measured flux densities are from SPIRE \citep{Sayers2013} and NIKA (this work).}}
\begin{center}
\begin{tabular}{c|cc|ccc|cc|cc}
\hline
\hline
 & & & \multicolumn{5}{c}{Measured flux densities (mJy)} & \multicolumn{2}{c}{NIKA prediction (mJy)} \\
Label & R.A.$^{(a)}$ & Dec.$^{(a)}$ & \multicolumn{3}{c}{SPIRE (GHz)$^{(a)}$} & \multicolumn{2}{c}{NIKA (GHz)} & \multicolumn{2}{c}{NIKA (GHz)} \\
 &  &  & 1200 & 857 & 600 & 260 & 150 & 260 (model) & 150 (model) \\
 \hline
SMG01 & +07:17:32.11 & +37:43:21.0 &     57.6$\pm$     0.9 &     30.1$\pm$     0.9 &      9.9$\pm$     1.2 &      1.0$\pm$     0.8 &      0.2$\pm$     0.2 &      0.9$\pm$     0.2 &      0.1$\pm$     0.1 \\ 
SMG02 & +07:17:25.40 & +37:47:05.7 &     25.2$\pm$     0.9 &     14.7$\pm$     0.9 &      1.3$\pm$     1.4 &      0.7$\pm$     1.3 &     -0.1$\pm$     0.3 &      0.5$\pm$     0.2 &      0.1$\pm$     0.1 \\ 
SMG03 & +07:17:30.92 & +37:46:50.6 &     15.2$\pm$     0.9 &     13.9$\pm$     0.9 &      6.2$\pm$     1.2 &      0.1$\pm$     0.9 &      0.6$\pm$     0.2 &      0.8$\pm$     0.2 &      0.1$\pm$     0.1 \\ 
SMG04 & +07:17:34.91 & +37:45:47.4 &     18.4$\pm$     0.9 &     10.6$\pm$     1.0 &      5.1$\pm$     1.7 &      0.3$\pm$     0.7 &      0.0$\pm$     0.2 &      0.6$\pm$     0.2 &      0.1$\pm$     0.1 \\ 
SMG05 & +07:17:33.58 & +37:44:01.9 &     14.3$\pm$     0.8 &     16.8$\pm$     0.9 &     13.3$\pm$     1.2 &      1.3$\pm$     0.7$^{(b)}$ &      0.2$\pm$     0.2$^{(b)}$ &      1.8$\pm$     0.3 &      0.4$\pm$     0.1 \\ 
SMG06 & +07:17:25.06 & +37:43:55.6 &      8.8$\pm$     1.1 &      7.3$\pm$     1.0 &      2.7$\pm$     1.3 &      0.1$\pm$     0.8 &      0.2$\pm$     0.2 &      0.4$\pm$     0.2 &      0.1$\pm$     0.1 \\ 
SMG07 & +07:17:37.28 & +37:45:34.1 &     12.1$\pm$     0.9 &     12.7$\pm$     1.0 &     14.1$\pm$     1.2 &      0.1$\pm$     0.7 &      0.4$\pm$     0.2 &      1.6$\pm$     0.3 &      0.3$\pm$     0.1 \\ 
SMG08 & +07:17:30.87 & +37:46:18.4 &      1.3$\pm$     1.1 &      1.3$\pm$     0.9 &      0.0$\pm$     1.1 &      0.1$\pm$     0.7 &     -0.1$\pm$     0.2 &      $<0.05$ &     $<0.02$ \\ 
SMG09$^{(c)}$ & +07:17:31.24 & +37:44:43.0 &      6.3$\pm$     1.0 &     11.9$\pm$     1.0 &     15.1$\pm$     1.0 &      2.5$\pm$     0.7$^{(b)}$ &      0.6$\pm$     0.2$^{(b)}$ &      3.0$\pm$     0.4 &      0.6$\pm$     0.2 \\ 
SMG10 & +07:17:40.43 & +37:45:26.7 &      4.2$\pm$     0.9 &      5.3$\pm$     0.8 &      3.1$\pm$     1.2 &     -0.3$\pm$     0.8 &      0.4$\pm$     0.2 &      0.5$\pm$     0.2 &      0.1$\pm$     0.1 \\ 
SMG11 & +07:17:34.95 & +37:44:33.1 &      4.7$\pm$     1.1 &      4.8$\pm$     0.8 &      4.9$\pm$     1.4 &      0.0$\pm$     0.7$^{(b)}$ &     -0.9$\pm$     0.2$^{(b)}$ &      0.5$\pm$     0.2 &      0.1$\pm$     0.1 \\ 
\hline
\end{tabular}
\end{center}
{\small {\bf Notes.} $^{(a)}$ From \cite{Sayers2013}. $^{(b)}$ Likely to be contaminated by SZ. $^{(c)}$ High redshift lensed galaxy candidate with HST counterpart.}
\label{tab:submm_sources}
\end{table*}

\subsection{Diffuse submillimeter emission from the ICM}\label{sec:Diffuse_sub-millimeter_emission_from_the_ICM}
While galaxies are the major contributors to the submillimeter emission in galaxy clusters \citep{Coppin2011}, diffuse dust emission associated to the ICM could also lead to submillimeter emission \citep[see, for example,][who study the overall contribution of dust in clusters]{Montier2005,Planck2016XLIII}. Such signal would likely be diffuse, at arcmin scales, and we expect it to be small compared to the galaxies, which give rise to emission on small scales. Using \textit{Herschel} SPIRE maps toward \mbox{MACS~J0717.5+3745}, we check that such signal is, at least, much lower than the one arising from point sources. Moreover, we do not observe any diffuse emission which would be correlated with the kSZ signal measured in Section \ref{sec:A_map_of_the_kinetic_Sunyaev_Zel_dovich_signal}. Therefore, the dust contribution associated with the ICM is neglected in the NIKA bands.

\subsection{Construction of a mask for point sources}\label{sec:Construction_of_a_mask_for_point_sources}
The spectral modeling of both radio and submillimeter point sources allows us for a good first estimate of the contamination expected in our data. However, the model uncertainties are relatively large and once subtracted to the respective channels, we observe small residuals in the case of radio sources, that are visible in particular for sources that are outside the region dominated by SZ signal. This is likely due to our modeling being too simplistic since it does not account for the steepening of the radio spectrum, as discussed in Section \ref{sec:Radio_sources}. Therefore, in addition to subtracting the point source model, we also construct a mask to avoid using pixels that we consider potentially biased when using the NIKA maps for photometry or for fitting purposes. This mask is constructed by applying a threshold (0.2 and 0.4 mJy/beam at 150 and 260 GHz, respectively) on the point source best-fit models at 150 and 260 GHz simultaneously. Such procedure is conservative and takes advantage of the high angular resolution of the NIKA data. The mask we use is given on the right panel of Figure~\ref{fig:point_source_modeling}.

\section{X-ray data analysis}\label{sec:X_ray_data_reduction}
In the work presented in this paper, we have complemented the SZ data with X-ray observations for two reasons. First, \mbox{MACS~J0717.5+3745} is known to be a very hot cluster \citep{Ma2009} such that SZ relativistic corrections are expected to be fairly significant and should be taken into account by using an extra estimate of the gas temperature, available from X-ray spectroscopy. Second, X-ray data are also needed in order to disentangle the line-of-sight gas velocity from the density and the temperature (see equation \ref{eq:yksz}). In this Section, we discuss the X-ray data reduction in terms of imaging and spectroscopy.

\subsection{Data preparation}
\mbox{MACS~J0717.5+3745} was observed by the XMM-\textit{Newton} telescope using the European Photon Imaging Camera (EPIC, \citealt{turner2001} and \citealt{struder2001}) for $195$ ks (obs-IDs $0672420101$, $0672420201$, and $0672420301$) and by the \textit{Chandra} Advanced CCD Imaging Spectrometer (ACIS, \citealt{garmire2003}) for $250$ ks in total (obs-IDs 1655, 4200, 16235, 16305). We processed XMM-\textit{Newton} datasets using the Science Analysis System (SAS) version $15.0.0$, applying the latest version of calibration files available in March $2016$. We processed Chandra datasets using the \textit{Chandra} Interactive Analysis of Observation (CIAO) version $4.6.5$ and calibration database version $4.7$. Both datasets suffer from contamination due to the high energy particle flux. To reduce this component, we removed from the XMM-\textit{Newton} datasets all the events which keyword PATTERN is $>4$ and $>13$ for MOS$1,2$ and PN cameras, respectively. We applied Very Faint\footnote{\url{cxc.harvard.edu/cal/Acis/Cal\_prods/vfbkgrnd}} (VF) mode filtering to obs-IDs 4200 and 16305. Since our background subtraction technique for \textit{Chandra} can be applied only to observations taken using the VF mode, we removed obs-IDs 1655 and 16235 from our analysis, thus reducing the effective observation time to $153$ ks.
 
To remove observation intervals affected by flare episodes we followed the light-curve filtering procedures described in \cite{pratt2007} and in the \textit{Chandra} COOKBOOK\footnote{\url{cxc.harvard.edu/contrib/maxim/acisbg/COOKBOOK}} for XMM-\textit{Newton} and \textit{Chandra} datasets, respectively. We removed from the analysis all the intervals where the count rate exceeded $3\sigma$ from the mean value. We find no flare contamination in the \textit{Chandra} dataset, so the full observation time of $153$ ks was used. For the XMM-\textit{Newton} dataset, the useful exposure times were $160$ ks and $116$ ks for MOS$1,2$ and PN cameras, respectively. We ran the \textit{wavdetect} algorithm \citep{freeman2002} to identify point sources on exposure corrected images in the $[0.3-2]$ keV and in the $[0.5-2 ; 2-8 ; 0.5-8]$ keV bands for XMM-\textit{Newton} and \textit{Chandra}, respectively. Point source lists thus produced were inspected by eye, merged, and used as mask to remove point source contribution from the analysis. Point sources only detected by \textit{Chandra} were masked from XMM-\textit{Newton} dataset, using a circular region with radius $15''$ within $3'$ from the aimpoint.

To perform spectral and imaging analysis we binned both datasets in sky coordinates and energy, creating an energy position photon cube for each dataset following the procedure described in \cite{bourdin2008}. To each event in the cube we assigned an effective area, which we use to correct for vignetting and exposure time, and the background noise. In the following, all the techniques described have been used on both \textit{Chandra} and XMM-\textit{Newton} datasets, unless otherwise stated.

The exposure- and vignetted-corrected \textit{Chandra} image in the $[0.5-2.5]$ keV band and its wavelet de-noised map are shown in the top left and right panel, respectively, of Figure \ref{fig:Xray_all_maps}. The wavelet filtering is performed following the method described in \cite{bourdin2013}. Note that the resolution of the wavelet map is not fixed. It depends on the image statistics as the filtering is constrained to detect structures above $3\sigma$ at different scales. As already found by several teams \citep[e.g.,][]{Zitrin2009,Ma2009}, the cluster exhibits a very complex morphology, with at least $4$ identifiable substructures. Because of the higher resolution of ACIS, we used the \textit{Chandra} data as the baseline for the X-ray imaging analyses.

\subsection{Background estimation}
The X-ray background can be divided into a sky and an instrumental component. For the latter we used the analytical models produced by \cite{bourdin2013} and \cite{bartalucci2014} for XMM-\textit{Newton} and \textit{Chandra}, respectively. We normalized both models, extracting the events from a region free from the cluster emission in the $[10-12]$ keV and in the $[9.5-10.6]$ keV band for XMM-\textit{Newton} and \textit{Chandra}, respectively. The sky background is due to foreground Galactic emission and an additional extra-Galactic component. For the former we used two absorbed thermal models produced by the Astrophysical Plasma Emission Code \citep[APEC,][]{smith2001} for which the temperatures are fixed to $0.248$ keV and $0.099$ keV \citep[see][]{kuntz2000}. The extra-Galactic component was modeled using an absorbed power law with $\gamma$ fixed to $1.42$, as proposed by \cite{lumb2002}. We determined the normalization of all the components by performing a joint fit of the spectrum extracted from a region free of source emission. We noticed the presence of a small excess in the $[0.3-1]$ keV band in the XMM-\textit{Newton} dataset, compatible with the solar wind charge exchange emission \citep{snowden2004}. We found that one absorbed APEC model with a temperature of $0.86$ keV is sufficient to reproduce this excess. We added this component in the joint fit. Once normalized, the background sky model is added to the instrumental model. Then, in all our spectral analysis procedures, we simply scale the resulting total background model by the ratio of the extraction areas.

\subsection{Spectroscopy analysis}\label{sec:Spectroscopy_analysis}
\begin{figure*}[h]
\centering
\includegraphics[width=0.9\textwidth]{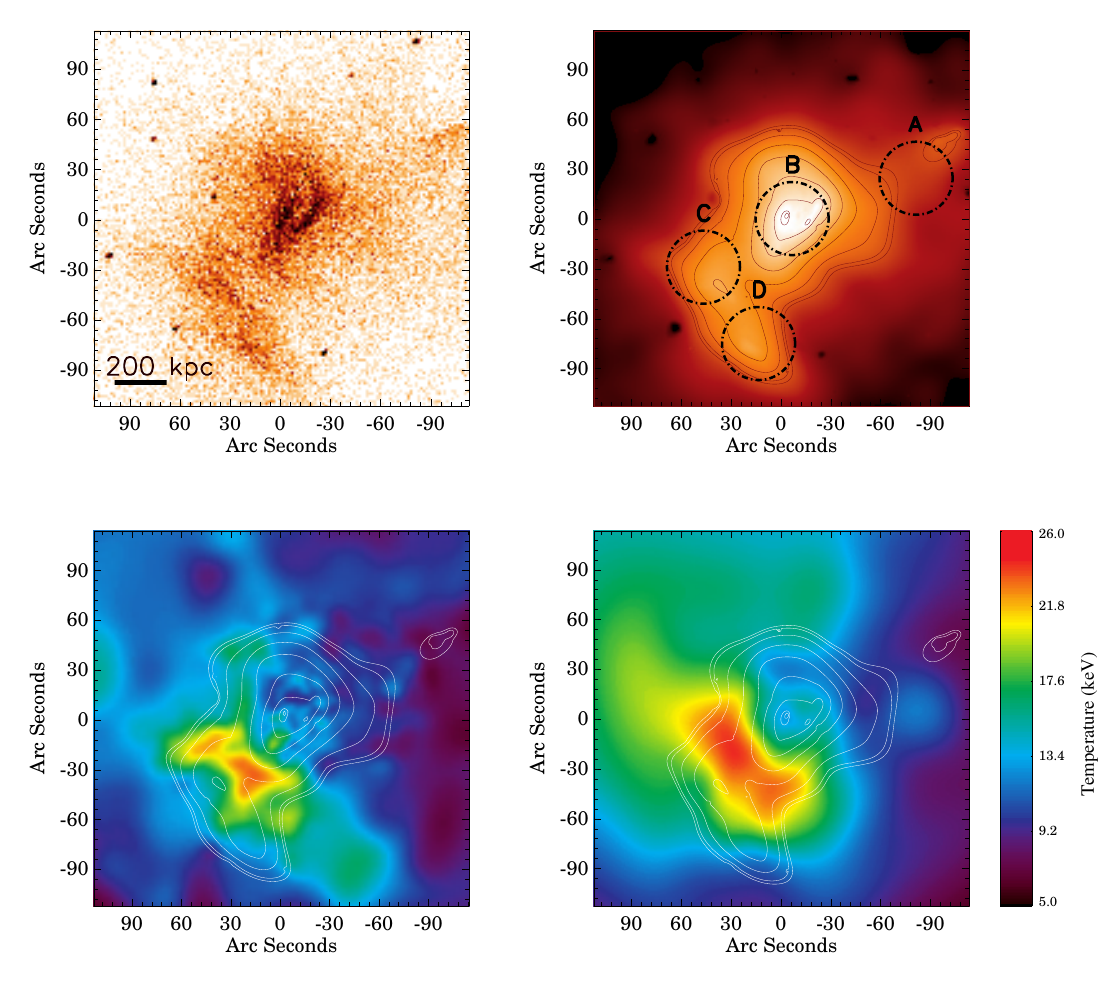}
\caption{\footnotesize{XMM-\textit{Newton} and \textit{Chandra} observations of \mbox{MACS~J0717.5+3745}. {\bf Top left panel}: Exposure- and vignetting-corrected \textit{Chandra} photon count image in the $[0.5-2.5]$ keV band. The image has a resolution of 3.1". {\bf Top right image}: wavelet de-noised map of the \textit{Chandra} image shown in the top left panel. Iso-contours are shown with black solid lines to enhance substructures. The four regions in Table \ref{tab:coord_region} are highlighted with black circles. \textit{Bottom left image}: XMM-\textit{Newton} wavelet de-noised temperature map, obtained using a threshold cut of $1\sigma$. Contours from the wavelet map image are overlaid using white contours. {\bf Bottom right image}: same as bottom left panel except that we show the temperature map computed using the \textit{Chandra} dataset. All the maps shown are centered on the X-ray peak.}}
\label{fig:Xray_all_maps}
\end{figure*}

All X-ray spectral analyses were performed following the scheme described in \cite{bourdin2008}. For a given region of interest, we add to the background model an absorbed APEC to account for the cluster emission, and fit it via $\chi^{2}$ minimization to the exposure- and vignetting-corrected spectrum extracted from the photon cube. The absorption is fixed to the Galactic value along the line-of-sight $n_{H} = 6.63 \times 10^{-20}$ cm$^{-2}$, as determined from the LAB survey \cite{kalberla2005}. All the models are convolved by the appropriate response matrix files, accounting for energy resolution loss due to instrumental effects. The source emission model and the sky background models are multiplied by the appropriate ancillary response file. From the fitting procedures performed in the $[0.3-10]$ keV and $[0.7-10]$ keV band for XMM-\textit{Newton} and \textit{Chandra}, respectively, we determined the APEC model temperature, abundance and normalization. 

We produced temperature maps applying the wavelet filtering algorithm describe in \cite{bourdin2004} and in \cite{bourdin2008}. Briefly, we sample the field of view in square grids where each element size increases by powers of $2$. The minimum size of the meta-pixel is defined so that it contains at least $200$ photons. In each meta-pixel we extract the spectrum and measure the temperature, fixing the abundance to $0.3$ solar, obtaining a first set of temperature and associated fluctuation maps. We then convolve these maps with a B$2$-spline function evaluated at different scales to obtain the corresponding wavelet coefficients. The `de-noised' temperature map is constructed using a $1\sigma$ threshold for the wavelet coefficient maps. The de-noised temperature maps that we obtained from the XMM-\textit{Newton} and \textit{Chandra} datasets, are shown in the left and right bottom panel, respectively, of Figure \ref{fig:Xray_all_maps}. Both temperature maps identify the same hot structure, with temperatures of the order of $\sim 24$ keV, and a cold region in the northwest sector associated with the most prominent structures. Note that the wavelet filtering is applied to maps where the element size is determined by the statistics. For this reason the XMM-\textit{Newton} dataset allows a finer element sampling, due simply to the larger number of counts in this observation.

To check the validity of the values in the temperature maps, following \cite{wang2016}, we extract the spectra from the four circular regions shown in the top right image of Figure \ref{fig:Xray_all_maps}, using circular radii of $15''$ and $30''$. We then measured the temperatures via our fitting procedure, which we report in Table \ref{tab:subcluster_Tx_measurements} together with their $1\sigma$ errors. Because of \textit{Chandra}'s shorter exposure time and poorer sensitivity, we measure the temperature only in a $30''$ circular bin aperture. The values reported for the two instruments are consistent with the values found in the respective temperature maps.
\begin{table}[h]
\caption{{\footnotesize X-ray spectroscopic temperatures of the four subclusters in \mbox{MACS~J0717.5+3745}, centered on the regions defined in Table \ref{tab:coord_region}. The temperatures are provided for both XMM-\textit{Newton} and \textit{Chandra} data for 30$"$ apertures radius and we also provide the values for 15$"$ radius in the case of XMM-\textit{Newton}.}}
\begin{center}
\begin{tabular}{cccc}
\hline
\hline
Subcluster & \multicolumn{2}{c}{XMM-\textit{Newton}} & \textit{Chandra} \\
 & 15$"$ & 30$"$ & 30$"$ \\
\hline
A & $  7.57^{+1.62}_{-1.23}$ keV & $  7.71^{+0.78}_{-0.68}$ keV & $  9.29^{+2.20}_{-1.48}$ keV \\
B & $10.70^{+1.12}_{-0.86}$ keV & $11.22^{+0.68}_{-0.61}$ keV & $13.33^{+1.59}_{-1.28}$ keV \\
C & $16.39^{+3.48}_{-2.60}$ keV & $18.09^{+2.01}_{-1.78}$ keV & $17.82^{+3.62}_{-2.73}$ keV \\
D & $12.55^{+2.75}_{-1.92}$ keV & $13.56^{+1.82}_{-1.38}$ keV & $16.31^{+4.18}_{-2.78}$ keV \\
\hline
\end{tabular}
\end{center}
\label{tab:subcluster_Tx_measurements}
\end{table}
As expected, on average \textit{Chandra} measures higher temperatures than XMM-\textit{Newton} (see, e.g., \citealt{martino2014} and \citealt{schellenberger2015}), but the values are consistent within the errors. Our results are also in agreement those of \cite{Sayers2013}, even if the regions we consider here are not exactly centered on the same coordinates. In the case of subcluster C, however, the temperature that we find are about the same for both X-ray observatories. Our \textit{Chandra} derived value is also lower for this subcluster than that reported in \cite{Sayers2013}, but is in agreement within the errors. Because of the better resolution and significance, we use the XMM-\textit{Newton} spectroscopic results as our baseline.

\section{A map of the kinetic Sunyaev-Zel'dovich signal}\label{sec:A_map_of_the_kinetic_Sunyaev_Zel_dovich_signal}
In the case of pure tSZ signal with small relativistic corrections, we expect the spatial distribution of the signal to be proportional at any frequency. However, this is clearly not the case for the NIKA data such that another contribution is necessary to explain our observations. Motivated by the results of \cite{Ma2009}, \cite{Mroczkowski2012} and \cite{Sayers2013}, we assume that the observed signal is due to the tSZ and the kSZ effects, and we use the NIKA data to disentangle the two. The main goal of this Section is to produce, for the first time, a resolved map of the kSZ effect. This constitutes the main result of this paper.

\subsection{Reconstruction of the kinetic Sunyaev-Zel'dovich signal}\label{sec:Reconstruction_of_the_kinetic_Sunyaev_Zeldovich_signal}
The two NIKA maps, cleaned of contamination (see Section \ref{sec:astrophysical_contamination_of_the_Sunyaev_Zel_dovich_signal}), provide a measurement of the surface brightness, $\Delta I_{\nu}$, in the different regions of the cluster. The sensitivity to the tSZ and the kSZ contributions are known from the coefficients given in Table \ref{tab:sz_coefficients}, and are corrected on each pixel of the sky for relativistic effects using the XMM-\textit{Newton} X-ray spectroscopic temperature map in Figure \ref{fig:Xray_all_maps}, under the assumption that the temperature is constant along the line-of-sight. We can therefore invert equation \ref{eq:dIsz} in order to separate the contribution of the tSZ signal, given by
\begin{equation}
	y_{\rm tSZ} = \frac{g(\nu_1, T_e) \Delta I_{\nu_2} - g(\nu_2, T_e) \Delta I_{\nu_1}}{I_0 g(\nu_1, T_e) f(\nu_2, T_e) - I_0 g(\nu_2, T_e) f(\nu_1, T_e)},
\label{eq:measurement_ytsz}
\end{equation}
and that of the kSZ signal,
\begin{equation}
	y_{\rm kSZ} = \frac{f(\nu_1, T_e) \Delta I_{\nu_2} - f(\nu_2, T_e) \Delta I_{\nu_1}}{I_0 f(\nu_1, T_e) g(\nu_2, T_e) - I_0 f(\nu_2, T_e) g(\nu_1, T_e)}.
\label{eq:measurement_yksz}
\end{equation}
We propagate the noise through Monte Carlo realizations as discussed in Section \ref{sec:NIKA_observations_and_data_reduction}, including the CIB contribution and its induced correlation between the two NIKA frequencies. The statistical errors on both tSZ and kSZ are largely dominated by the noise at 260~GHz due to the intrinsic lower sensitivity of this band, with respect to the 150 GHz one. 

\subsection{Reconstruction of the kinetic signal toward subclusters}\label{sec:Reconstruction_of_the_kinetic_signal_toward_sub_clusters}
\begin{figure*}[h]
\centering
\includegraphics[width=0.45\textwidth]{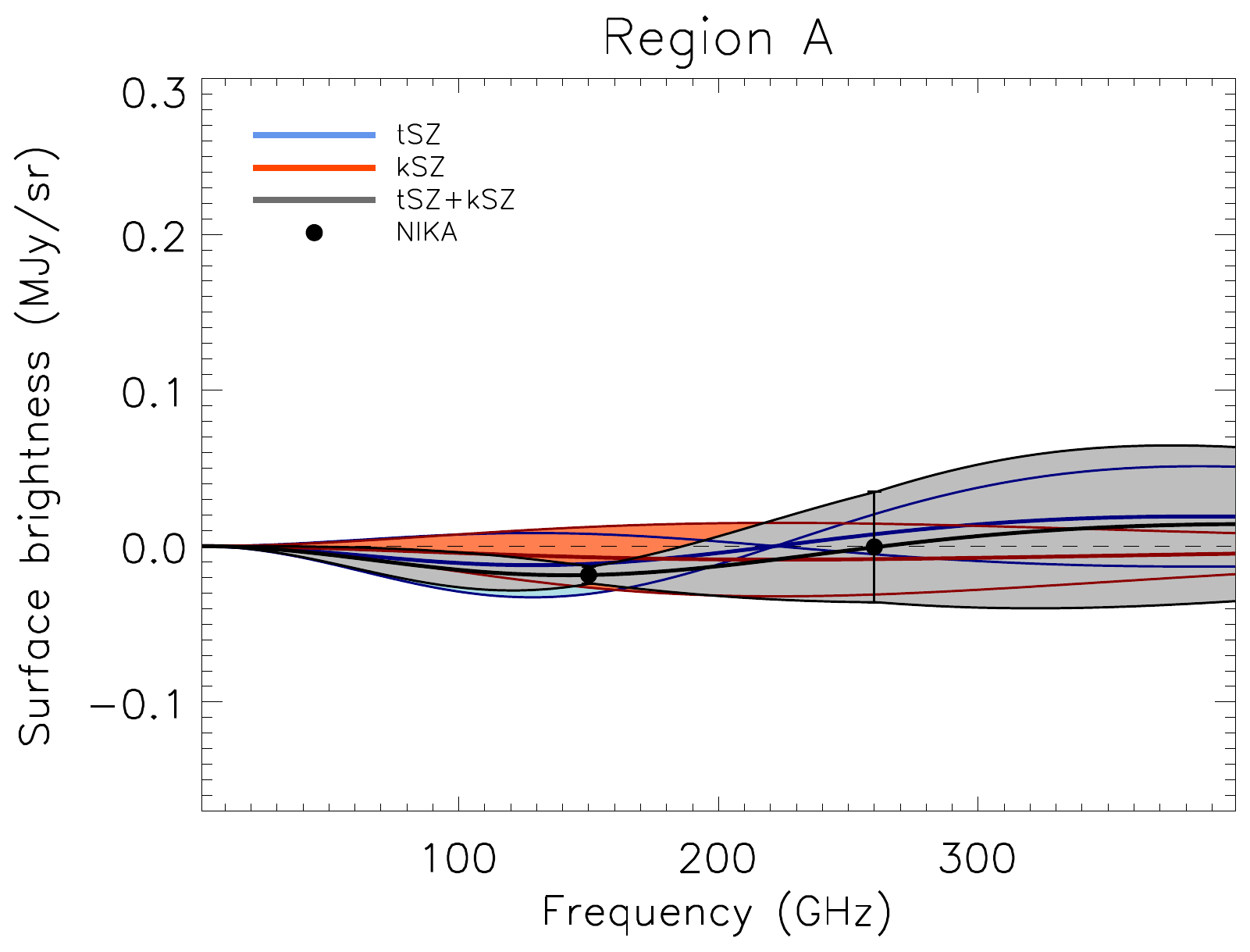}
\includegraphics[width=0.45\textwidth]{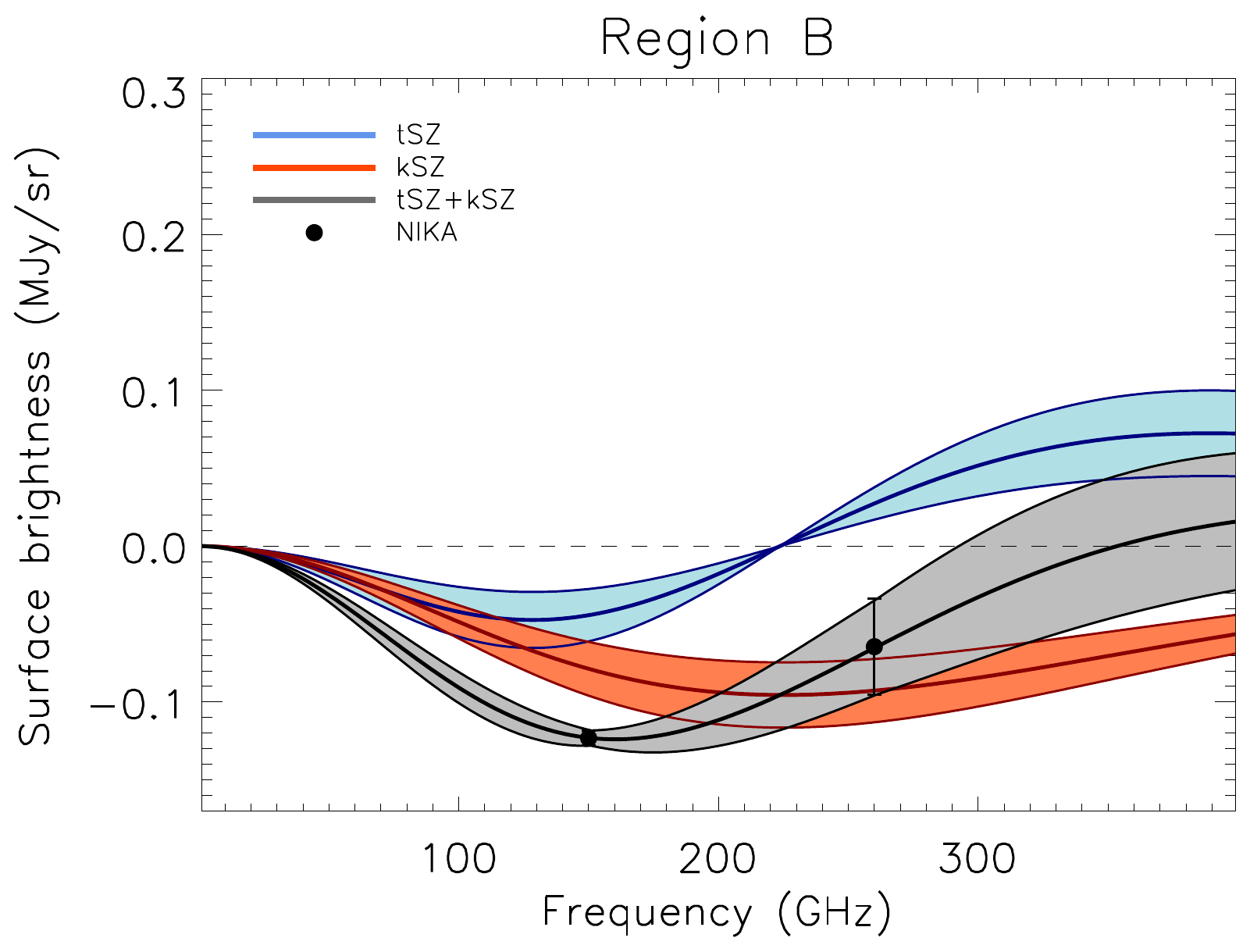}
\includegraphics[width=0.45\textwidth]{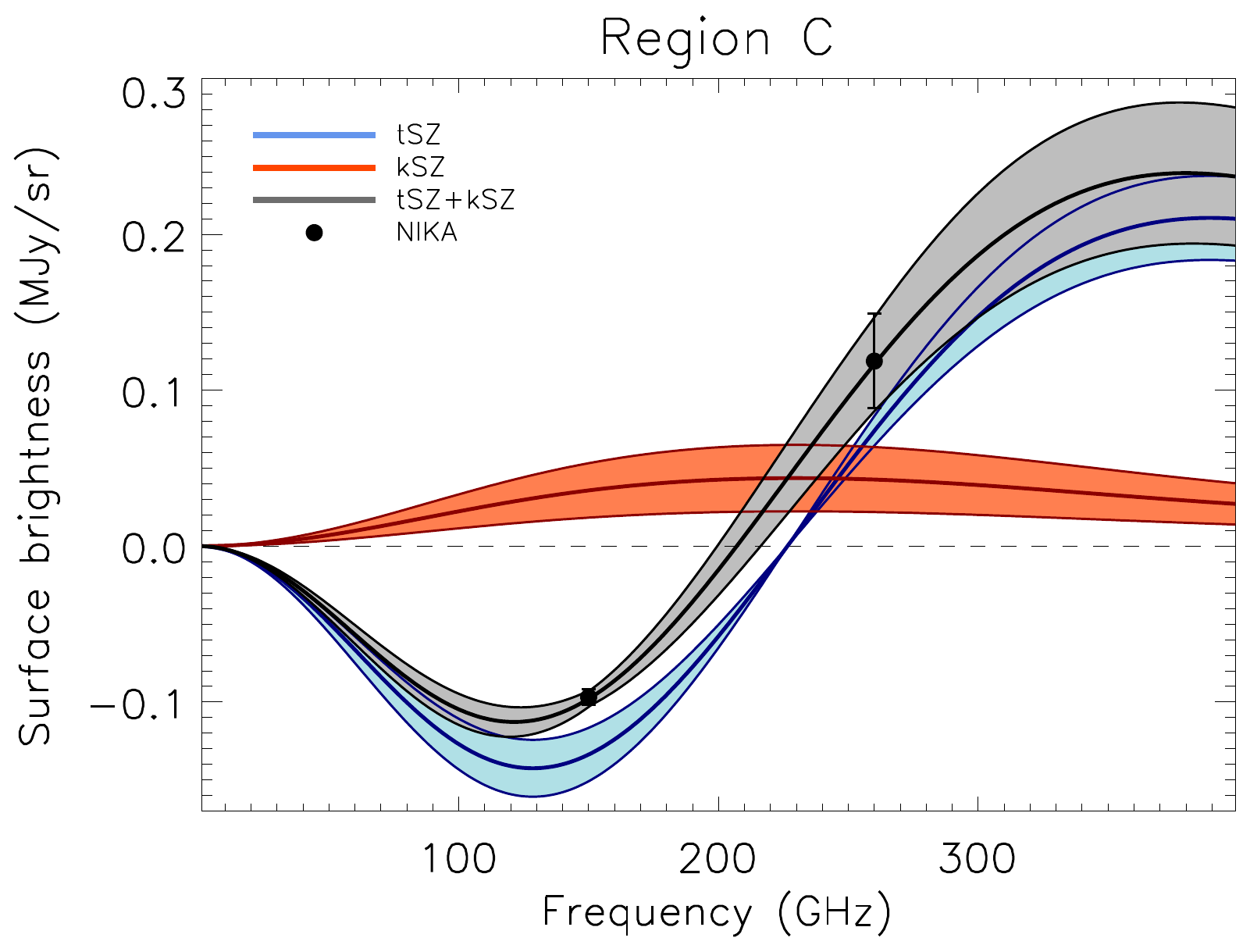}
\includegraphics[width=0.45\textwidth]{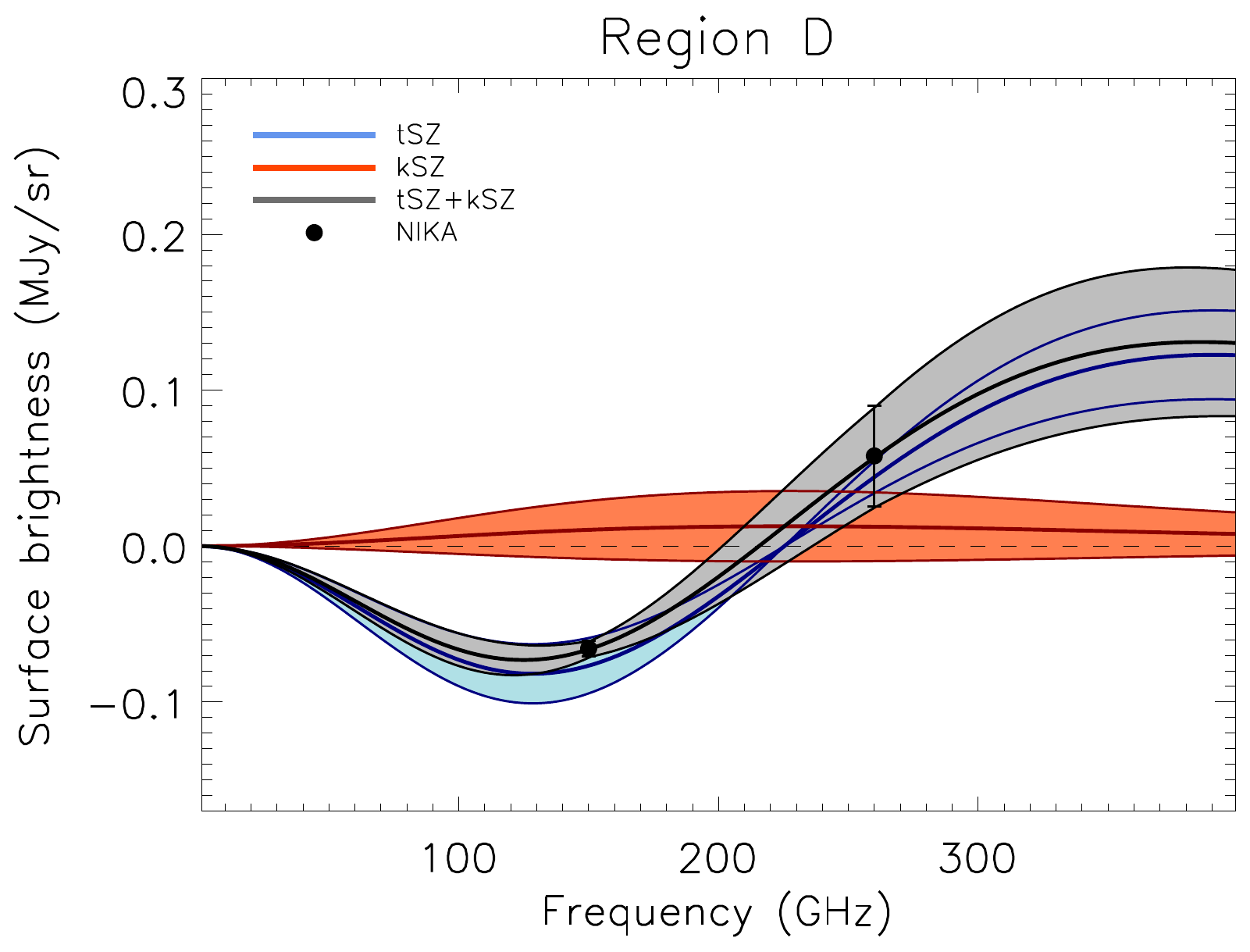}
\caption{\footnotesize{Constraints on the kSZ and the tSZ amplitudes toward the regions associated with the four subclusters (from A to D). The black data points provide the average brightness measured by NIKA in the 22 arcsec radius regions with center coordinates given in Table \ref{tab:coord_region} and shown in Figure \ref{fig:NIKA_raw_maps}. The shaded areas give the 68\% confidence level constraints on the kSZ (red), the tSZ (blue) and the sum of the two (gray). Possible point source residuals were masked when computing the average surface brightness.}}
\label{fig:spectra_in_regions}
\end{figure*}

We first consider the mean SZ surface brightness measured in the individual regions provided in Table \ref{tab:coord_region}, i.e., within 22 arcsec radius disks. The point source mask of Figure \ref{fig:point_source_modeling} was applied before computing the brightness, so that pixels that are potentially affected by mis-subtraction of the point sources are rejected. Following Section \ref{sec:Reconstruction_of_the_kinetic_Sunyaev_Zeldovich_signal}, we use these measurements to constrain the amplitude of both the tSZ and kSZ spectra. The results are presented in Figure \ref{fig:spectra_in_regions} for the four subclusters. From equations \ref{eq:dIsz} and \ref{eq:yksz}, we can see that a positive kSZ amplitude corresponds to a negative gas line-of-sight velocity, which means that the cluster is moving toward the observer, with respect to the CMB reference frame. The signal-to-noise ratio is too low in region A and both tSZ and kSZ amplitudes are compatible with zero. However, at 150 GHz the SZ signal is significantly detected, and we assume this to be dominated by tSZ. In region B, we detect tSZ and kSZ signal at $2.6 \sigma$ and $-4.6 \sigma$, respectively. Moreover, we find that the kSZ induced brightness is slightly more negative than that of the tSZ, at the scales probed by NIKA. The kSZ signal is negative, which explains the decrement observed in this region at 260~GHz with respect to the 150~GHz expectation. In region C, we significantly detect tSZ signal and find indication for kSZ signal at $7.8 \sigma$ and $2.1 \sigma$, respectively, but the situation is reversed with respect to region B and the kSZ signal is positive. The tSZ signal strength is very high, in agreement with the expectation from the merger scenario, responsible for gas heating in this region. This is also in agreement with our X-ray derived temperature map. Finally, tSZ signal is detected in region D at $4.3 \sigma$ and the kSZ significance is $0.6 \sigma$.

\subsection{Mapping of the kSZ and tSZ signal}
\begin{figure*}[h]
\centering
\includegraphics[width=0.49\textwidth]{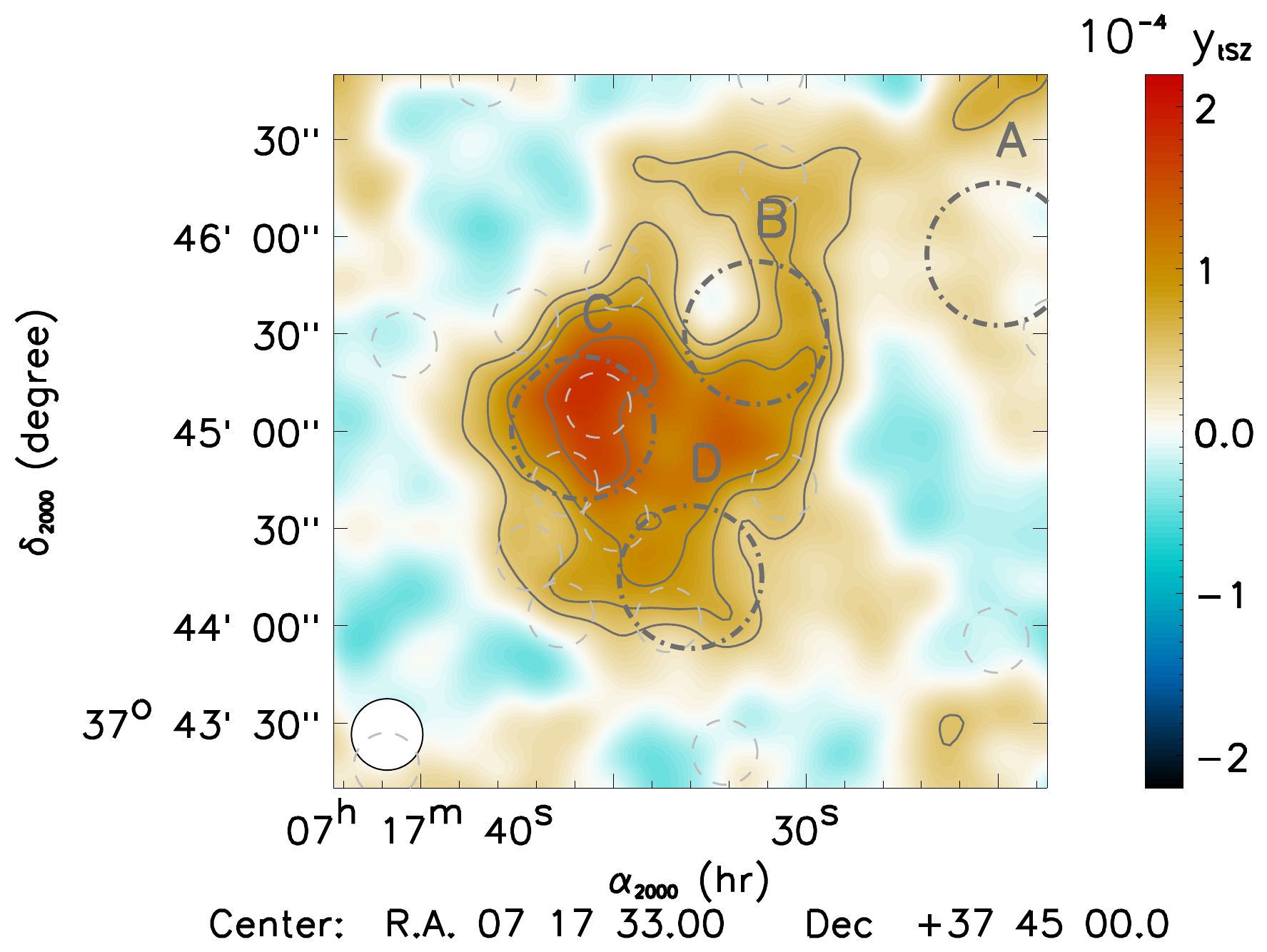}
\includegraphics[width=0.49\textwidth]{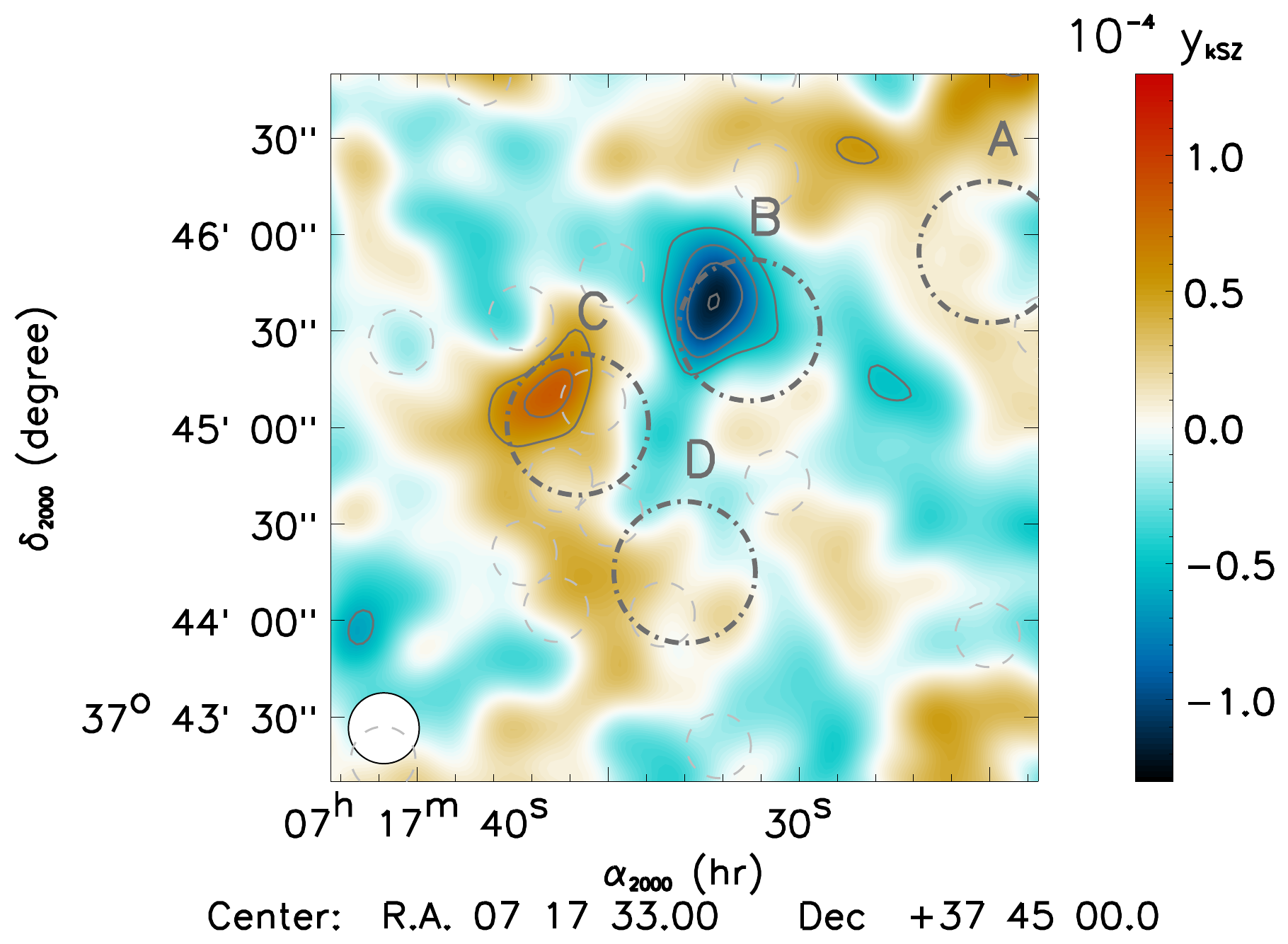}
\caption{\footnotesize{Map of the thermal SZ effect, $y_{\rm tSZ}$ (left), and the kinetic SZ effect, $y_{\rm kSZ}$ (right). Gray contours are multiples of $1 \sigma$, starting at $\pm 2 \sigma$. The map effective resolution, 22 arcsec, is shown as a white circle on the bottom left corner. Subcluster regions are represented in gray. The positions of the identified point sources (see Section \ref{sec:astrophysical_contamination_of_the_Sunyaev_Zel_dovich_signal}), from which mis-subtraction residuals remain a potential local contaminant to the signal, are represented in light gray.}}
\label{fig:tSZ_kSZ_maps}
\end{figure*}

The results of Section \ref{sec:Reconstruction_of_the_kinetic_signal_toward_sub_clusters} provide quantitative illustration of our constraints, but they directly depend on the choice of the region coordinates and their aperture size. Therefore, we now take advantage of the resolved nature of the NIKA data and consider the full surface brightness maps (in addition to the XMM-\textit{Newton} temperature map for relativistic corrections) and we use them to compute maps of the tSZ and kSZ amplitude, i.e., maps of $y_{\rm tSZ}$ and $y_{\rm kSZ}$, that are independent of our baseline regions. Before combining the two data sets, we convolve the NIKA maps to the same angular resolution, i.e., 22 arcsec. The resulting maps are provided in Figure \ref{fig:tSZ_kSZ_maps}. The residuals of point sources are not considered here, but the obtained maps are compared to the mask to check that the kSZ signal we detect is not affected by point sources. The coordinates of the point sources identified in Section \ref{sec:Radio_sources} and \ref{sec:Submillimeter_point_sources} are also represented in Figure \ref{fig:tSZ_kSZ_maps}. While these regions are potentially biased, they remain local and do not spatially correlate with the main significant kSZ structure that we observe. We observe a tSZ peak coincident with the region of subcluster C. The peak signal-to-noise ratio reaches $7.3 \sigma$, which is more than twice smaller than on the 150 GHz map due to the noise introduced by the 260~GHz data. The signal is more diffuse in the other regions but it covers the cluster extent on arcmin scale. The kSZ map shows two distinct peaks of opposite signs and similar amplitudes, reaching $-5.1$ and $+3.4 \sigma$, respectively. The negative peak is almost coincident with region B and the positive peak is located at the northeast edge of region C. No kSZ signal is significantly detected near subclusters A or D. Similarly, we do not observe any diffuse large scale kSZ signal that would result from the overall motion of the cluster and subclusters.

\subsection{Systematic effects}
In addition to the noise and astrophysical contaminants, which we account for as discussed in Section \ref{sec:astrophysical_contamination_of_the_Sunyaev_Zel_dovich_signal}, our measurements are affected by three main systematic effects, which we discuss here.

The first is the absolute calibration uncertainties of the NIKA data. A change in the absolute calibration directly impacts the kSZ measurement since it changes the relative weights of the two NIKA bands. Assuming a typical gas temperature of 10 keV, equations \ref{eq:measurement_ytsz} and \ref{eq:measurement_yksz} can be expressed as $y_{\rm tSZ} \propto -0.38 \times \Delta I_{\nu_2} + 0.62 \times \Delta I_{\nu_1}$ and $y_{\rm kSZ} \propto 0.20 \times \Delta I_{\nu_2} + 0.80 \times \Delta I_{\nu_1}$, where the numerical coefficients indicates the fraction of the maps at 260~GHz and 150~GHz that contribute to the tSZ and kSZ maps. We can see that changing our 150~GHz calibration by 7\% (the calibration uncertainty), leads to a change of 2.7\% for the tSZ map, and 1.4\% for the kSZ map. Similarly, changing our 260~GHz calibration by 12\% leads to a change of 7.4\% for the tSZ and 9.6\% for the kSZ map. While the absolute calibration uncertainty does not significantly affect the significance of detection, it can slightly change the relative strength of the signal in the different regions. If we focus on region B and C, we notice that the 150~GHz surface brightness is negative in both regions, while it changes sign at 260~GHz, being negative in region B. On the other hand, the two linear coefficients that are used to compute the kSZ map from the primary NIKA maps (0.20 and 0.80 as written above) are both positive. Therefore, if the calibration changes in the same direction at both frequencies, the changes will accumulate in region B, and they will oppose each other in region C. Similarly, if the calibration changes are opposite in sign, they will oppose each other in region B, and they will accumulate in region C. Therefore, summing the calibration uncertainties, the kSZ signal in region B can increase (or decrease) by 11.0\% while it increases (decrease) only by 8.2\% in region C. Symmetrically, the same is true when inverting B and C. Note that we apply a simple sum of the calibration errors instead of a quadratic sum because they are likely to be strongly correlated. This effect will propagate linearly to the constraint on the gas line-of-sight velocity in Section \ref{sec:constraint_on_the_gas_line_of_sight_velocity_distribution_of_MACSJ0717}, but it is small compared to the statistical uncertainties. Similar arguments apply to the tSZ map.

The second systematic effect is due to the uncertainty in the derived X-ray temperature, and the assumption that the temperature is constant along the line-of-sight. In order to estimate how changes in the temperature affect our results, we reproduce the kSZ map applying $\pm 25$\% change in the normalization of the temperature map (i.e., about twice the typical difference between XMM-\textit{Newton} and \textit{Chandra}). This allows us to test the systematic effect responsible for the temperature difference between \textit{Chandra} and XMM-\textit{Newton}. We also add Gaussian noise of amplitude 3 keV at a resolution of 22 arcsec. Due to the fact that the relativistic corrections are close to proportional in the two NIKA bands, the changes in the temperature map lead to only small changes in the kSZ significance, i.e. less than $0.15 \sigma$ for the signal peaks. The absolute amplitude of the signal, however, is affected by up to $0.4 \sigma$ changes at the positive peak, which correspond to hot gas in region C, where relativistic effects are important.

The third systematic effect is due to the large angular scale filtering that affects the NIKA data. The filtering is the same in the two bands, and as the quantities $y_{\rm tSZ}$ and $y_{\rm kSZ}$ are a linear combination of the two NIKA maps, they are affected by the same filtering. Therefore, the tSZ and kSZ reconstructed signals are smoothly apodized at scales larger than $\sim 2$ arcmin. The zero level of the kSZ map is directly given by the respective zero levels of the 150 and 260 GHz map, being set to zero in the external regions of the map, as detailed in Section \ref{sec:NIKA_observations_and_data_reduction}. While an arbitrary change in the brightness zero levels could change the kSZ signal zero level, the peak to peak difference between the positive and negative observed signal will remain unchanged. In fact, the kSZ signal we observe is much more compact than that of the tSZ effect, so we expect filtering and zero level effects to be less important for kSZ observations. The transfer function is nonetheless accounted for when fitting our data with a model as detailed in Section \ref{sec:constraint_on_the_gas_line_of_sight_velocity_distribution_of_MACSJ0717}.

\subsection{Comparison to maps at other wavelengths}\label{sec:Qualitative_comparison_to_other_wavelengths}
\begin{figure*}[h]
\centering
\includegraphics[trim=4.6cm 2.3cm 3cm 0cm, clip=true,height=5.8cm]{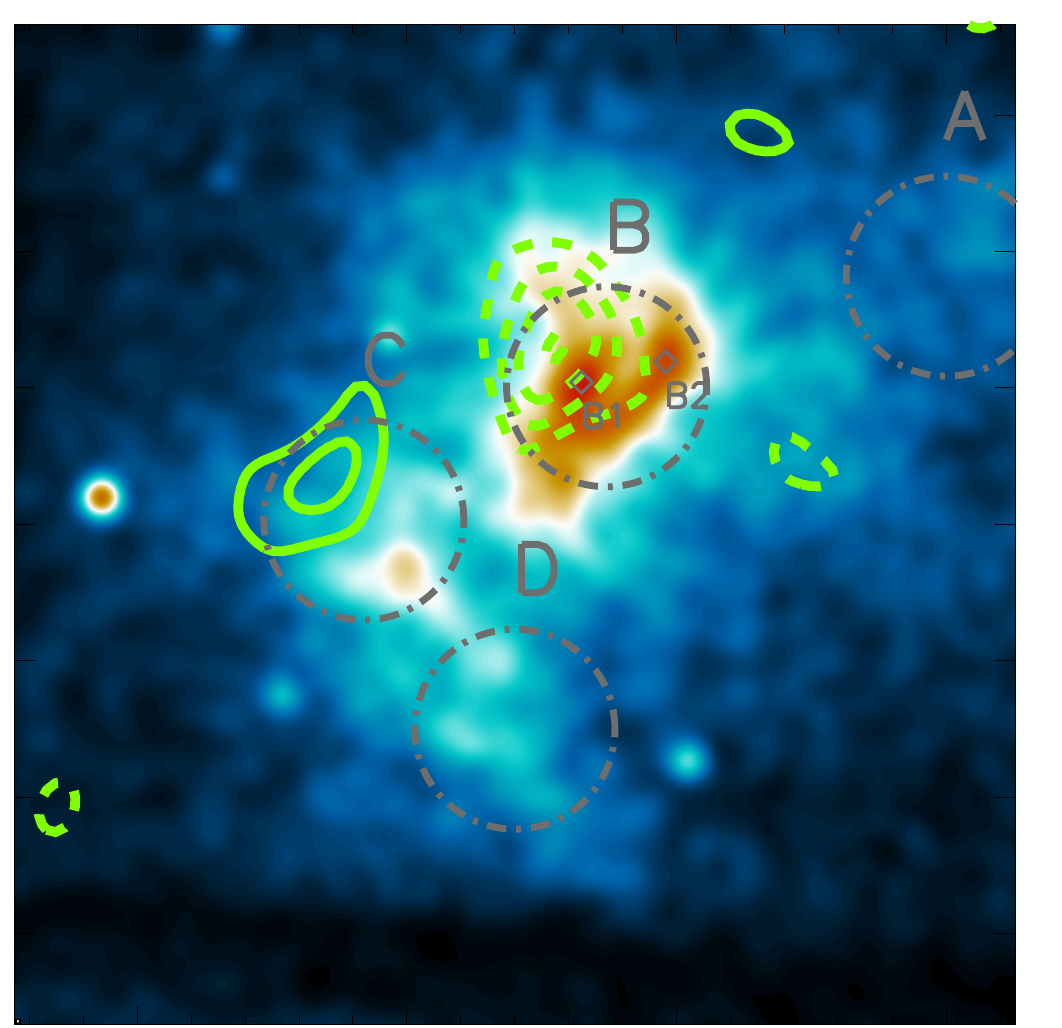}
\includegraphics[trim=4.6cm 2.3cm 3cm 0cm, clip=true,height=5.8cm]{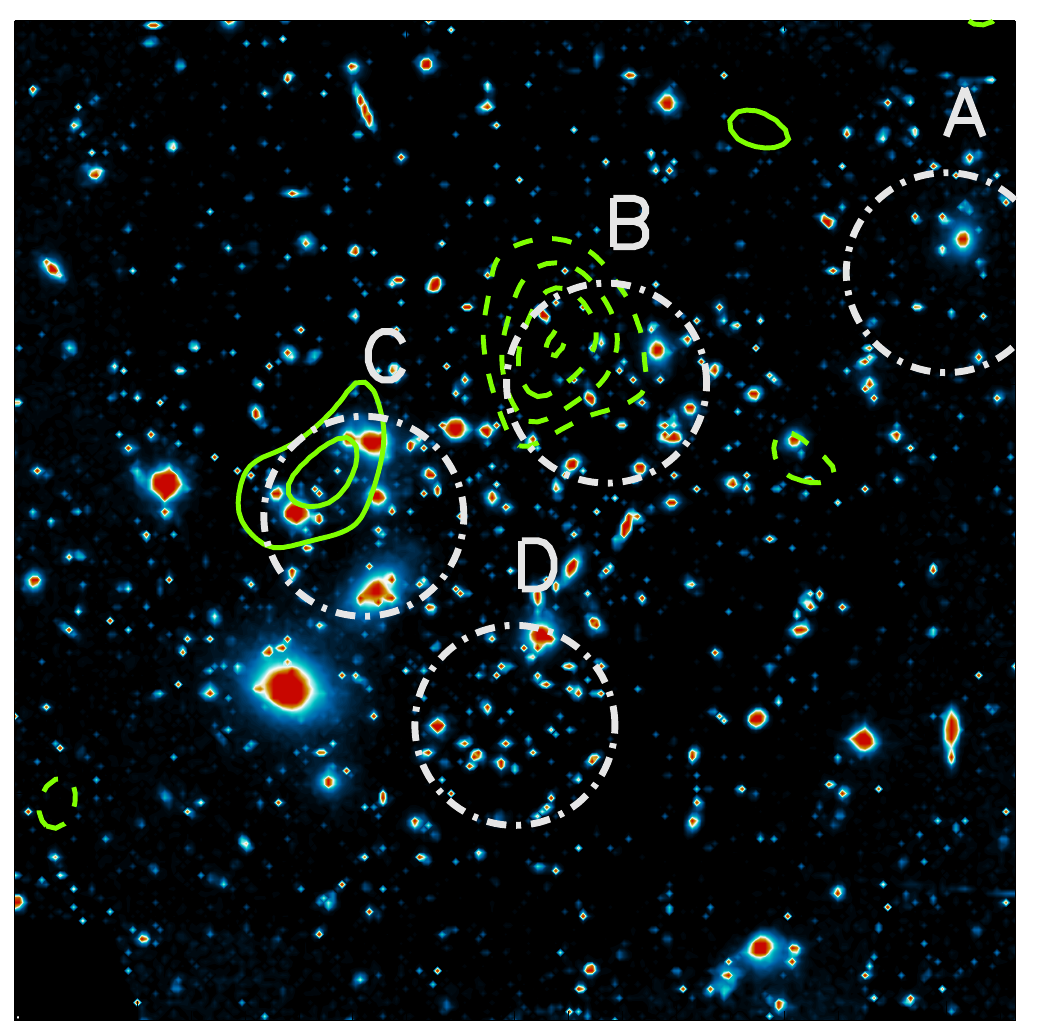}
\includegraphics[trim=4.6cm 2.3cm 3cm 0cm, clip=true,height=5.8cm]{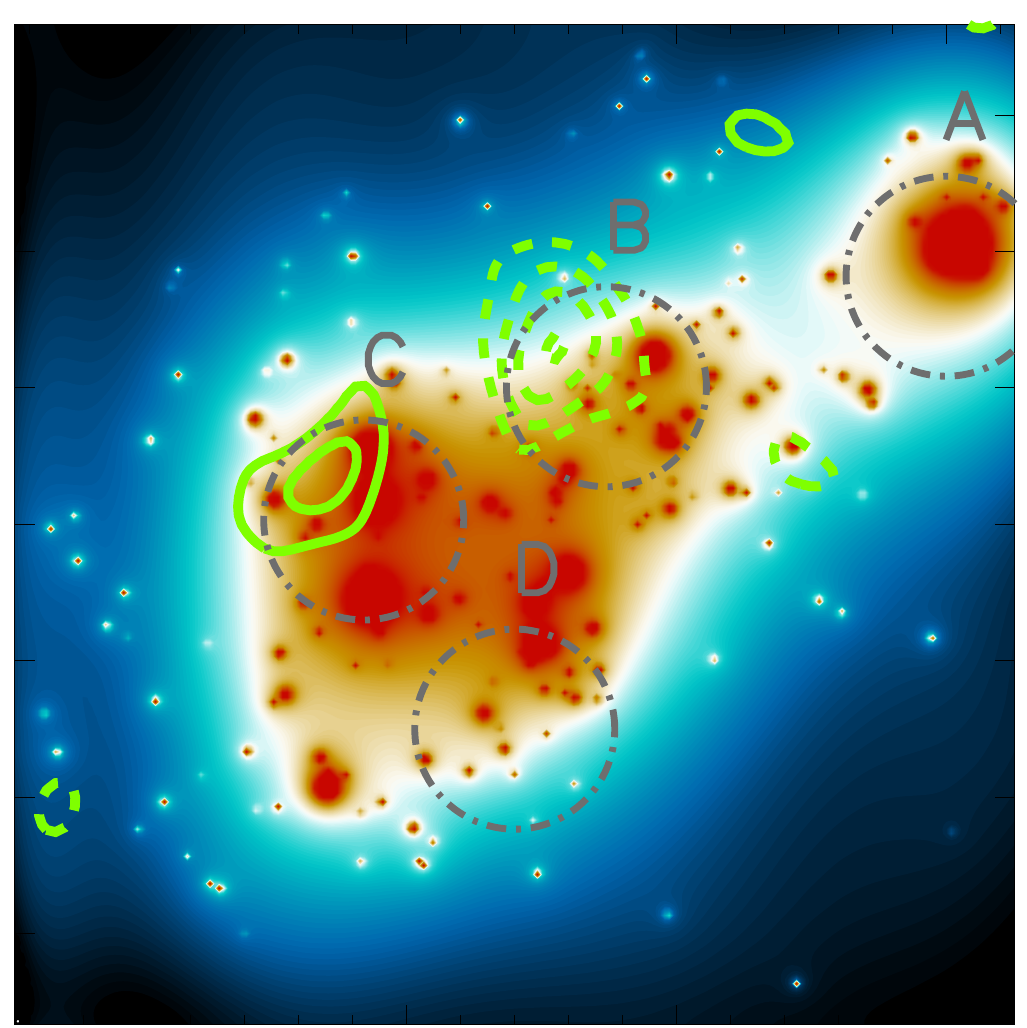}
\caption{\footnotesize{Comparison between the kSZ signal and other probes of the cluster components, shown on saturated linear scales. The astrometries are the same as those in Figure \ref{fig:tSZ_kSZ_maps}, and the green contours reproduce the kSZ signal-to-noise ratio, given as a solid line if positive (+2 and +3 $\sigma$) and dashed line if negative (-2, -3, -4 and -5 $\sigma$). Our reference regions are provided as gray circles. The positions of B1 and B2 are also shown as gray diamonds on the left image. {\bf Left:} \textit{Chandra} X-ray photon count image, smoothed to 7 arcsec FWHM. {\bf Middle:} HST optical combined image from the CLASH data \citep{Postman2012}. {\bf Right:} Lensing mass ($\kappa$) from \cite{Zitrin2011}.}}
\label{fig:tSZ_kSZ_multiL}
\end{figure*}
In Figure \ref{fig:tSZ_kSZ_multiL}, we compare the kSZ map obtained with NIKA to maps at other wavelengths, taken as tracers of the different cluster components. We use the \textit{Chandra} X-ray photon counts as a proxy for the gas distribution ($\propto n_e^2$), HST optical imaging \citep[CLASH data,][]{Postman2012} to identify the galaxy spatial distribution, and the strong lensing mass reconstruction to probe the dark matter \citep{Zitrin2011}. 

The X-ray image shows two main peaks in region B. These two peaks are also seen in SZ observations by MUSTANG at 90 GHz with a 13 arcsec effective angular resolution \citep{Mroczkowski2012}. The most significant peak is located on the east side of the region (hereafter B1), with the second on the west side (hereafter B2), the two being separated by about 20 arcsec. The group of galaxies associated with subcluster B is on average more coincident with B2, even if galaxies are also observed around B1. The dark matter map follows the distribution of the galaxies, and is therefore better aligned with B2. The negative kSZ signal coincides well with B1, but is significantly offset from B2. The morphology of clump B2 is that of an arc (or a tail) around the dense core, similar to what is observed in mergers on the plane of the sky \citep[e.g.,][for the bullet cluster]{Clowe2006}. This could suggest, if this is the case, that B2 is an independent subcluster (with respect to B1) moving mostly perpendicular to the line-of-sight, and therefore, not responsible for the kSZ signal. The signal associated to B2 could also be due to the stripped gas of the main core B1, which would therefore move slower than B1 and is not significantly detected in kSZ. In both cases, our kSZ data, together with X-ray imaging, suggest that subcluster B is a complex object in itself. The main structure, B1, is rapidly moving along the line-of-sight, while the second one, B2, is moving either mostly perpendicular to the line-of-sight, or in the same direction than B1, but slower.

The positive kSZ signal coincides well with the northern half of the galaxies associated with subcluster C, while the other half of the galaxies are coincident with no kSZ detection. The X-ray peak is centered on the southern part of the galaxy group and the offset between the kSZ peak and the X-ray peak is about 20 arcsec. As this subcluster is highly disturbed, we can expect strong inhomogeneities in the gas velocity distribution, but any further interpretation would require higher signal-to-noise ratio kSZ imaging. Nonetheless, it is worth noting that the galaxy velocity dispersion within this subcluster is very high \citep[$1761^{+234}_{-607}$ km/s for 10 redshifts, see][]{Ma2009}, which could also suggest that the underlying distribution is multimodal.

No kSZ signal is detected around subcluster D but we observe an offset between the gas and the brightest galaxies and dark matter, the gas density peak being about 25 arcsec shifted to the south with respect to the other component. Subcluster A is not clearly identified in the X-ray data, but the galaxy and dark matter locations are consistent. Finally, the foreground radio galaxy, F, is visible in X-ray and in optical.

\section{Constraints on the gas line-of-sight velocity distribution of \mbox{MACS~J0717.5+3745}}\label{sec:constraint_on_the_gas_line_of_sight_velocity_distribution_of_MACSJ0717}
The detection of the kSZ effect implies the detection of gas motion, and opens up the possibility of studying the velocity distribution of the gas contained within the cluster, i.e. gas substructure. We detect a positive line-of-sight gas velocity in the direction of subcluster B, i.e., the gas is moving away from us in the CMB reference frame, and negative gas line-of-sight velocity toward subcluster C, i.e. the gas is moving toward us. However, measuring the absolute line-of-sight velocity from our kSZ map requires us disentangling it from the integrated line-of-sight electron density (see equations \ref{eq:dIsz} and \ref{eq:yksz}), accounting for projection effects, and this cannot be done from our SZ observations alone. In this Section, we model the cluster gas distribution in terms of electron density and gas line-of-sight velocity, and we fit our data using extra constraints available from the X-ray spectroscopic temperature. Unlike the results presented in Section \ref{sec:A_map_of_the_kinetic_Sunyaev_Zel_dovich_signal}, the results we detail in this Section are therefore dependent on the model we use to describe the cluster. In particular, we stress that any mis-modeling of the gas density, which is intrinsically subject to strong assumptions for such a complex object, will be reflected in the optical depth, that is itself degenerate with the gas line-of-sight velocity.

\subsection{Physical modeling of the gas distribution of \mbox{MACS~J0717.5+3745}}\label{sec:Physical_modeling_of_the_gas_distribution}
As discussed in Section \ref{sec:Introduction}, \mbox{MACS~J0717.5+3745} is known to be multimodal with four main subclusters. Therefore, we model the cluster with a set of four spherically symmetric subclusters describing the gas distribution associated to each subcluster. We model each electron density profile with a $\beta$--model \citep{Cavaliere1978},
\begin{equation}
	n_e^{(i)}(r) = n_{e0}^{(i)} \left[1+\left(\frac{r}{r_c^{(i)}}\right)^2 \right]^{-3 \beta^{(i)} /2},
\label{eq:beta_model}
\end{equation}
where $i$ labels each subcluster, $n_{e0}$ is the central electron density, $r_c$ the core radius and $\beta$ is the outer slope of the profile. We also assume that the subclusters are isothermal, with a gas temperature $T_{\rm x}^{(i)}$ estimated from XMM-\textit{Newton} within 30 arcsec radius apertures, as reported in Table \ref{tab:subcluster_Tx_measurements}. The line-of-sight gas velocity of each subcluster is also assumed to be a constant, i.e. each subcluster has its own bulk velocity. The choice of the $\beta$--model is motivated by its simplicity and by the limited number of parameters needed to describe the gas density. Moreover, it is not guaranteed that a more complex radial profile would provide a better description of the data since it is the spherical symmetry assumption that is the major limitation of our model.

The NIKA surface brightness maps are the primary observables to which we compare our model, and they are modeled as
\begin{eqnarray}
\frac{\Delta I_{\nu}^{\rm model}}{I_0} & = & \sigma_{\rm T} \sum\limits_{i} f_{\nu}(T_{\rm x}^{(i)}) \frac{k_B T_{\rm x}^{(i)}}{m_e c^2} \int n_e^{(i)} dl \nonumber \\
       & + & \sigma_T \sum\limits_{i} g_{\nu}(T_{\rm x}^{(i)}) \frac{-v_z^{(i)}}{c} \int n_e^{(i)} dl, \nonumber \\	        
\label{eq:Inu_model}
\end{eqnarray}
where $T_{\rm x}^{(i)}$ and $v_z^{(i)}$ are scalar quantities. The line-of-sight integrations are computed analytically for each sky pixel as
\begin{equation}
	\int^{+\infty}_{-\infty} n_e^{(i)} dl = \sqrt{\pi} \ n_{e0}^{(i)} r_c^{(i)}\frac{\Gamma\left(\frac{3}{2} \beta^{(i)} -\frac{1}{2}\right)}{\Gamma\left(\frac{3}{2} \beta^{(i)}\right)} \left[1+\left(\frac{R^{(i)}}{r_c^{(i)}}\right)^2 \right]^{\frac{1}{2}-\frac{3 \beta^{(i)}}{2}},
\label{eq:beta_model_integ}
\end{equation}
where $R$ is the projected radius from each subcluster centers and the quantity $\Gamma$ represents the Gamma function. Since the SZ surface brightness depends linearly on the electron density, the contribution of each subcluster is summed regardless of cross-cluster terms that would appear otherwise (e.g., for the X-ray surface brightness, which depends on $n_e^2$). This allows us to ignore the exact coordinate of each subcluster along the line-of-sight. The projected location of the subclusters centers, however, are parameters of our model.

\subsection{Fitting algorithm}\label{sec:Fitting_algorithm}
The fitting approach we use consists of processing test simulations to incorporate the same observational effects that affect our data (see Section \ref{sec:NIKA_observations_and_data_reduction}), in order to predict the observable signal that we can directly compare to the SZ surface brightness maps. The SZ maps, obtained for each point of the parameter space that is explored are convolved to the beam and the transfer function of the NIKA processing. We also account for the overall zero level of the maps individually using two nuisance parameters, because they are not constrained by the NIKA observations\footnote{This corresponds to the transfer function being zero at angular wavenumber $k=0$.}. The zero level is, however, not correlated to the constraint on the velocity, corresponding to the fact that the kSZ map does not show any signal on large scales. Indeed, while the value of the $\chi^2$ would change with the zero level, the local peaks observed in the kSZ signal of Figure \ref{fig:tSZ_kSZ_maps}, which are related to the constraints on the velocity we obtain, cannot be accounted for by any offset in the maps. The parameter space is sampled using Monte Carlo Markov Chains (MCMC). The Metropolis-Hasting algorithm \citep[e.g.,][]{Chib1995} is used to define the evolution of the chains, by using the Gaussian log likelihood computed as
\begin{equation}
\mathcal{L} \propto \sum\limits_{m,n} \left(\Delta I_{\rm NIKA}^{\rm data} - \Delta I_{\rm NIKA}^{\rm model}\right)_m \left(C_{\rm NIKA}^{-1}\right)_{mn} \left(\Delta I_{\rm NIKA}^{\rm data} - \Delta I_{\rm NIKA}^{\rm model}\right)_n,
\label{eq:likelihood1}
\end{equation}
where $m$ and $n$ label sky pixels and subscript NIKA stands for the joint 150 GHz and the 260 GHz data. The quantity $C_{\rm NIKA}$ is the full noise covariance matrix of the NIKA surface brightness maps, including the noise correlation between the two bands (see Section \ref{sec:Diffuse_galactic_emission_and_cosmological_background}). Pixels that are potentially contaminated by point source residuals are rejected using the mask of Figure \ref{fig:point_source_modeling}, so that they do not contribute to the likelihood, and therefore, do not affect the results.

Our model includes a total of 26 parameters ($4 \times 3$ for the density, 4 for the line-of-sight velocity, $2 \times 4$ for the coordinates, and 2 for the zero level of the maps). Even if the model describing each subcluster is relatively simple, the overall complexity of the cluster, and in particular the degeneracies between the characteristic radius, slope parameters, and subcluster centroids (which are not well determined), does not allow us to let all parameters free. In particular, the slopes of the density profile are not well constrained by our data. Therefore they are varied but limited to the range [1/3, 3]\footnote{The lower limit is required from equation \ref{eq:beta_model_integ}, since the argument of the Gamma function, $\frac{3}{2} \beta^{(i)} -\frac{1}{2}$, must be larger than zero.}, in order to marginalize over their uncertainties and thereby avoid diverging chains in the MCMC. We also apply a 10 arcsec standard deviation Gaussian prior on the coordinates of the subclusters, centered on the coordinates given in Table~\ref{tab:coord_region}. The model is symmetric under the permutation of subclusters, and this ensures that we avoid exchanges between the different subclusters during the evolution of the chains, but we check a posteriori that the best-fit coordinates are consistent with our prior values.

In the case of subcluster A, the observed signal is almost consistent with noise at the $3 \sigma$ level at 150 GHz and is not detected at 260 GHz. The velocity, which is fully degenerate with the gas density, is therefore constrained at the $3 \sigma$ level. This prevents the MCMC sampling of the posterior likelihood from converging. Therefore, we do not fit for the velocity of subcluster A and set $v_z^{(A)} = 0$.

The convergence of the chains are verified using the Gelman--Rubin convergence criteria \citep{Gelman1992}. After convergence, we remove a burn-in phase and account for the correlation length of the chains. Finally, the chains histograms provide the probability density function in parameter space.

\subsection{X-ray prior on the gas density}
The fitting algorithm discussed in Section \ref{sec:Fitting_algorithm} does not use X-ray imaging information because it suffers from projection effects that are not accounted for in our model. Our model would require additional parameters to predict the expected X-ray surface brightness because, unlike for SZ images, X-ray images requires the knowledge of the relative coordinates of the subclusters along the line-of-sight. This is due to cross-terms between subclusters, which appear when squaring the density, and would be responsible for additional emission on large scales. However, NIKA SZ data alone are subject to strong degeneracies between the line-of-sight velocity and the optical depth, which can be broken with additional information from X-rays. This can significantly improve the constraint on the gas velocity, but at the cost of requiring extra assumptions on the line-of-sight geometry of the cluster.

In the following, we therefore consider two cases when fitting the model of Section \ref{sec:Physical_modeling_of_the_gas_distribution} to our data.
\begin{enumerate}
\item As our baseline, hereafter F1, we do not consider X-ray imaging. The fit relies only on SZ imaging and X-ray estimates of the gas temperature (see Table \ref{tab:subcluster_Tx_measurements}). In this case, we do not use all the available information to constrain the velocity, but we are not significantly affected by assumptions concerning the line-of-sight distribution of the gas.
\item The second fit, hereafter F2, makes use of the extra information from X-ray imaging. In this case, we consider the deprojected XMM-\textit{Newton} density profile centered on the X-ray peak \citep[extracted as in][]{Adam2016}, i.e. within a few arcsec of region B. We fit the profile with a $\beta$--model and use the constraint obtained on $n_{e0}$ (the parameter that is the most degenerate with $v_z$) as a prior in the fit described in Section \ref{sec:Fitting_algorithm}. Such an approach is only possible toward subcluster B because any density profile extraction in the other regions, i.e. away from the X-ray peak, would be strongly affected by deviations from spherical symmetry.
\end{enumerate}
While fit F1 is conservative, fit F2 allows us to break the $v_z$ -- $\tau$ degeneracy, and to check the consistency between our SZ-based constraint and X-ray data.

\subsection{Constraints on the velocity}\label{sec:Constraints_on_the_velocity}
\begin{figure*}[h]
\centering
\includegraphics[trim=0cm 0.7cm 3.33cm 0cm, clip=true, totalheight=4cm]{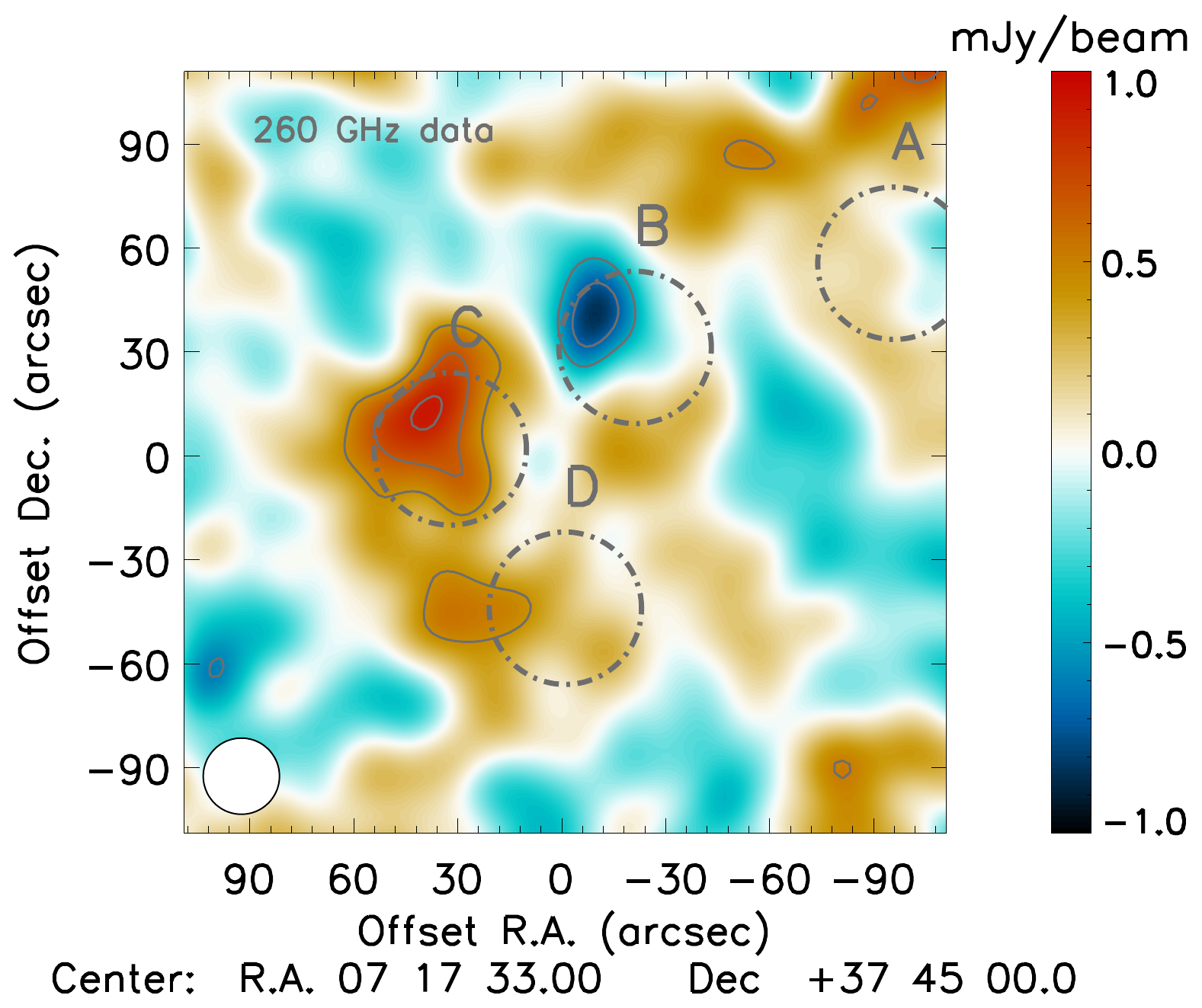}
\includegraphics[trim=2.3cm 0.7cm 3.33cm 0cm, clip=true, totalheight=4cm]{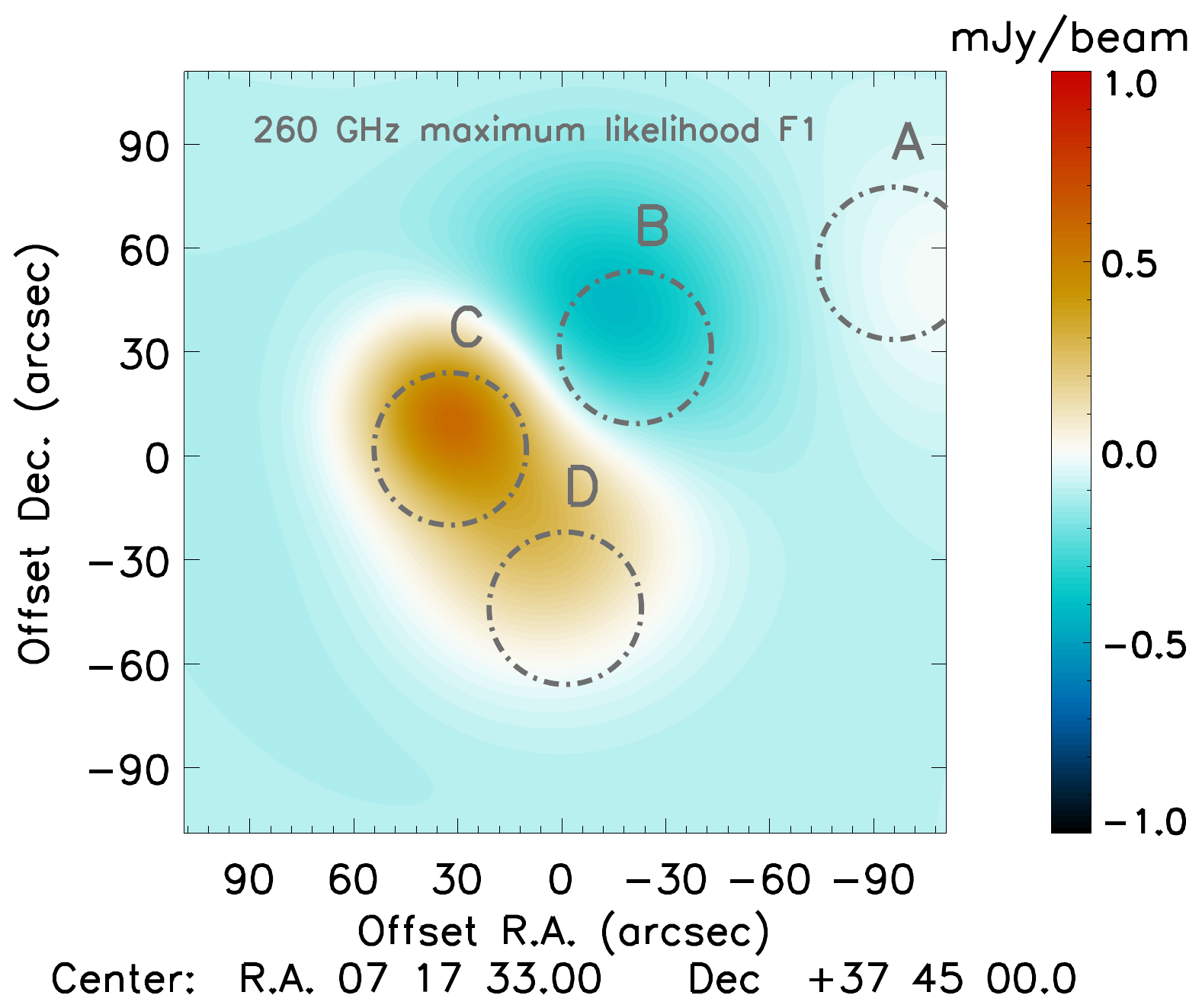}
\includegraphics[trim=2.3cm 0.7cm 3.33cm 0cm, clip=true, totalheight=4cm]{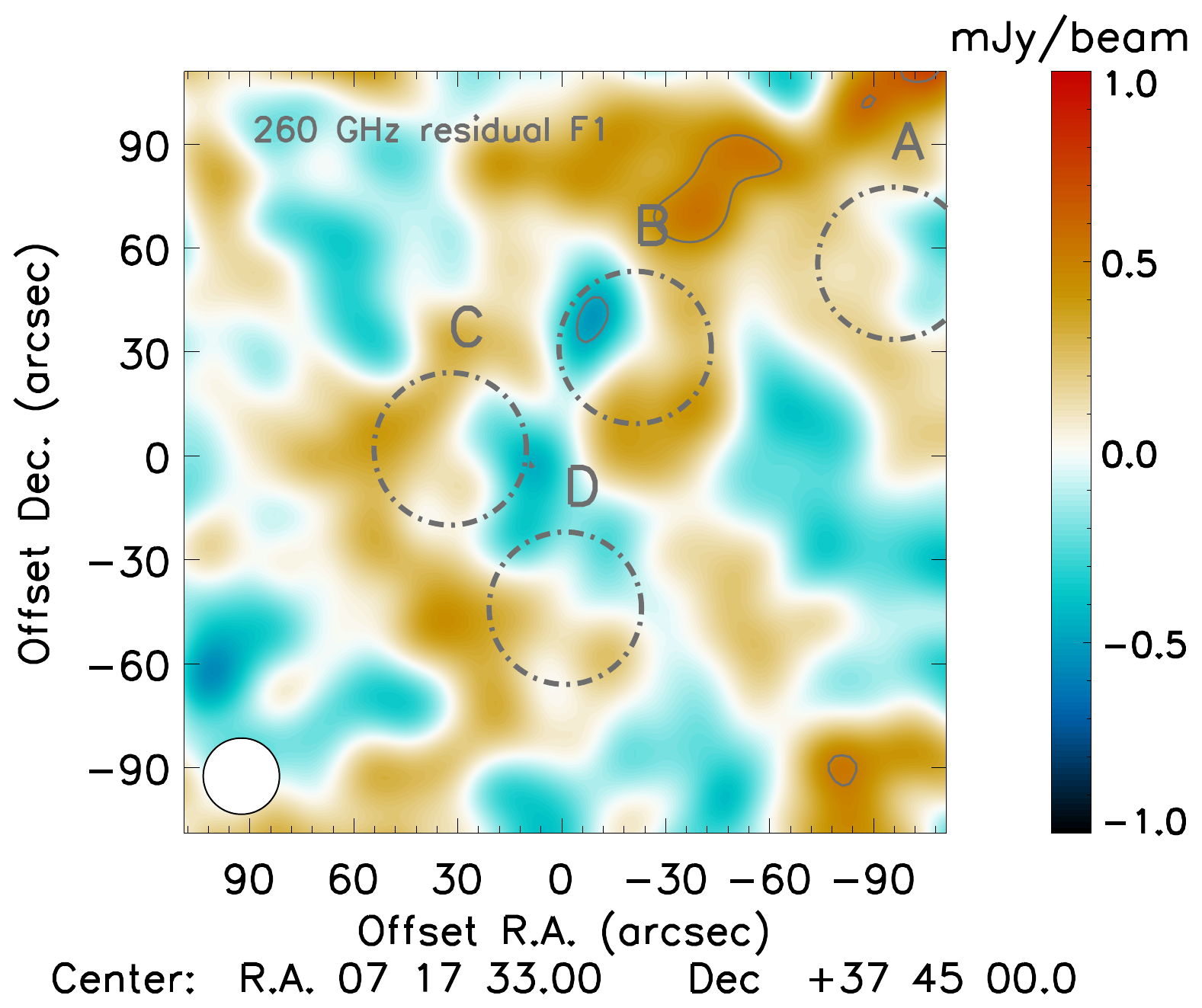}
\includegraphics[trim=2.3cm 0.7cm 3.33cm 0cm, clip=true, totalheight=4cm]{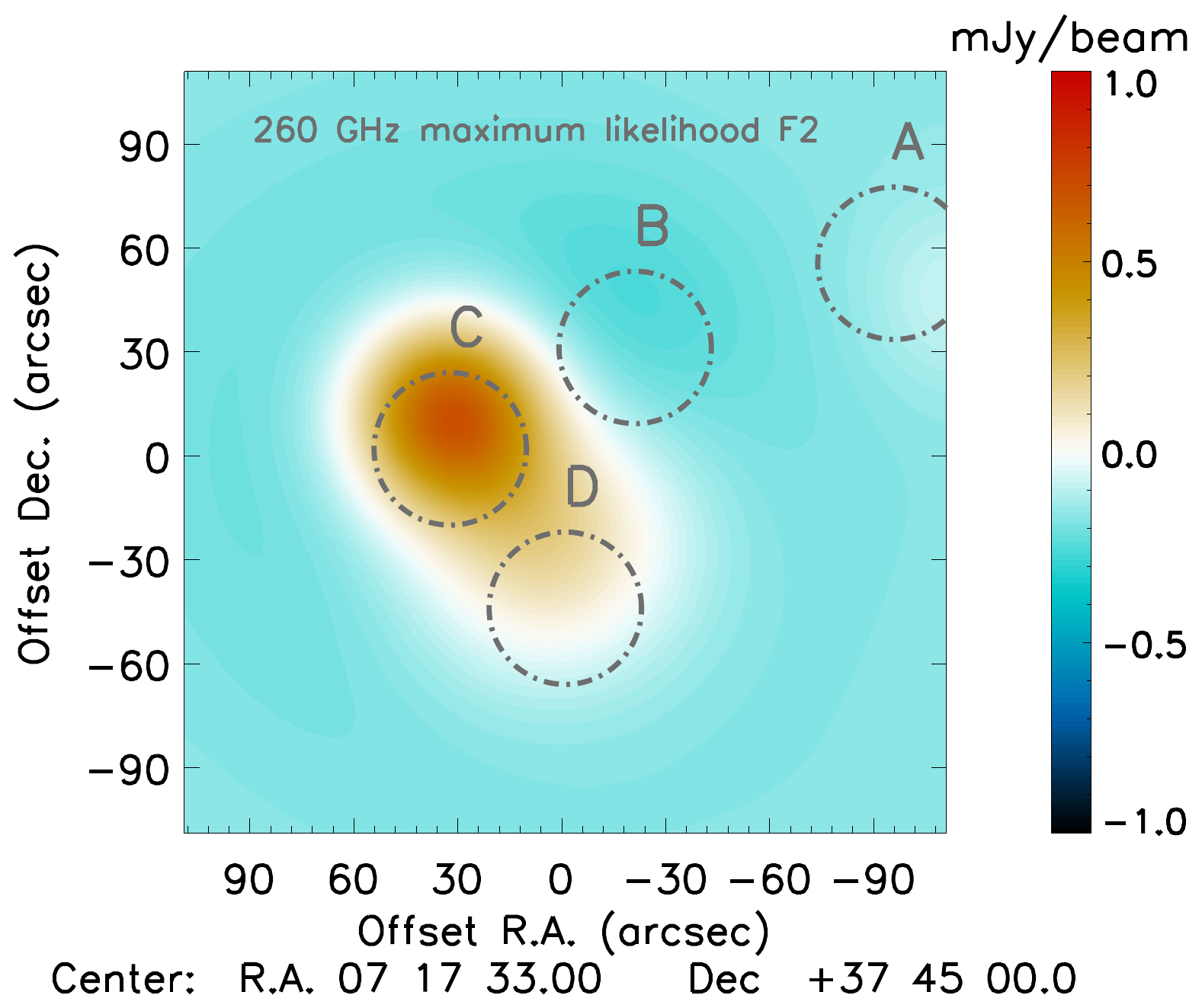}
\includegraphics[trim=2.3cm 0.7cm 0cm 0cm, clip=true, totalheight=4cm]{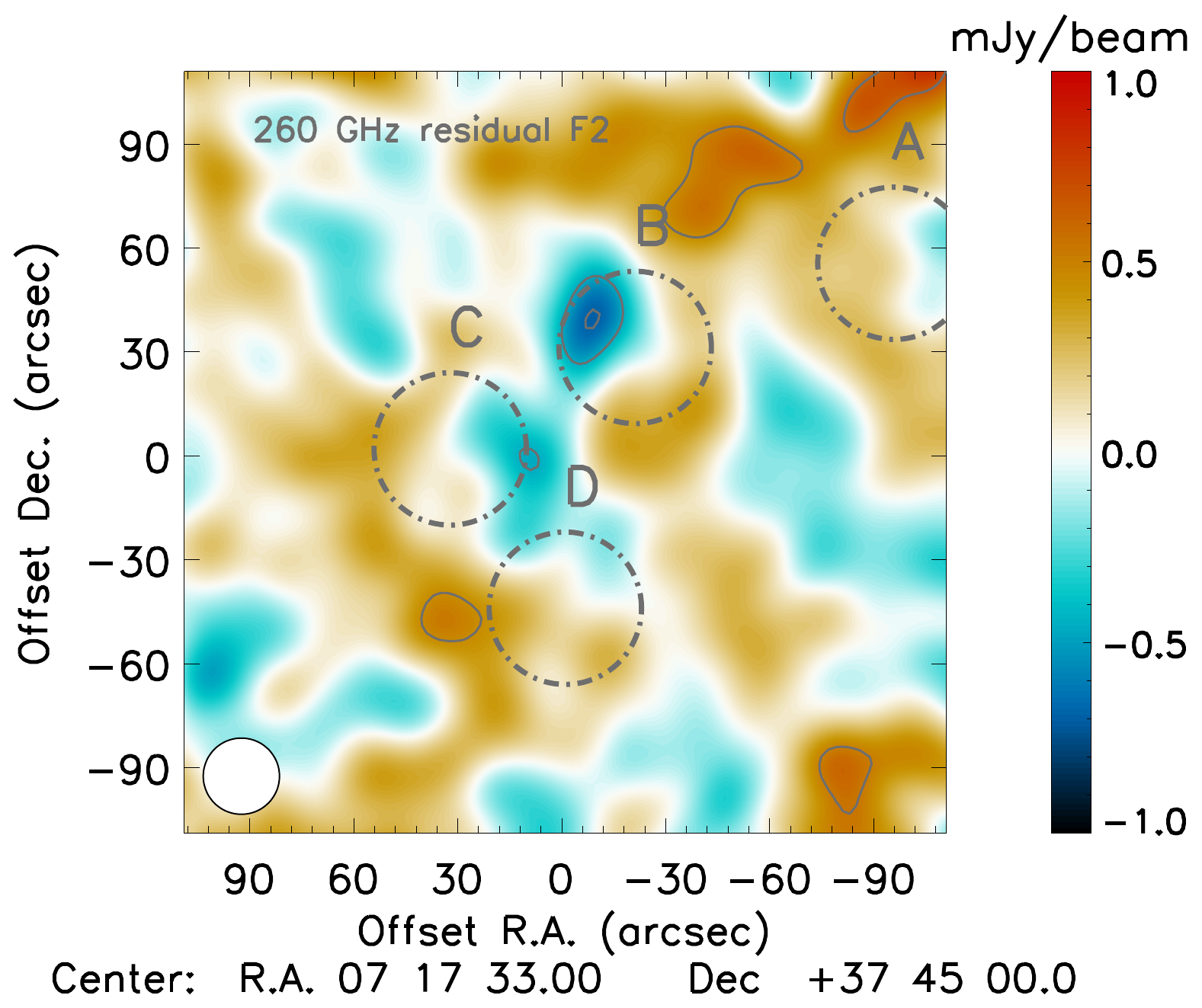}
\includegraphics[trim=0cm 0.7cm 3.33cm 0cm, clip=true, totalheight=4cm]{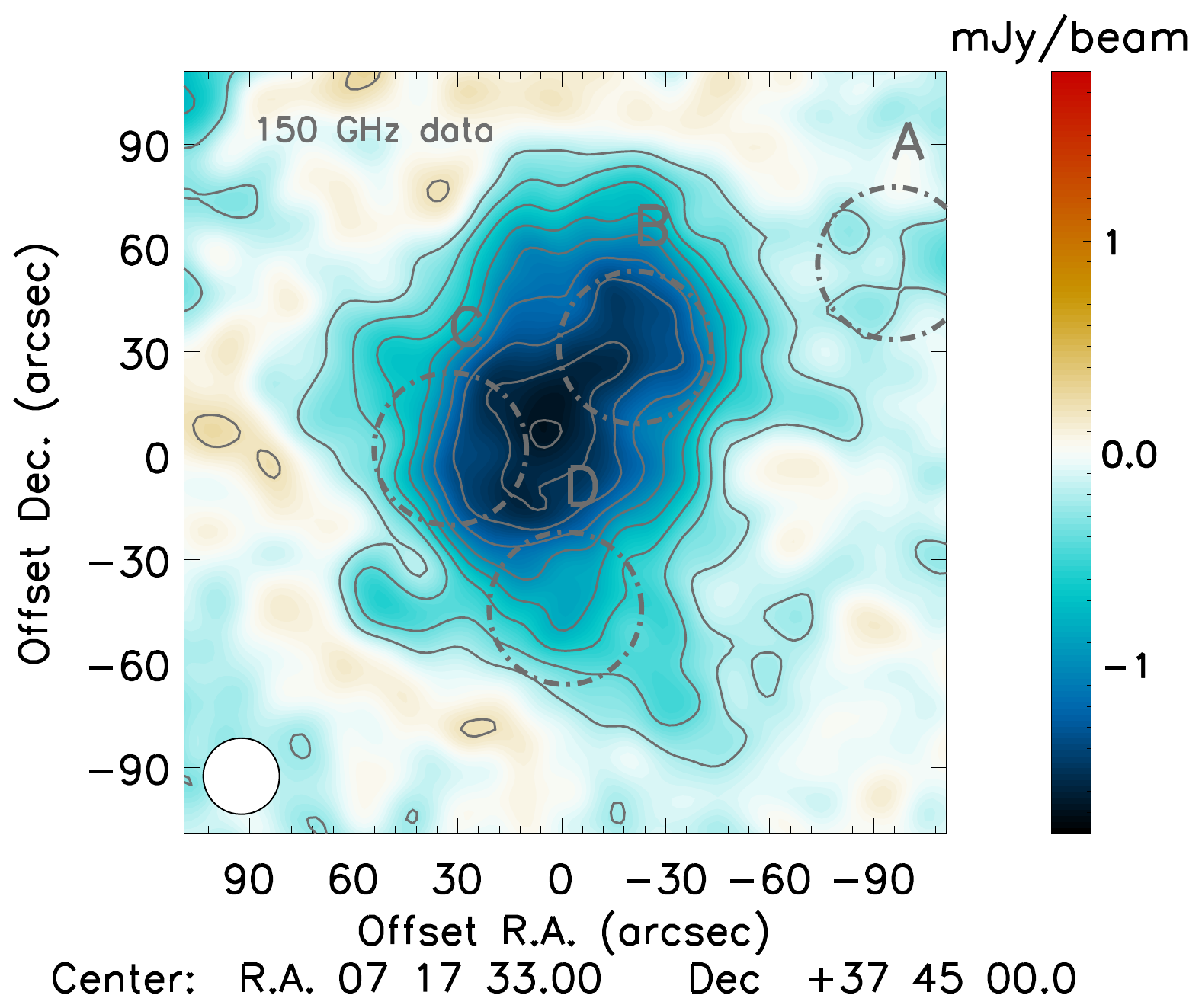}
\includegraphics[trim=2.3cm 0.7cm 3.33cm 0cm, clip=true, totalheight=4cm]{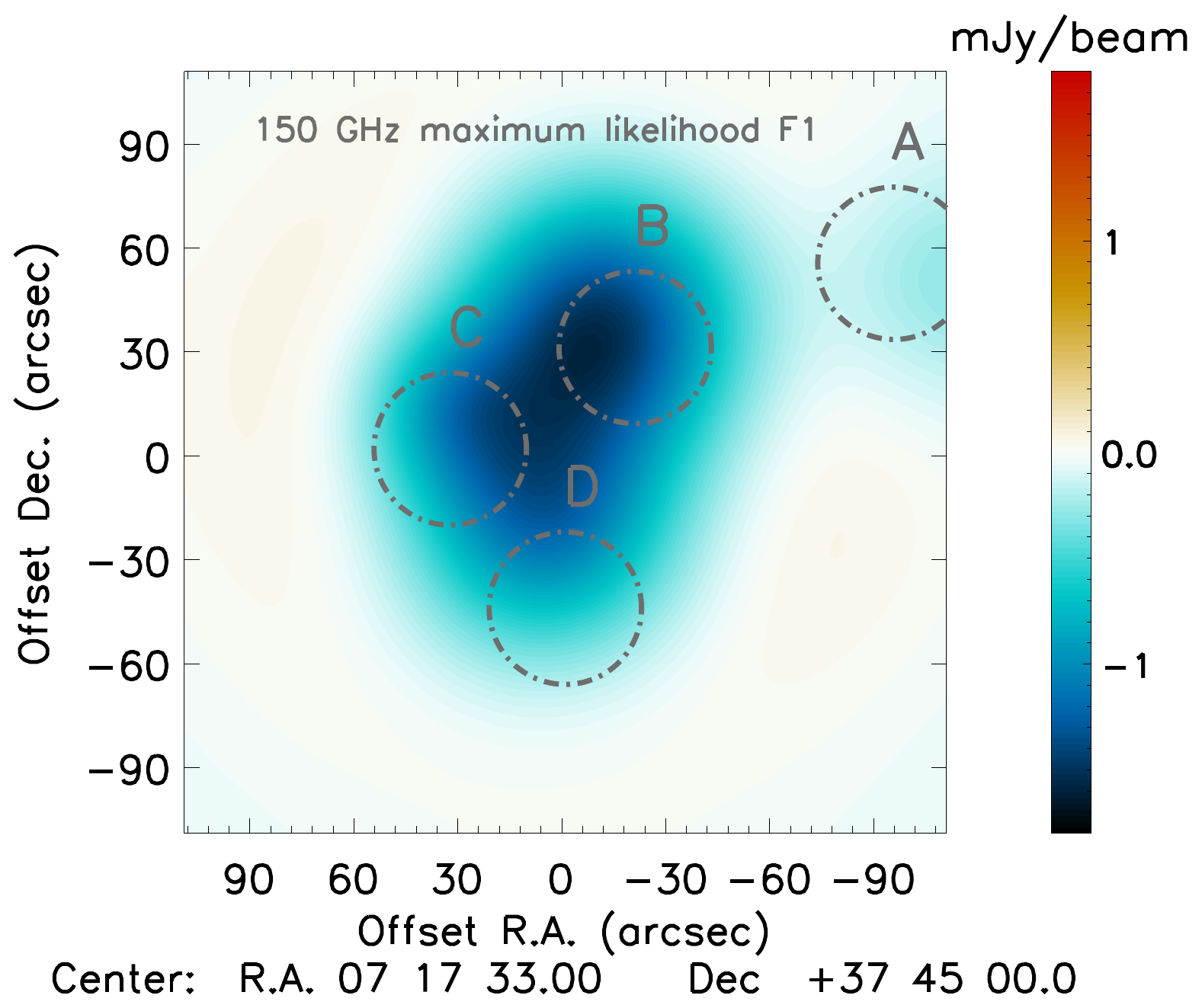}
\includegraphics[trim=2.3cm 0.7cm 3.33cm 0cm, clip=true, totalheight=4cm]{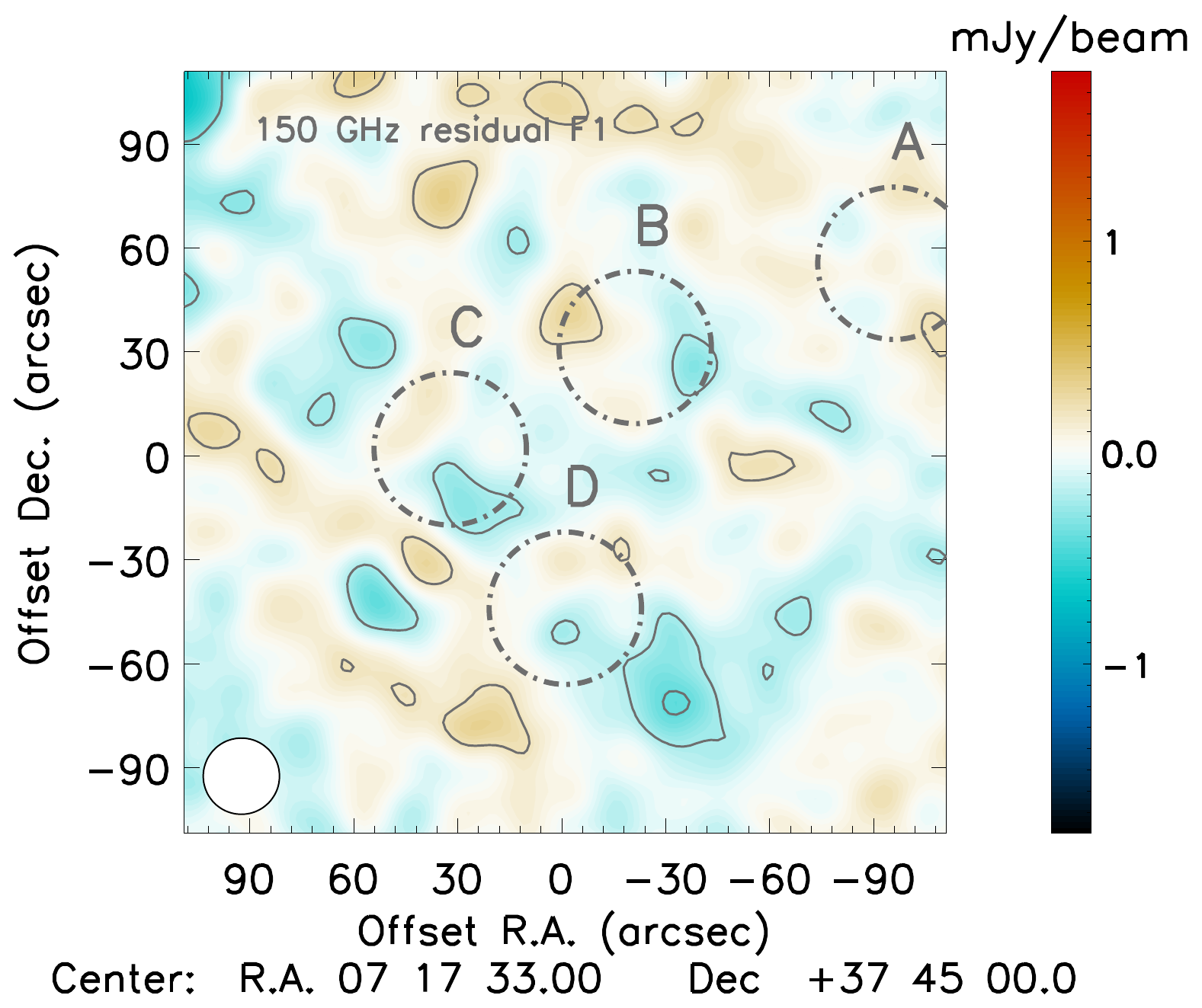}
\includegraphics[trim=2.3cm 0.7cm 3.33cm 0cm, clip=true, totalheight=4cm]{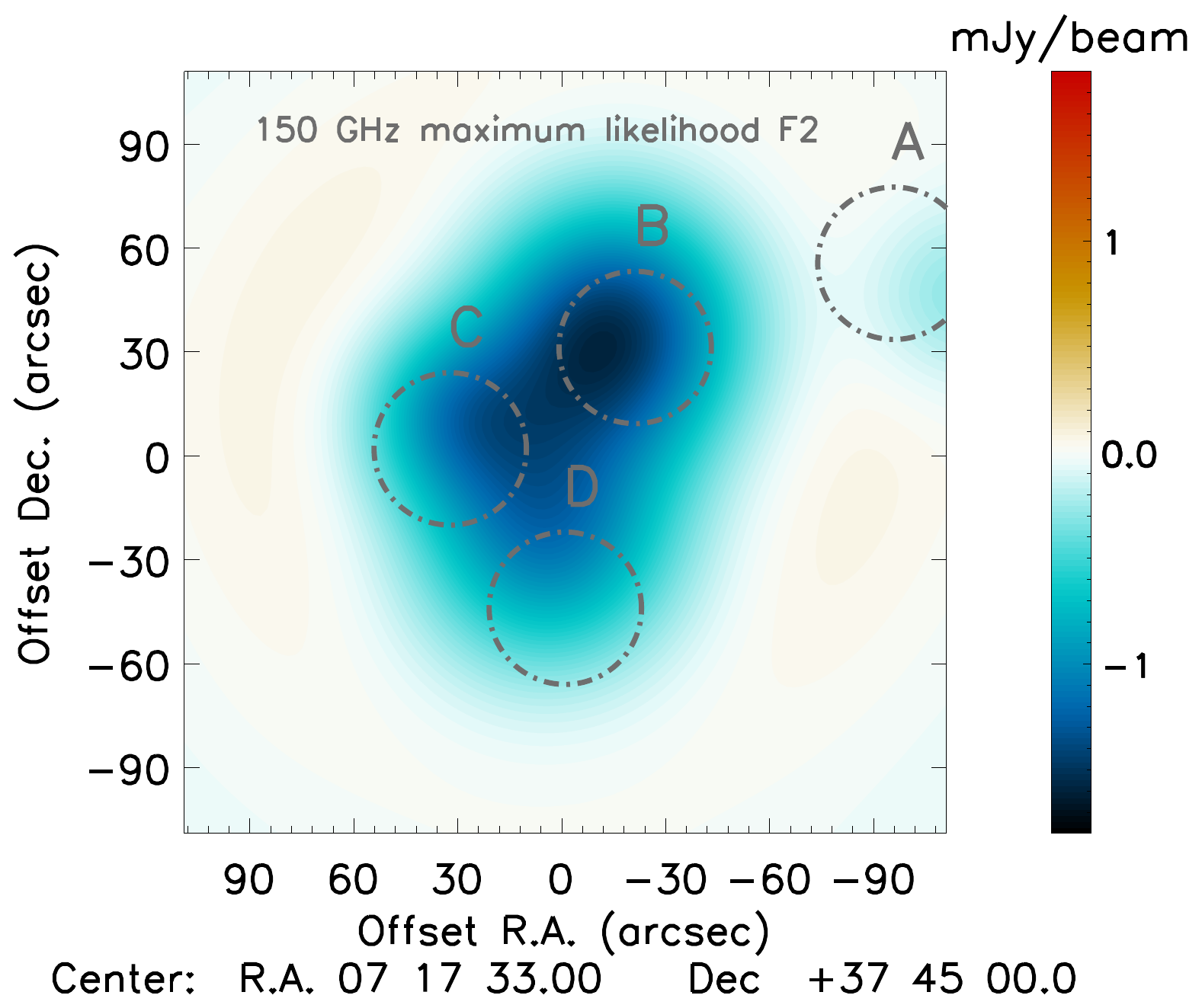}
\includegraphics[trim=2.3cm 0.7cm 0cm 0cm, clip=true, totalheight=4cm]{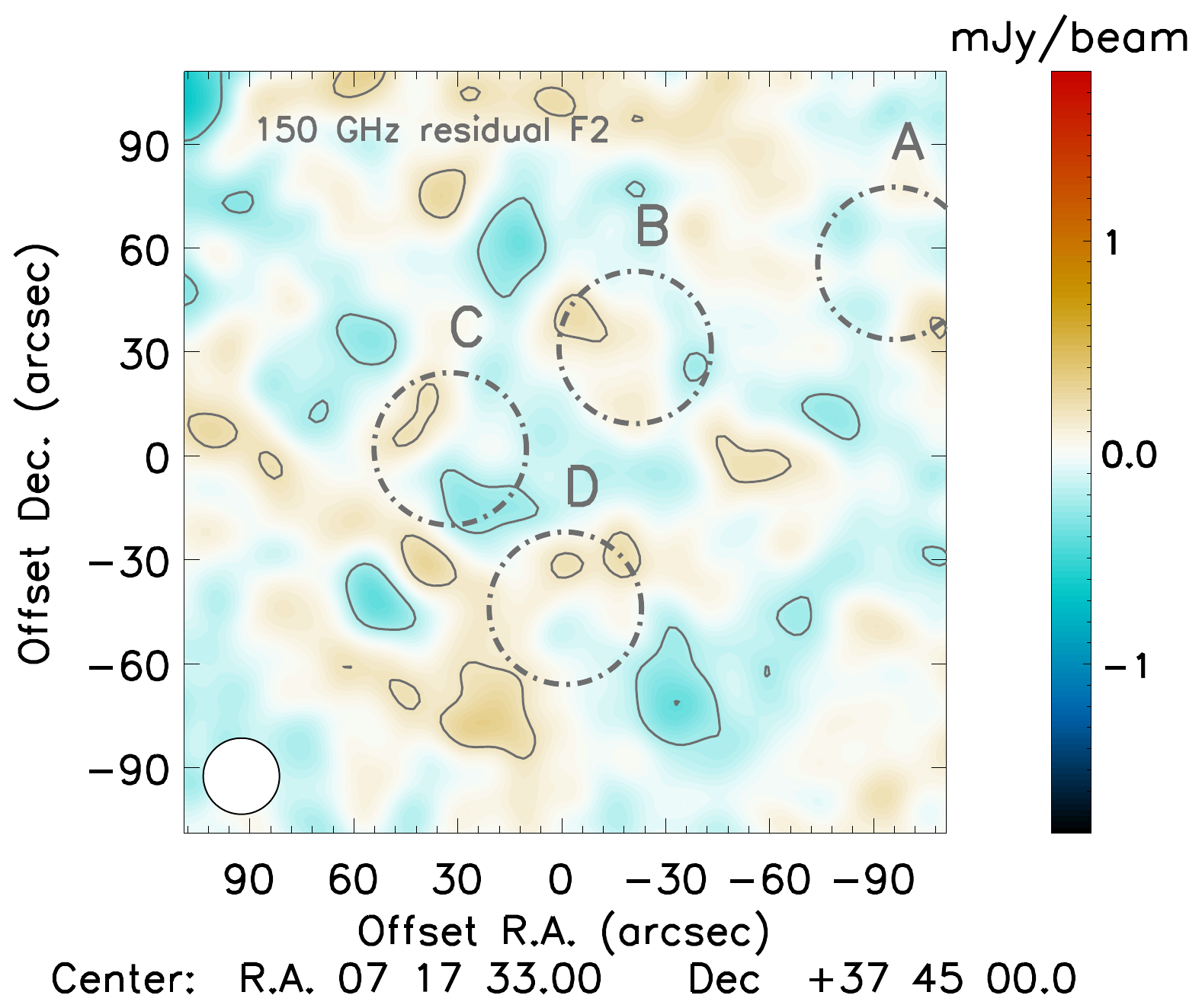}
\caption{\footnotesize{Comparison between the data and the best-fit model. From left to right, the figure shows the input data subtracted from point sources, the best-fit model F1, the residual F1, the best-fit model F2, and the residual F2. The top row provides the 260 GHz data and the bottom row the 150 GHz data. Contours are multiples of $2 \sigma$ at 150 GHz and $1 \sigma$ at 260 GHz, starting at $\pm 2 \sigma$.}}
\label{fig:best_fit_maps}
\end{figure*}

The results of the MCMC fit F1 and F2 are shown in Figure \ref{fig:best_fit_maps}, including the input data, the maximum likelihood model, and the residual between the two. Despite the complexity of the cluster, a simple multi-component $\beta$--model is able to reproduce the SZ surface brightness data fairly well in both cases.

The 150 GHz residual map is consistent with noise over most of the cluster extent for both F1 and F2, but we observe several features exceeding $2 \sigma$. In the southwest region, an excess larger than $4 \sigma$ is observed for both F1 and F2, indicating significant deviation from spherical symmetry for subcluster D. We also observe a significant negative excess in the direction of the foreground galaxy F, indicating that our radio model predicts a larger flux than measured, perhaps because we have neglected the possible steepening of its spectrum at high frequencies. The corresponding pixels are masked in the fit (see Section \ref{sec:Construction_of_a_mask_for_point_sources}).

\begin{figure*}[h]
\centering
\includegraphics[width=0.33\textwidth]{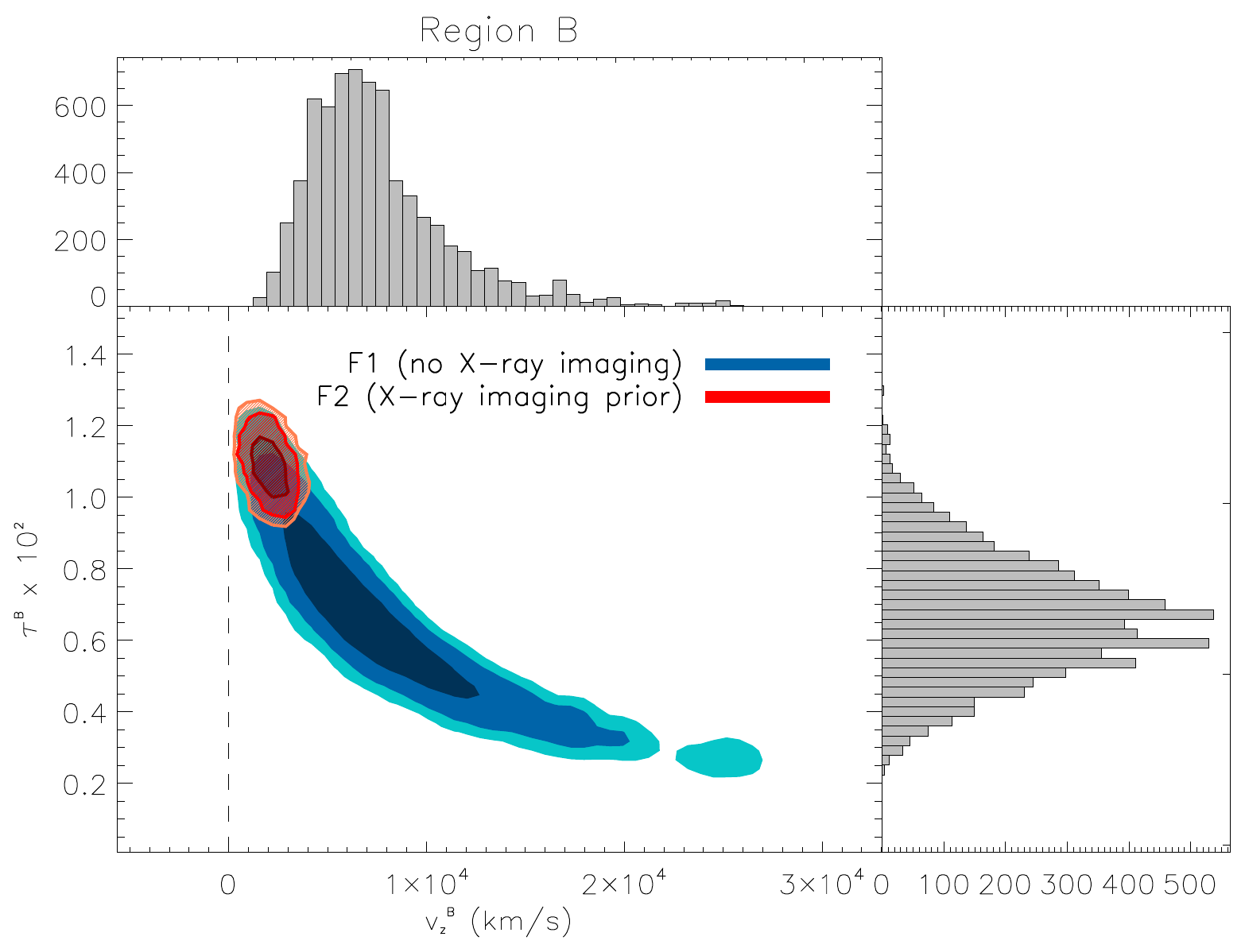}
\includegraphics[width=0.33\textwidth]{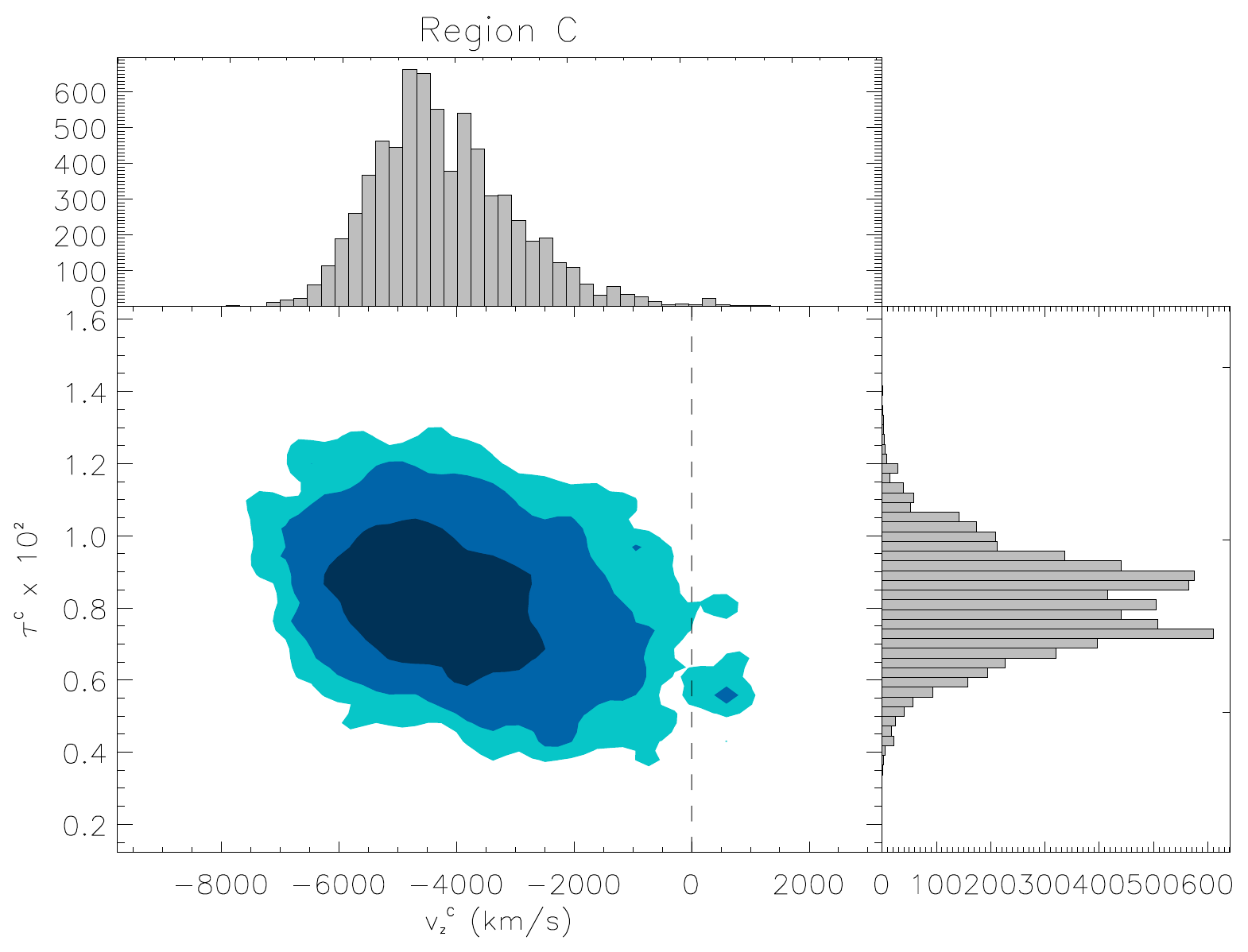}
\includegraphics[width=0.33\textwidth]{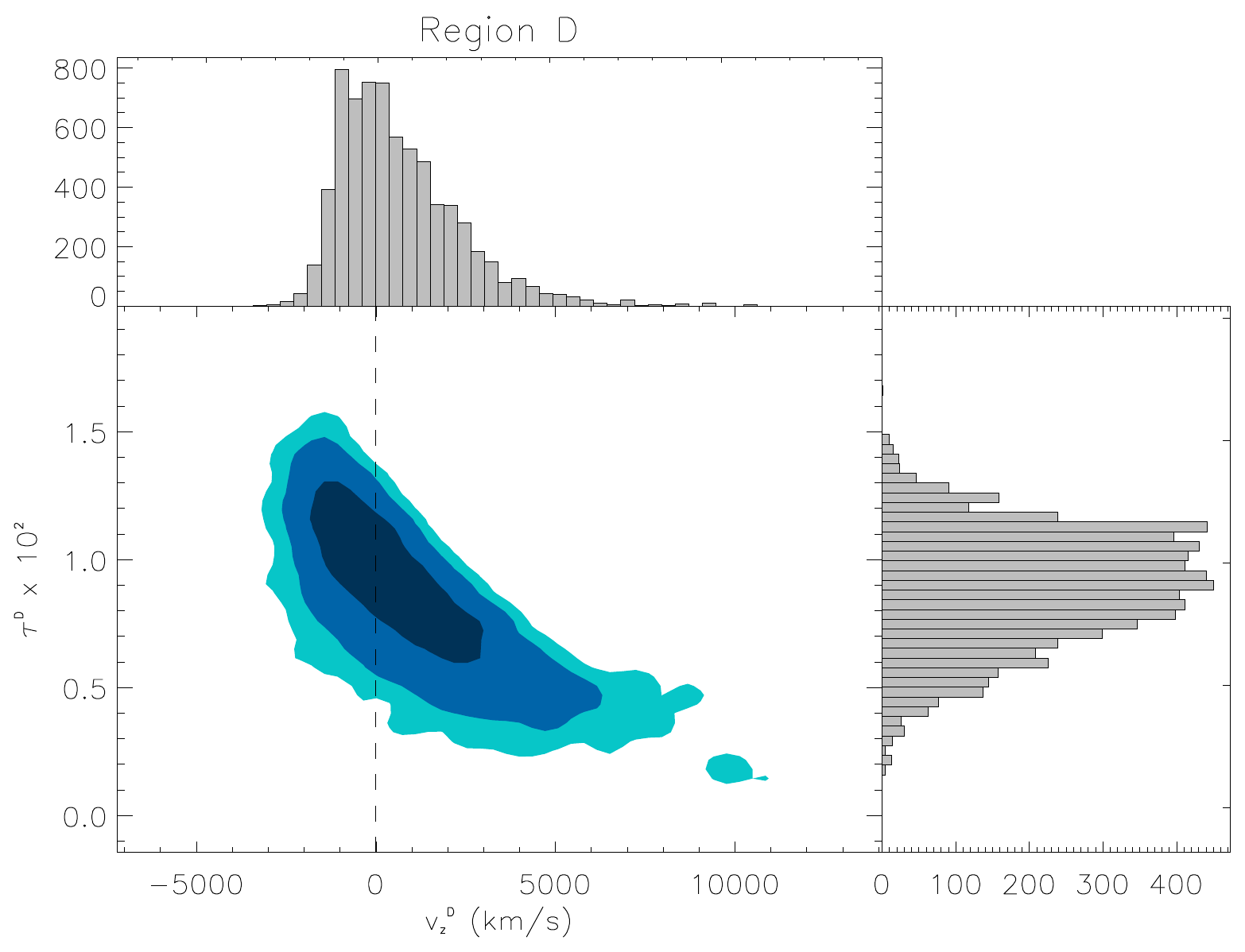}
\caption{\footnotesize{Constraints of the model on the ICM distribution in the plane line-of-sight velocity -- central optical depth, $v_z^{(i)}$ -- $\tau_0^{(i)}$, for subclusters B, C and D. The purple, dark blue and light blue are constraints contours at 68, 95 and 99\% confidence limit in the case of our baseline fit (F1, without X-ray imaging prior on the gas density). For subcluster B, the red contours are similar to the blue ones in the case of the fit F2 (with the X-ray imaging prior). The marginalized probability density distribution are also given by the histograms in the case of F1. The dashed line give the zero location of the velocity axis. As discussed in the text, given the limitation of the models to describe the cluster, we stress that the error contours do not necessarily reflect the true underlying uncertainties.}}
\label{fig:velocity_constraint}
\end{figure*}

The 260~GHz residual maps are on overall consistent with noise because of the lower signal-to-noise ratio at this frequency, but we notice that the shape of our model for subcluster B is too extended (or not peaked enough) compared to the 260~GHz data. It results in a positive residual around the kSZ peak, and a negative residual at the peak location, reaching $-2.3 \sigma$ for F1 and $-3.1 \sigma$ for F2. This could be due to the fact that subcluster B is itself made of two substructures, as discussed previously. Both B1 and B2 give rise to tSZ signal so that the signal at 150 GHz is relatively extended, while the kSZ, dominant at 260 GHz, is only seen toward B2, which would lead to a more compact signal than the one at 150 GHz. Alternatively, the constant temperature assumption could be invalid. If subcluster B is a cool-core system, as expected from \cite{Ma2009}, the pressure profile would be more extended than the density profile. The best-fit density profile would therefore be too extended because it is partly driven by the tSZ signal, i.e. the pressure. This is what we observe, and it is certainly one of the limitations of our model. The fact that the residual is larger for model F2 in the direction of subcluster B could also indicate that all the gas, as measured from the X-ray density, does not contribute to the kSZ signal we observe. Indeed, a lower amount of moving gas is preferred to explain simultaneously our 150 and 260 GHz data, assuming the given XMM-\textit{Newton} temperature (see also Figure \ref{fig:velocity_constraint}). Therefore, our data favor (with low significance though) the hypothesis in which subcluster B is made of two separate subclusters moving with different velocities along the line-of-sight.

The MCMC chains are used to compute the optical depth at the center of each subcluster, expressed as
\begin{equation}
\tau_0^{(i)} = \sqrt{\pi} \ \sigma_{\rm T} \ n_{e0}^{(i)} \ r_c^{(i)} \ \frac{\Gamma\left(\frac{3}{2} \beta^{(i)} -\frac{1}{2}\right)}{\Gamma\left(\frac{3}{2} \beta^{(i)}\right)}.
\label{eq:tau0}
\end{equation}
This is the quantity that best reflects the degeneracy between the ICM density distribution and the line-of-sight velocity (see equation \ref{eq:yksz}). Figure \ref{fig:velocity_constraint} provides the posterior probability density function in the line-of-sight velocity versus central optical depth plane for each subcluster B, C and D. The one dimensional marginalized distributions are also provided as histograms. 

As expected, in Figure \ref{fig:velocity_constraint} we observe a strong degeneracy between $\tau_0^{(i)}$ and $v_z^{(i)}$ for all the subclusters in our baseline fit F1 (in blue), even though it is slightly weaker for subcluster C. This degeneracy is broken in the case of F2 when using priors (subcluster B only) on the X-ray density, and the constraint are significantly improved (in red). While the line-of-sight velocity is fully compatible with zero for subcluster D, subcluster B and C show velocity distributions that are significantly different from zero.

The constraints from F1 and F2 are consistent, and the residual with respect to the best-fit model is not significantly degraded in case of F2 (see Figure \ref{fig:best_fit_maps}). Nevertheless, we observe a small tension (less than $2 \sigma$) in the case of subcluster B between the fits F1 and F2, which could be due to subcluster B being made of two subcomponents with different line-of-sight velocity, as discussed in the case of Figure \ref{fig:best_fit_maps}.

\begin{figure*}[h]
\centering
\includegraphics[width=0.49\textwidth]{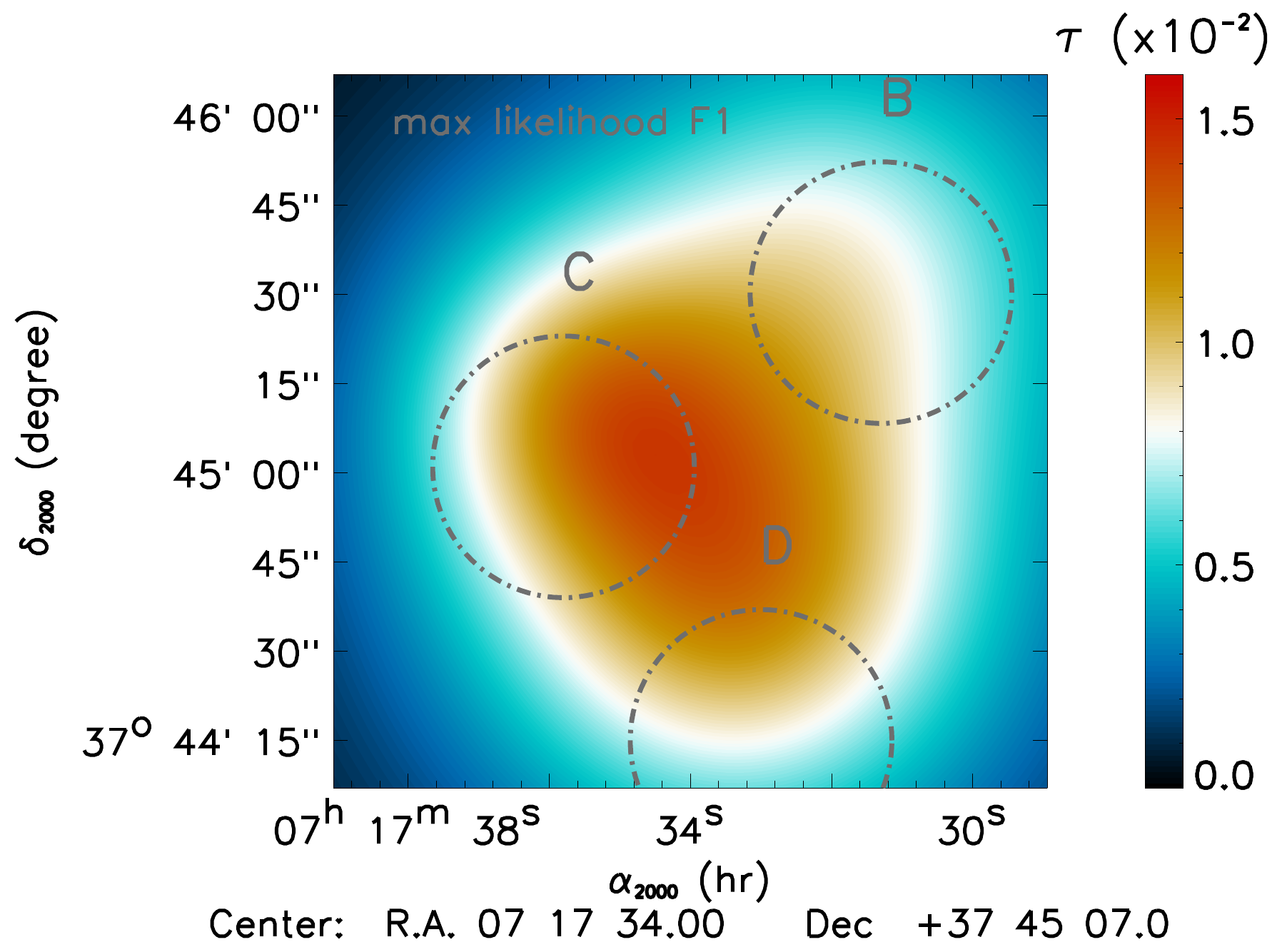}
\includegraphics[width=0.49\textwidth]{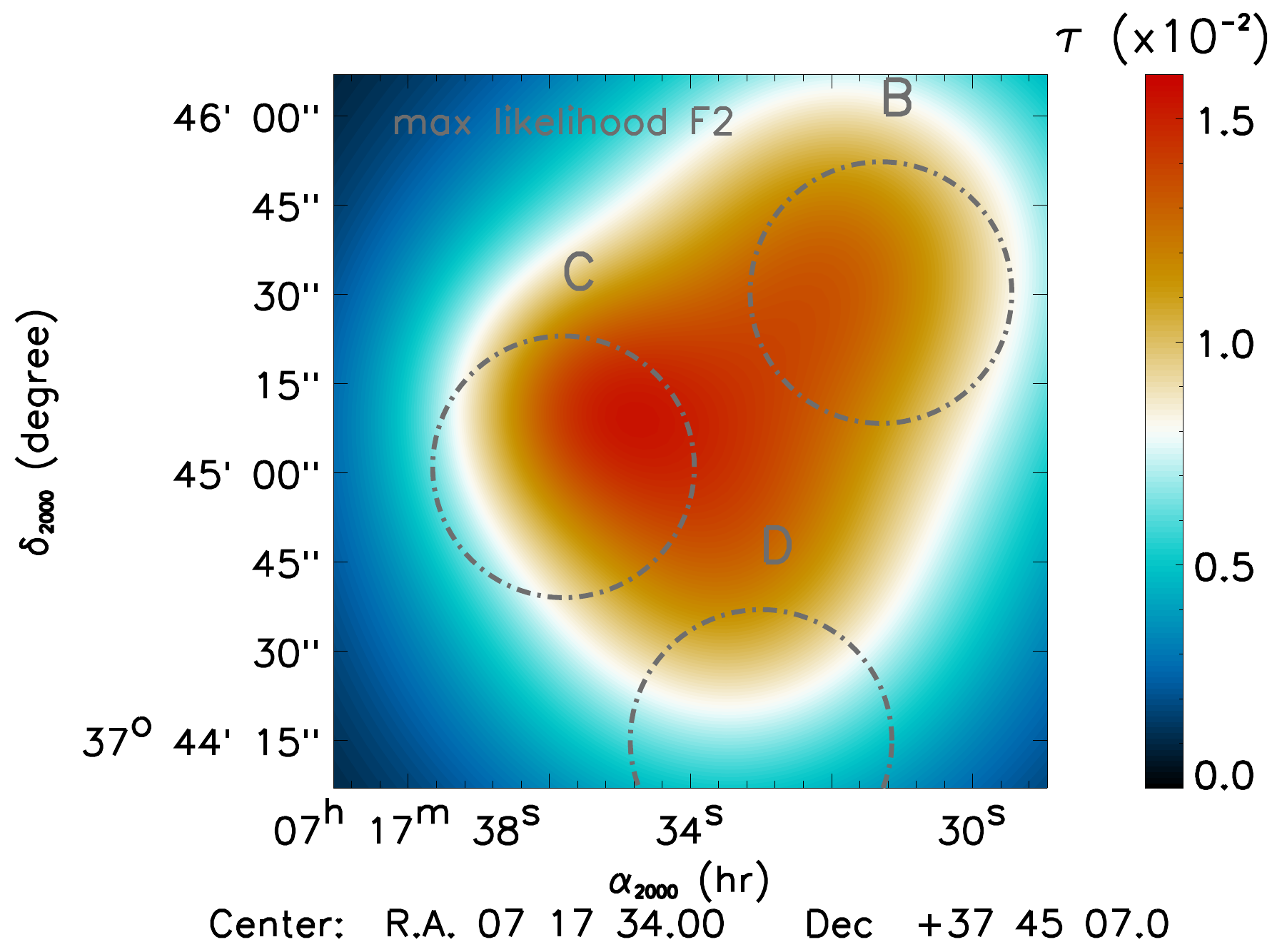}
\includegraphics[width=0.49\textwidth]{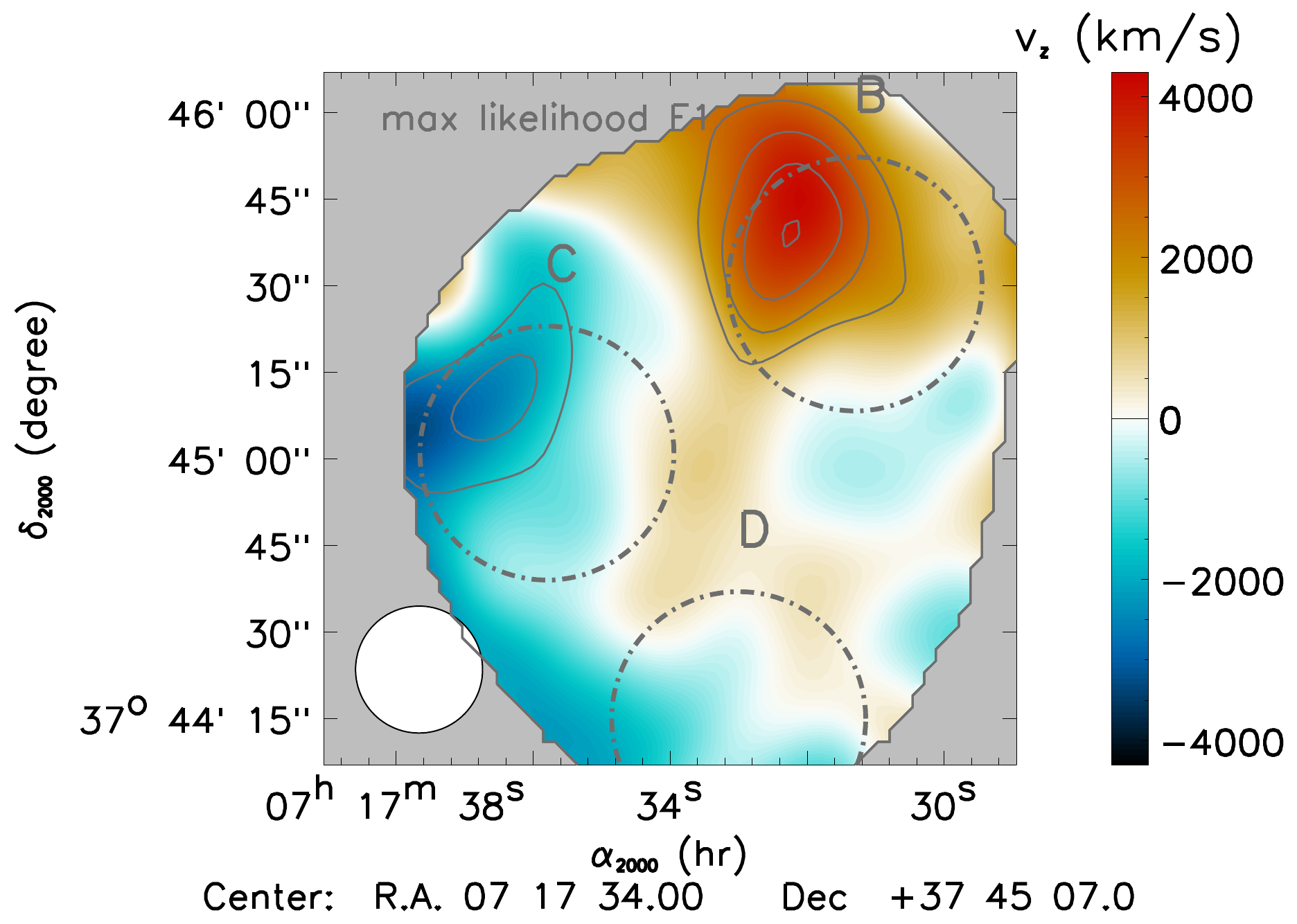}
\includegraphics[width=0.49\textwidth]{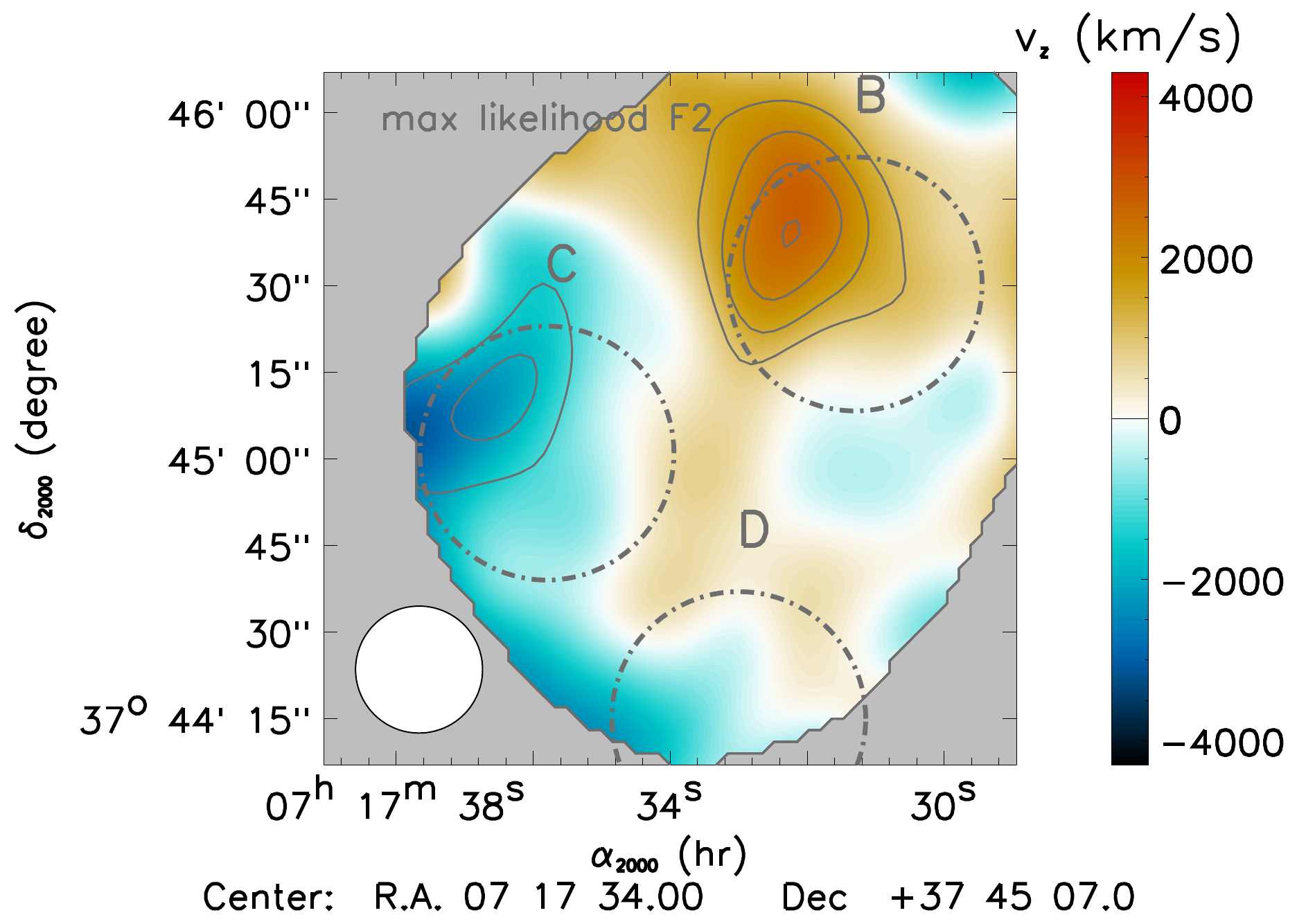}
\caption{\footnotesize{{\bf Top:} map of the best-fit optical depth model. {\bf Bottom:} projected gas line-of-sight velocity map. The resolution of the map is 22 arcsec FWHM as provided by the white circle in the bottom left corner. The left panels correspond to fit F1 and the right panels to fit F2. Regions where the rms of the noise is larger than 3 times the minimum value have been masked for display purpose. Contours show the detection level of the velocity (thus proportional to that of the kSZ map of Figure \ref{fig:tSZ_kSZ_maps}) and are spaced by $1 \sigma$, starting at $\pm 2 \sigma$. We stress that they do not reflect the true uncertainty on the velocity because they do not account for the statistical and systematic effect of the optical depth models. The field shown in here is smaller than the other ones displayed in this paper.}}
\label{fig:velocity_tau_map}
\end{figure*}

In addition to the central optical depths, our model allows us to compute the optical depth map of the cluster. This map is not independent of the best-fit velocities because of the $\tau_0^{(i)}$ -- $v_z^{(i)}$ degeneracies, but it does not depend significantly on the way the velocity is modeled because the density model is mostly driven by the tSZ signal. We use this $\tau$ model together with our kSZ map to obtain a gas line-of-sight velocity map via equation \ref{eq:yksz}, which we use to identify structures in the velocity and assess the validity of our assumption about the velocity being constant in each region. The optical depth map and the line-of-sight velocity map are shown in Figure \ref{fig:velocity_tau_map} for both fit F1 and F2. The kSZ map is subject to statistical uncertainties, while the optical depth model imposes systematic uncertainties not accounted for in the velocity map. Therefore, the error contours provide the signal-to-noise ratio but they do not reflect the true uncertainties of the map because it depends on the model (the signal-to-noise contours are thus identical for the two models, and to that of the kSZ map of Figure \ref{fig:tSZ_kSZ_maps}). In particular, the constant temperature assumption can affect the shape and amplitude of the subclusters in the optical depth map. As the uncertainties of the velocity map increase with decreasing optical depth, we mask pixels for which the noise rms is boosted by a factor larger than three, with respect to the minimum noise rms of the map.

In the case of our baseline fit, F1 (Figure \ref{fig:velocity_tau_map}, left panels), the optical depth presents a smooth structure which is maximum in region C. This is counter intuitive with respect to the X-ray imaging where the peak is clearly located toward subcluster B. The X-ray imaging being sensitive to $n_e^2$, this could indicate that subcluster B is highly peaked while subcluster C is more diffuse and extends further in space. Our best-fit model F1 being limited by degeneracies, we check this result by computing the best-fit model F2 (right panels). In this case, the optical depth map increases toward subcluster B by about 40\%, and its value between subcluster B, C and D stays almost constant. Nonetheless, we do not observe a peak toward subcluster B and our model suggests that it is more compact than subclusters C and D.

On the velocity map, we observe a dipolar distribution associated to subclusters B and C. The map is mostly negative between regions B and D but it is not significant. In the case of the fit F2, the velocity toward subcluster B reduces by about 40\%. The dipolar structure becomes therefore more symmetrical in terms of amplitude. We also note that unlike the optical depth model, the velocity map is convolved by the NIKA transfer function similarly to the kSZ map. We report our line-of-sight velocities from our two approaches, alongside velocities from other studies in Table \ref{tab:comparison_velocity_measurements}.

In summary, the best-fit residuals indicates that subcluster B is more complex than a single isothermal $\beta$-model. Our constraints indicate large velocities, but they are limited by degeneracies with the optical depth. They can be broken using X-ray imaging at the cost of extra assumptions. We finally extract a gas velocity map, showing a dipolar structure, using the kSZ map and our optical depth model.

\subsection{Comparison to previous results and discussion}
The first kSZ detection toward \mbox{MACS~J0717.5+3745} was made by Bolocam using 140 and 268 GHz observation \citep{Sayers2013}, following constraints on the velocity of galaxies by \cite{Ma2009}, as discussed in Section \ref{sec:Introduction}. Table \ref{tab:comparison_velocity_measurements} summaries velocity measurements performed toward \mbox{MACS~J0717.5+3745}.

The raw NIKA and Bolocam maps are difficult to directly compare because they probe different angular scales as a result of beam smoothing and large scale filtering. However, the best-fit model of \cite{Sayers2013} and our best-fit model are deconvolved from the respective transfer functions and can be quantitatively compared. While we consider a physical parametric description of the ICM, \cite{Sayers2013} use X-ray data to predict a tSZ template, to which they add a $\beta$-model component to include the kSZ signal. Each model has its advantages and disadvantages and both are limited in the case of a complex system like \mbox{MACS~J0717.5+3745}. Our model is clearly an idealization of the cluster, but it allows the building of a self-consistent description of the ICM. The model of \cite{Sayers2013}, on the other hand, does not assume any parametrization of the ICM for the tSZ signal, but it does use a $\beta$--model for the kSZ, and the two are not necessarily related to the same underlying gas density distribution. The tSZ template built by \cite{Sayers2013} is based on X-ray data and is therefore likely to better reproduce the projected geometry of the cluster than our model. However, it requires the conversion of $\int n_e^2 dl$ to $\int n_e dl$, which is done via an effective line-of-sight extent of the ICM, $\ell_{\rm eff}$. The latter is left as a free parameter, but is assumed to be a constant for the entire cluster extent. This is a strong assumption and our baseline model is not sensitive to such an effect. In fact, $\ell_{\rm eff}$ is directly related to the clumping of the gas, $< n_e^2> / <n_e>^2$, which is expected to be of the order of 1.4-1.6 for disturbed clusters \citep[e.g.,][]{Zhuravleva2013}. At the scales probed by NIKA, this would lead to local inaccuracies in the tSZ template ($\propto \sqrt{\ell_{\rm eff}}$) that could be mis-interpreted as kSZ signal. \cite{Sayers2013} assume that the temperature is constant along the line-of-sight, while we assume isothermal subclusters, both being based on X-ray spectroscopy. Finally, \cite{Sayers2013} use fixed coordinates for their $\beta$--model center and check a posteriori the effect of changing them within reasonable uncertainties. Due to the higher angular resolution of NIKA, we fully marginalize over the subcluster coordinates in our fit.

\cite{Sayers2013} also consider a nearly model independent approach, in which they use the direct SZ flux integration of their maps in the regions B and C (corrected for a model-based zero level), together with the isothermal assumption. Using this direct approach (referenced as "direct" in Table \ref{tab:comparison_velocity_measurements}), they obtain consistent results with respect to their model dependent approach (referenced as "model" in Table \ref{tab:comparison_velocity_measurements}), but with reduced significance.

Despite the fact that the cluster is modeled differently, our constraints on the gas line-of-sight velocity are compatible with the results from \cite{Sayers2013}. In the case of subcluster C, however, the two results are in mild tension ($\gtrsim 2 \sigma$). While this could easily be attributed to mis-modeling, it could also be due to differences in the observations. Indeed, our data are sensitive to scales as low as $\sim 20$ arcsec, and the constraint in region C is driven by the positive peak seen at 260 GHz at small scales. Due to Bolocam's larger beams, the corresponding scales are likely to be diluted over the cluster extent, leading to velocity averaged over a larger extension, and hence smaller. We also use our model to infer the integrated Compton parameter in the regions considered by \cite{Sayers2013}. We compute $Y(\Omega) = \int y_{\rm tSZ} d\Omega$, where $\Omega$ is taken as a 60 arcsec diameter circular region centered on subclusters B and C, using the coordinates of \cite{Sayers2013}. We obtain $\left(0.173 \pm 0.024\right) \times 10^{-3}$ arcmin$^2$ for region B ($\left(0.221 \pm 0.015\right) \times 10^{-3}$ arcmin$^2$ in the case of F2), and $\left(0.242 \pm 0.024\right) \times 10^{-3}$ arcmin$^2$ for region C. These values are in agreement within $1 \sigma$ ($2 \sigma$ in the case of F2) of the values measured by \cite{Sayers2013}.

Our constraints are also roughly compatible with the optical spectroscopic measurement of \cite{Ma2009}, within the error bars. However, as discussed in Section \ref{sec:Qualitative_comparison_to_other_wavelengths}, our data hint at substructure within subclusters B and C themselves. As a possible consequence, our measurement in the direction of subcluster C, which is dominated by the northern part of the region, indicates higher velocities than those measured in optical wavelengths. Additional optical spectroscopy would be necessary to better address the comparison between the galaxy velocities and the gas velocity.

The X-ray photon count image presents two main peaks, associated with subclusters B and C, which is elongated to subcluster D (Figure \ref{fig:Xray_all_maps}). The presence of hot adiabatically compressed gas between B and C/D suggests a pre-merger scenario between the two main structures. While the morphology of the temperature, forming a bar (see Figure \ref{fig:Xray_all_maps}), indicates that the merger is likely oriented on the plane of the sky, our kSZ data and the constraints we set on the gas velocity of each subcluster shows that there is a strong line-of-sight component to the merger axis. Our best-fit velocities depend on the gas density model and the error contours are large due to degeneracies, but our detection level is stable, and our data allow us to exclude null velocity at $5.1 \sigma$ for subcluster B, and $3.4 \sigma$ for subcluster C. Subcluster B is likely to be falling onto C \citep[the most massive of the subclusters, e.g.,][]{Limousin2015}. We note that the projected merger axis (northwest/southeast) coincides with the filament observed on the southeast of subcluster C \citep[e.g.,][]{Medezinski2013}. If the filament axis has a non-zero line-of-sight component, it could contribute to an explanation for the high velocity of B, via its gravitational interaction with the subcluster. While the main merger scenario supported by our data is relatively clear, the detail of each subcluster remains difficult to interpret because of complex structure within the subclusters themselves.

The root mean square of cluster peculiar velocities is expected to be of the order of 250 km/s \citep[e.g.,][]{Monteagudo2010}. The velocities we measure in \mbox{MACS~J0717.5+3745} deviate significantly from the bulk velocity flow. However, the free fall of massive subclusters onto one another from large distances is expected to lead to relative velocities of several thousands of km/s \citep{Sarazin2002}. The velocities measured in \mbox{MACS~J0717.5+3745} are high, but in line with expectations. Our results suggest that other well known mergers could be ideal targets for kinematic studies using kSZ mapping.

\begin{table*}[h]
\caption{{\footnotesize Comparison of the constraints on the velocity of each subclusters in \mbox{MACS~J0717.5+3745}. The NIKA results do not include calibration uncertainties or errors in the X-ray temperature of each subcluster. In addition, as discussed in the text, given the limitation of the models to describe the cluster, we stress that the error bars do not necessarily reflect the true underlying uncertainties.}}
\begin{center}
\begin{tabular}{c|c|c|c|c|c}
\hline
\hline
Subcluster & \multicolumn{5}{c}{$v_z$ (km/s)} \\
\hline
 & Galaxies & Gas (kSZ) & Gas (kSZ) & Gas (kSZ) & Gas (kSZ) \\
 & Optical \citep{Ma2009} & "Model" \citep{Sayers2013} & "Direct" \citep{Sayers2013} & NIKA (F1) & NIKA (F2) \\
\hline
A & $278^{+295}_{-339}$ & -- & -- & -- & -- \\
B & $3238^{+252}_{-242}$ & $3450 \pm 900$ & $2550 \pm 1050$ & $6607_{-2409}^{+3212}$ & $2058_{-447}^{+486}$ \\
C & $-733^{+486}_{-478}$ & $-550^{+1350}_{-1400}$ & $-500^{+1600}_{-1550}$ & $-4106_{-1104}^{+1594}$ & -- \\
D & $831^{+843}_{-800}$ & -- & -- & $150_{-1208}^{+1510}$ & -- \\
\hline
\end{tabular}
\end{center}
\label{tab:comparison_velocity_measurements}
\end{table*}

\section{Summary and conclusions}\label{sec:conclusions} 
The merging cluster, \mbox{MACS~J0717.5+3745}, was observed using the NIKA camera at the IRAM 30m telescope. The diffuse SZ signal is detected in the two NIKA bands at 150 and 260 GHz. We report the detection of radio and submillimeter point sources that contaminates our data. Using radio data from the literature and submillimeter data from SPIRE, in addition to the NIKA constraints themselves, we model the contaminant sources and extrapolate them to the NIKA bands to clean our data. One of the submillimeter sources detected with NIKA coincides with a high redshift lensed galaxy detected by HST. If confirmed, this would illustrate the strength of NIKA SZ observations in detecting high redshift sources lensed by clusters of galaxies by combining NIKA and \textit{Herschel} data \citep[see also][]{Adam2015,Adam2016}. In addition to point source removal, we take advantage of the high resolution of the NIKA observations to mask possible residuals.

To effectively use our data, we account for the different sources of noise including uncorrelated instrumental noise, atmospheric and electronic correlated noise residuals and a significant contribution from the CIB. While the non-astrophysical noise reduces by increasing the integration time, this is not the case for the CIB. Since our kSZ mapping is limited by statistics, itself dominated by the noise in our 260~GHz band, the CIB is already non negligible for the data we present (13.1 hours on target). This work thus shows that the CIB is likely to be a limiting factor for future high angular resolution kSZ mapping.

By combining the two NIKA bands, under the assumption that the SZ signal is the sum of a tSZ and a kSZ contribution, we extract the first map of the kSZ effect toward a cluster of galaxies. We account for relativistic corrections using an X-ray temperature map derived from XMM-{\it Newton} data. The kSZ map is dominated by a dipolar structure with peak signal in the direction of regions B (negative) and C (positive). This is in agreement with the merger scenario in which subcluster B is a compact core falling onto the main cluster, C. However, our data indicate that subcluster B and C are themselves likely to be made of subcomponents. Our results complement that of \cite{Sayers2013} at different scales. Apart from the relativistic corrections, which assume a constant temperature along the line-of-sight, the kSZ map we present in this paper is model independent.

We use the NIKA surface brightness maps in addition to X-ray temperature estimates of each subcluster to constrain the gas density and velocity distribution in \mbox{MACS~J0717.5+3745}. The gas density model assumes that each subcluster is described by a $\beta$--model with constant line-of-sight velocity and temperature. We perform a MCMC fit and provide constraints on the gas velocity toward each subcluster. While subcluster D is consistent with null line-of-sight velocity, subcluster B and C are moving away from us and toward us in the CMB reference frame, respectively. The relative significance of our measurement is stable and model independent, but the absolute velocity depends on our assumptions in the structure of the gas of \mbox{MACS~J0717.5+3745}. Our constraints are limited by degeneracies between the optical depth and velocity of the subclusters, and by the simplicity of our model, which is a strong idealization for a system known to be an extremely complex cluster of galaxies.

The results presented in this paper allow us to study the details of the gas motion in the cluster and complement other observations of \mbox{MACS~J0717.5+3745}. This shows the feasibility to study the distribution of the gas velocity within galaxy clusters, which is essential to our understanding of the dynamics of mergers at the substructure level.

\begin{acknowledgements}
We are thankful to the anonymous referee for useful comments that helped improve the quality of the paper.
We are grateful to Jack Sayers for helping us obtain the \textit{Herschel} SPIRE catalog we use in this paper.
We would like to thank the IRAM staff for their support during the campaigns. 
The NIKA dilution cryostat has been designed and built at the Institut N\'eel. In particular, we acknowledge the crucial contribution of the Cryogenics Group, and in particular Gregory Garde, Henri Rodenas, Jean Paul Leggeri, Philippe Camus. 
This work has been partially funded by the Foundation Nanoscience Grenoble, the LabEx FOCUS ANR-11-LABX-0013 and the ANR under the contracts "MKIDS", "NIKA" and ANR-15-CE31-0017. 
This work has benefited from the support of the European Research Council Advanced Grant ORISTARS under the European Union's Seventh Framework Programme (Grant Agreement no. 291294).
We acknowledge fundings from the ENIGMASS French LabEx (B. C. and F. R.), the CNES post-doctoral fellowship program (R. A.), the CNES doctoral fellowship program (A. R.) and the FOCUS French LabEx doctoral fellowship program (A. R.).
E.P. acknowledges the support of the French Agence Nationale de la Recherche under grant ANR-11-BS56-015.
This research has received funding from the European Research Council under the European Union’s Seventh Framework Programme (FP7/2007−2013)/ERC grant agreement No. 340519.
\end{acknowledgements}

\bibliography{biblio_MACSJ0717}

\end{document}

%% file: listeauthors.tex
\author{R.~Adam\inst{\ref{inst1}, \ref{inst8}}\thanks{Corresponding author: R\'emi Adam, \url{remi.adam@oca.eu}}	
\and I.~Bartalucci\inst{\ref{inst2}}
\and G.W.~Pratt\inst{\ref{inst2}}
\and P.~Ade\inst{\ref{inst3}}		
\and P.~Andr\'e\inst{\ref{inst2}}		
\and M.~Arnaud\inst{\ref{inst2}}
\and A.~Beelen\inst{\ref{inst4}}		
\and A.~Beno\^it\inst{\ref{inst5}}	
\and A.~Bideaud\inst{\ref{inst3}}	
\and N.~Billot\inst{\ref{inst6}}		
\and H.~Bourdin\inst{\ref{inst7}}
\and O.~Bourrion\inst{\ref{inst8}}	
\and M.~Calvo\inst{\ref{inst5}}		
\and A.~Catalano\inst{\ref{inst8}}	
\and G.~Coiffard\inst{\ref{inst9}}	
\and B.~Comis\inst{\ref{inst8}}		
\and A.~D'Addabbo\inst{\ref{inst5}, \ref{inst10}}
\and M.~De Petris\inst{\ref{inst10}}
\and J.~D\'emocl\`es\inst{\ref{inst2}}
\and F.-X.~D\'esert\inst{\ref{inst11}}	
\and S.~Doyle\inst{\ref{inst3}}		
\and E.~Egami\inst{\ref{inst12}}
\and C.~Ferrari\inst{\ref{inst1}}
\and J.~Goupy\inst{\ref{inst5}}		
\and C.~Kramer\inst{\ref{inst6}}	
\and G.~Lagache\inst{\ref{inst13}}
\and S.~Leclercq\inst{\ref{inst9}}	
\and J.-F.~Mac\'ias-P\'erez\inst{\ref{inst8}}	
\and S.~Maurogordato\inst{\ref{inst1}}
\and P.~Mauskopf\inst{\ref{inst3}, \ref{inst14}}	
\and F.~Mayet\inst{\ref{inst8}}		
\and A.~Monfardini\inst{\ref{inst5}}	
\and T.~Mroczkowski\inst{\ref{inst15}}
\and F.~Pajot\inst{\ref{inst4}}		
\and E.~Pascale\inst{\ref{inst3}}	
\and L.~Perotto\inst{\ref{inst8}}		
\and G.~Pisano\inst{\ref{inst3}}		
\and E.~Pointecouteau\inst{\ref{inst16}, \ref{inst17}}
\and N.~Ponthieu\inst{\ref{inst11}}	
\and V.~Rev\'eret\inst{\ref{inst2}}	
\and A.~Ritacco\inst{\ref{inst8}}	
\and L.~Rodriguez\inst{\ref{inst2}}	
\and C.~Romero\inst{\ref{inst9}}	
\and F.~Ruppin\inst{\ref{inst8}}		
\and K.~Schuster\inst{\ref{inst9}}	
\and A.~Sievers\inst{\ref{inst6}}	
\and S.~Triqueneaux\inst{\ref{inst5}}
\and C.~Tucker\inst{\ref{inst3}}		
\and M.~Zemcov\inst{\ref{inst18},\ref{inst19}}
\and R.~Zylka\inst{\ref{inst9}}}		

\institute{
Laboratoire Lagrange, Universit\'e C\^ote d'Azur, Observatoire de la C\^ote d'Azur, CNRS, Blvd de l'Observatoire, CS 34229, 06304 Nice cedex 4, France
  \label{inst1}
    \and
Laboratoire de Physique Subatomique et de Cosmologie, Universit\'e Grenoble Alpes, CNRS/IN2P3, 53, avenue des Martyrs, Grenoble, France
  \label{inst8}
  \and
Laboratoire AIM, CEA/IRFU, CNRS/INSU, Universit\'e Paris Diderot, CEA-Saclay, 91191 Gif-Sur-Yvette, France 
  \label{inst2}
  \and
Astronomy Instrumentation Group, University of Cardiff, UK
  \label{inst3}
\and
Institut d'Astrophysique Spatiale (IAS), CNRS and Universit\'e Paris Sud, Orsay, France
  \label{inst4}
\and
Institut N\'eel, CNRS and Universit\'e Grenoble Alpes, France
  \label{inst5}
\and
Institut de RadioAstronomie Millim\'etrique (IRAM), Granada, Spain
  \label{inst6}
\and
Dipartimento di Fisica, Universit\`a degli Studi di Roma 'Tor Vergata', via della Ricerca Scientifica, 1, I-00133 Roma, Italy
  \label{inst7}
\and
Institut de RadioAstronomie Millim\'etrique (IRAM), Grenoble, France
  \label{inst9}
\and
Dipartimento di Fisica, Sapienza Universit\`a di Roma, Piazzale Aldo Moro 5, I-00185 Roma, Italy
  \label{inst10}
\and
Institut de Plan\'etologie et d'Astrophysique de Grenoble (IPAG), CNRS and Universit\'e Grenoble Alpes, France
  \label{inst11}
    \and
Steward Observatory, University of Arizona, 933 North Cherry Avenue, Tucson, AZ 85721, USA
  \label{inst12}
  \and
Aix Marseille Universit\'e, CNRS, LAM (Laboratoire d'Astrophysique de Marseille) UMR 7326, 13388, Marseille, France
  \label{inst13}
  \and
School of Earth and Space Exploration and Department of Physics, Arizona State University, Tempe, AZ 85287
  \label{inst14}
  \and
European Organization for Astronomical Research in the Southern hemisphere, Karl-Schwarzschild-Str. 2, D-85748 Garching b. M\"unchen, Germany 
  \label{inst15}
  \and
Universit\'e de Toulouse, UPS-OMP, Institut de Recherche en Astrophysique et Plan\'etologie (IRAP), Toulouse, France
  \label{inst16}
\and
CNRS, IRAP, 9 Av. colonel Roche, BP 44346, F-31028 Toulouse cedex 4, France 
  \label{inst17}
\and
Center for Detectors, School of Physics and Astronomy, Rochester Institute of Technology, Rochester, NY 14623, USA
  \label{inst18}
  \and
Jet Propulsion Laboratory, 4800 Oak Grove Drive, Pasadena, CA 91109, USA
    \label{inst19}    
}